\documentclass[aip, jcp, citeautoscript, amsmath, amssymb, reprint,
]{revtex4-2}
\usepackage[title]{appendix}
\usepackage{graphicx, tikz, orcidlink}
\usepackage{epstopdf}
\usepackage{ascmac}
\usepackage{ulem}
\usepackage{cancel}
\usepackage{here}
\usepackage{physics}
\usepackage{bm}
\usepackage{comment}
\usepackage{booktabs}

\usepackage{amsthm,color}
\newtheoremstyle{query}%
{}{}
{\color{red}}
{}
{\sffamily\bfseries}{:}{12pt}
{}
\theoremstyle{query}
\newtheorem{aq}{Author Query/Comment}

\newcommand{\baq}{\begin{aq}}
\newcommand{\eaq}{\end{aq}}

\begin{document}
\title{System-Bath Modeling in Vibrational Spectroscopy via Molecular Dynamics: A Machine Learning Framework for Hierarchical Equations of Motion (HEOM)}
\date{Last updated: \today}

\author{Kwanghee Park\orcidlink{0009-0005-5223-8234}}
\email[Author to whom correspondence should be addressed: ]{park.kwanghee.75n@st.kyoto-u.ac.jp}
\affiliation{Department of Chemistry, Graduate School of Science,
Kyoto University, Kyoto 606-8502, Japan}
\author{Ju-Yeon Jo\orcidlink{0000-0003-1646-7310}}
\affiliation{Graduate School of Energy Science,Kyoto University, Kyoto 606-8502, Japan}

\author{Yoshitaka Tanimura\orcidlink{0000-0002-7913-054X}}
\email[Author to whom correspondence should be addressed: ]{tanimura.yoshitaka.5w@kyoto-u.jp}
\affiliation{Department of Chemistry, Graduate School of Science,
Kyoto University, Kyoto 606-8502, Japan}

\begin{abstract}
{Molecular vibrations in solutions, especially OH stretching and bending in water, drive ultrafast energy relaxation and dephasing in chemical and biological systems}. 
We present a machine learning approach for constructing system-bath models  {of intramolecular vibrations in solution, compatible with quantum simulations via the hierarchical equations of motion (HEOM). Using classical molecular dynamics trajectories generated with a force field specifically developed for quantum molecular dynamics, the model captures anharmonic mode coupling and non-Markovian dissipation through spectral distribution functions (SDFs). These features, in turn, enable quantum mechanical treatment of ultrafast energy relaxation, vibrational dephasing, and thermal excitation within the HEOM framework.. The trained model yields physically interpretable parameters, validated against infrared spectra. Notably, combining Brownian oscillator and Drude SDFs---representing inter- and intramolecular vibrational modes---significantly improves learning performance and supports rigorous simulation of nonlinear vibrational spectroscopy.}
\end{abstract}
\maketitle

\section{Introduction}
\label{sec:intro}
Modern molecular laser spectroscopy involves the {sequential irradiation of a sample with ultrafast laser pulses at precisely} controlled time intervals. {The resulting spectroscopic observables are governed by nonlinear response functions and reflect complex intermolecular and intramolecular dynamics. Interpreting these spectra remains a persistent challenge due to the intricate nature of the underlying molecular interactions.}\cite{mukamel1999principles,Cho2009,Hamm2011ConceptsAM,HammPerspH2O2017,JansenChoShinji2DVPerspe2019}

{While} molecular dynamics (MD) simulations {hold promise for capturing intricate spectral signatures---particularly in two-dimensional (2D) spectroscopy---their foundation in classical mechanics inherently constrains the precision of peak positions and line shapes.\cite{Shinji2DRaman2006,HT06JCP,LHDHT08JCP,HT08JCP,Wei2015Nagata2DRamanTHz,IHT14JCP,IHT16JPCL} To accurately capture} nonlinear phenomena such as the 2D infrared (IR) echo spectrum, it is essential to incorporate the quantum entanglement {between molecular motion and its surrounding environment,\cite{SkinnerCPL2004,IT08CP,TI09ACR} herein referred to as  ``bathentanglement.''\cite{T20JCP} Thus,}  a quantum mechanical framework---such as the hierarchical equations of motion (HEOM)---is indispensable for elucidating the underlying dynamics that manifest as observable spectral features.\cite{JianLiu2018H2OMP,Imoto_JCP135,medders2015irraman,Paesan2018H2OCMD}

Model-based approaches have emerged as practical, flexible, and effective tools for simulating nonlinear vibrational spectra.\cite{SkinnerCPL2004,IT08CP,TI09ACR} By introducing a thermal bath that reflects environmental influences on primary vibrational modes, these models enable the computation of various nonlinear spectra. 
One such strategy describes vibrational relaxation and dephasing using Brownian, exciton, and stochastic models.\cite{ChoOhmine1994,mukamel1999principles,Mukamel2009,SkinnerStochs2003,Skinner2005,JansenSkinnerJCP2010} These models incorporate energy states and noise spectral distribution functions (SDFs) that are obtained from MD simulations{ and spectroscopic experiments.}

While conventional approaches often struggle to capture intricate vibrational mode couplings and non-Markovian environmental effects, the multi-mode anharmonic Brownian model provides a versatile framework grounded in the theory of open quantum systems. This model systematically incorporates anharmonic interactions among vibrational modes, along with both homogeneous and inhomogeneous spectral broadening.\cite{ST11JPCA,IIT15JCP,IT16JCP,TT23JCP1,TT23JCP2,HT25JCP1,HT25JCP2} Its dynamics are computed using the hierarchical equations of motion (HEOM), a numerically ``exact'' formalism that rigorously accounts for the effects of thermal environmental beyond perturbative and Markovian limits.\cite{T06JPSJ,T20JCP}

Spectral simulations based on the HEOM formalism have been successfully applied to a range of problems, including 2D vibrational spectroscopies. \cite{ST20JPSJ,TS20JPSJ,T06JPSJ,T20JCP} The model description and computational accuracy have been validated, but the overall performance depends critically on the choice of model parameters.

To date, {the selection of parameters} and SDFs within this framework has relied predominantly on empirical tuning or MD simulations, with the primary objective of reproducing experimentally observed spectral features across a range of modalities---including infrared absorption, off-resonant Raman, two-dimensional (2D) Raman,\cite{TS20JPSJ} 2D terahertz-Raman,\cite{IIT15JCP,HT25JCP2} 2D infrared-Raman,\cite{IT16JCP,TT23JCP1} and 2D infrared spectroscopy.\cite{TT23JCP2,HT25JCP1} While this {heuristic approach} has yielded qualitative agreement in many cases, it remains inherently unsystematic and computationally intensive. Furthermore, its generalizability is constrained by a fundamental limitation: the intensity of spectroscopic observables does not necessarily correlate with the intrinsic strength of individual vibrational modes. This ambiguity becomes especially pronounced in scenarios where distinct vibrational modes exhibit degenerate frequencies or when spectroscopically silent or dark modes are present, thereby obfuscating the interpretation of spectral signatures.

This approach has been demonstrated using liquid water,\cite{IIT15JCP,IT16JCP,TT23JCP1,TT23JCP2,HT25JCP1,HT25JCP2} a system for which extensive 2D spectroscopic data are available from both experiments and simulations. Our method complements ongoing developments in classical and quantum HEOM-based computational schemes.

In this study, we employ machine learning (ML) framework to directly extract model parameters and SDFs of thermal baths from MD trajectories.\cite{UT20JCTC,UT21JCTC} Previous {efforts to construct such models 
resulted in SDFs that were too intricate to be incorporated into the HEOM framework, thereby precluding spectral simulations.}\cite{UT20JCTC} Here, we {retain the model structure used} in earlier studies but refine the parameters by constraining the SDFs to forms compatible with the HEOM formalism.
{We demonstrate this approach using} liquid water, a system for which extensive 2D spectroscopic data are available from both experiments and simulations. Our method complements ongoing developments in classical and quantum HEOM-based computational schemes.

This paper is structured as follows. Section~\ref{sec:theory} introduces the multimode anharmonic Brownian model with nonlinear system-bath (S-B) interactions, with particular emphasis on molecular liquids as the target system. The machine learning algorithm used to extract the model parameters is also described. Section~\ref{sec:result} presents the evaluation and analysis of the model parameters for three intramolecular vibrational modes of liquid water. Finally, Sec.~\ref{sec:conclusion} provides concluding remarks.

\section{Theory}
\label{sec:theory}

\subsection{Multimode anharmonic Brownian (MAB) model}
\label{sub:MMBO}

To simulate both linear and nonlinear vibrational spectra of molecules in condensed phases, we 
{adopt} the multimode anharmonic Brownian (MAB) model. In this framework, anharmonic intramolecular vibrational modes are nonlinearly coupled to surrounding molecular modes, which are treated as multiple bath systems. Each bath is represented by an ensemble of harmonic oscillators. This model provides a versatile and systematic approach to incorporating anharmonic mode–mode coupling in the context of open quantum dynamics theory.

Homogeneous and inhomogeneous broadening effects are {accounted for via} nonlinear and non-Markovian system-bath (S-B) interactions.\cite{KT02JCP1,KT04JCP,IT06JCP,OT97PRE} 
The HEOM formalism enables the direct computation of 2D vibrational spectra for {a range of} molecular liquids, including water.\cite{ST20JPSJ,TS20JPSJ,T06JPSJ,T20JCP}

The total Hamiltonian of the MAB model is formulated as follows:\cite{ST11JPCA,IIT15JCP,IT16JCP,TT23JCP1,TT23JCP2,HT25JCP1,HT25JCP2}
\begin{eqnarray}
  \hat{H}_\mathrm{tot}&&= \sum_{s} \qty( \hat{H}_{A}^{(s)} + \sum_{s>s'} \hat{U}_{ss'}\qty(\hat{q}_s, \hat{q}_{s'}))  \nonumber  \\
  &&+ \sum_{s} \sum_{j_s}\qty[\frac{\hat{p}_{j_s}^{2}}{2m_{j_s}}+\frac{m_{j_s}\omega_{j_s}^{2}}{2}\left(\hat{x}_{j_s}-{\frac{\alpha_{j_s} \hat{V}_s(\hat{ q}_s)}{m_{j_s}\omega_{j_s}^2} } \right)^2], \nonumber  \\
  \label{eqn:H_total}
\end{eqnarray}
where the Hamiltonian for the $s$th mode is defined as
\begin{eqnarray}
\hat{H}_{A}^{(s)}= \frac{\hat{p}_s^{2}}{2m_s} +\hat U_s(\hat{q}_s)
\label{eqn:SH}
\end{eqnarray}
with a mass $m_s$, a coordinate ${\hat{q}_s}$, and a momentum ${\hat p_s}$. The anharmonic potential for the $s$th mode is given by
\begin{eqnarray}
{\hat U_s(\hat{q}_s)= \frac{1}{2} m_s \omega_s^2 \hat{q}_s^2 +\frac{1}{3!}g_{s^3}\hat{q}_{s}^3,}
\label{eqn: Potenential s}
\end{eqnarray}
where $\omega_s$ is the vibrational frequency and $g_{s^3}$ denotes the cubic anharmonicity.

The interaction potential between the $s$th and $s'$th vibrational modes is formulated as
\begin{eqnarray}
\hat{U}_{ss'}(\hat{q}_s, \hat{q}_{s'}) &&= g_{s{s'}}\hat{q}_s\hat{q}_{s'} \nonumber \\
&&+ \frac{1}{6}  \qty(g_{s^2s'}\hat{q}_s^2 \hat{q}_{s'} + g_{s{s'}^2} \hat{q}_s \hat{q}_{s'}^2 ),
\label{eqn: Potential ss'}
\end{eqnarray}
where $g_{s{s'}}$ denotes the second-order anharmonic coupling coefficient, while $g_{s^2s'}$ and $g_{s{s'}^2}$ characterize the third-order contributions.  
Each oscillator in the $s$th bath, labeled by index $j_s$, is characterized by its momentum ${p}_{j_s}$, coordinate ${x}_{j_{s}}$, mass $m_{j_{s}}$, frequency $\omega _{{j_s}}$, and coupling strength $\alpha _{j_s}$. To preserve the system's translational invariance, a counter term is incorporated into each bath, as illustrated in Ref.~\onlinecite{TW91PRA}.

While 2D spectroscopy has {elucidated the roles of vibrational relaxation and dephasing as key mechanisms driving molecular motion,\cite{SkinnerI,SkinnerII,SkinnerIII,SkinnerIV,SkinnerV,SkinnerVI,SkinnerVII,YagasakiSaitoJCP20082DIR,YagasakiSaitoJCP2011Relax,Yagasaki_ARPC64,ImotXanteasSaitoJCP2013H2O,Imotobend-lib2015} incorporating these effects into theoretical models requires careful treatment of non-Markovian S–B interactions---particularly those of the linear–linear (LL) and square–linear (SL) types\cite{KT02JCP1,KT04JCP,IT06JCP,OT97PRE}---as well as anharmonic mode–mode couplings.}
Accordingly, we describe the system component of the S–B interaction,
$\hat{V}_{s}({\hat{q}_s})$  in terms of LL and SL contributions as
\begin{eqnarray}
  \hat{V}_{s}(\hat{q}_s)\equiv \hat{V}^{(s)}_{\mathrm{LL}}\hat{q}_s+ \frac{1}{2} \hat{V}^{(s)}_{\mathrm{SL}}\hat{q}_s^{2},
  \label{eqn:LLSL}
\end{eqnarray}
where $V^{(s)}_{\mathrm{LL}}$ and $V^{(s)}_{\mathrm{SL}}$ denote the respective coupling strengths.\cite{T06JPSJ,TI09ACR}
{While Eq. \eqref{eqn:H_total} has been used to describe the collective coordinates of the molecular liquid, this study adopts a single-molecule perspective. Accordingly, we introduce the re-oriented bath coordinate, $\tilde{x}_{j_s} = x_{j_s} - ({\alpha_{j_s}V_s({q_s})})/({2m_{j_s}\omega^2_{j_s}})$ and rewrite the total Hamiltonian Eq. \eqref{eqn:H_total} as follows:
\begin{align}
\hat{H}_\mathrm{tot}&=\hat{H}_{S}  - V_{s}(\hat {q_s})\sum _{j_s}\alpha _{j_s}{\hat x}_{j_s}   \nonumber \\ 
&+ \sum _{j_s}\left(\frac{p_{j_s}^2}{2m_{j_s}}+\frac{m_{j_s}\omega _{j_s}^2\tilde{x}_{j_s}^2}{2} \right)  \label{eq:h_bathint}
\end{align}
where $\hat{H}_{S}\equiv \sum_s \hat{H}_{A}^{(s)} + \sum_{s>s'} \hat{U}_{ss'}\qty(\hat{q}_s, \hat{q}_{s'})$.
The dynamics of the baths can be characterized via the SDFs and the inverse temperature $\beta=1/k_{\mathrm{B}}T$, where $k_{\mathrm{B}}$ is  Boltzmann's constant and $T$ is the thermodynamic temperature. These quantities {enter the theory} through the symmetrized correlation function and the relaxation function of the collective coodinate of the bath associated with the $s$th mode defined as $\hat{X}_{s} \equiv \sum_{j_s} \alpha_{j_s} \hat{x}_{j_s}$. The antisymmetric and symmetric correlation functions of $\hat{X}_{s}$ are then expressed as
 $iL_1^{(s)} (t) = i\langle {\hat X}_{s}(t) {\hat X}_{s}  -{\hat X}_{s} {\hat X}_{s}  (t) \rangle_B/\hbar$ and $L_2^{(s)} (t) =  \langle {\hat X}_{s}(t) {\hat X}_{s} +{\hat X}_{s} {\hat X}_{s}  (t)  \rangle_B/2$, where $\hat{X}_{s}(t)$ denotes the Heisenberg representation of $\hat{X}_{s}$ under the bath Hamiltonian $\hat{H}_B^{(s)}$, and $\langle \cdots \rangle_B$ indicates the thermal average over the bath degrees of freedom.\cite{T06JPSJ,TK89JPSJ1} }
 
The SDF for the $s$th mode is defined as
\begin{eqnarray}
J_{\text{s}}(\omega)= \sum_{j_s} \frac{\alpha_{j_s}^2}{2m_{j_s} \omega_{j_s}} \delta(\omega - \omega_{j_s}).
\end{eqnarray}
{In terms of SDF, we have $i L_1^{(s)} (t) = 2 i \int d\omega J_{s}(\omega) \sin ( \omega t )$ and $L_2^{(s)} (t) = \hbar \int d\omega J_{s}(\omega) \coth(\beta\hbar \omega / 2 ) \cos ( \omega t)$.}
In general, SDFs can exhibit {intricate} structures.\cite{UT20JCTC} However, the HEOM {framework imposes limitations on} the functional form of SDFs that can be accommodated.\cite{T20JCP} In this work, we consider two representative forms:
\begin{itemize}
  \item[(a)] \textbf{Drude SDF}, widely employed in 2D spectral simulations\cite{IIT15JCP,HT25JCP2,IT16JCP,TT23JCP1,ST11JPCA,TT23JCP2,HT25JCP1} and supported by several source codes,\cite{TT23JCP2,HT25JCP2} is expressed as
  \begin{equation}
  J_s^{\rm D}(\omega) = \frac{m_s \zeta_s^{\rm D}}{2\pi} \frac{(\gamma_s^{\rm D})^2 \omega}{\omega^2 + (\gamma_s^{\rm D})^2},
  \label{eq:drude}
  \end{equation}
  where $\zeta_s^{\rm D}$ is the S-B coupling strength and $\gamma_s^{\rm D}$ characterizes the spectral width, which is inversely related to the vibrational dephasing time $\tau_s = 1/\gamma_s^{\rm D}$.

  \item[(b)] \textbf{Brownian Oscillator (BO) + Drude SDF}, which incorporates both Drude and underdamped BO components,\cite{T12JCP} is given by
  \begin{align}
  J_s(\omega) &= \frac{m_s \zeta_s^{\rm D}}{2\pi} \frac{(\gamma_s^{\rm D})^2 \omega}{\omega^2 + (\gamma_s^{\rm D})^2} \nonumber \\
  &+ \frac{m_s \zeta_s^{\rm B}}{2\pi} \frac{(\gamma_s^{\rm B})^2 (\omega_s^{\rm B})^2 \omega}{[(\omega_s^{\rm B})^2 - \omega^2]^2 + (\gamma_s^{\rm B})^2 \omega^2},
  \label{eq:drudePlusBO}
  \end{align}
where $\zeta_s^{\rm B}$ and $\gamma_s^{\rm B}$ denote the coupling strength and inverse correlation time of the BO bath, respectively, while $\omega_s^{\rm B}$ represents its central frequency.\cite{TT09JPSJ,TT10JCP,DT15JCP} The BO component typically accounts for spectrally inactive silent modes that lie outside the observation window.
\end{itemize}

The dipole moment and polarizability operators are defined as
\begin{equation}
  \hat{\mu} = \sum_{s} \mu_s \hat{q}_s + \sum_{s,s'} \mu_{ss'} \hat{q}_s \hat{q}_{s'}
  \label{eq:mu}
\end{equation}
and
\begin{equation}
  \hat{\Pi} = \sum_{s} \Pi_s \hat{q}_s + \sum_{s,s'} \Pi_{ss'} \hat{q}_s \hat{q}_{s'},
  \label{eq:Pi}
\end{equation}
respectively, where $\mu_s$ and $\mu_{ss'}$ denote the linear and nonlinear components of the dipole moment, {respectively, and} $\Pi_s$ and $\Pi_{ss'}$ denote the corresponding elements of the polarizability.

The vibrational modes interact via mechanical anharmonic coupling (MAHC), characterized by the coefficients $g_{s^2 s'}$ and $g_{ss'^2}$, and electric anharmonic coupling (EAHC), described by the nonlinear dipole and polarizability terms $\mu_{ss'}$ and $\Pi_{ss'}$.\cite{IT16JCP}

\subsection{Constructing the MAB Model via a Machine Learning (ML) Approach}

\begin{figure*}[!t]
  \centering
  \includegraphics[scale=0.9]{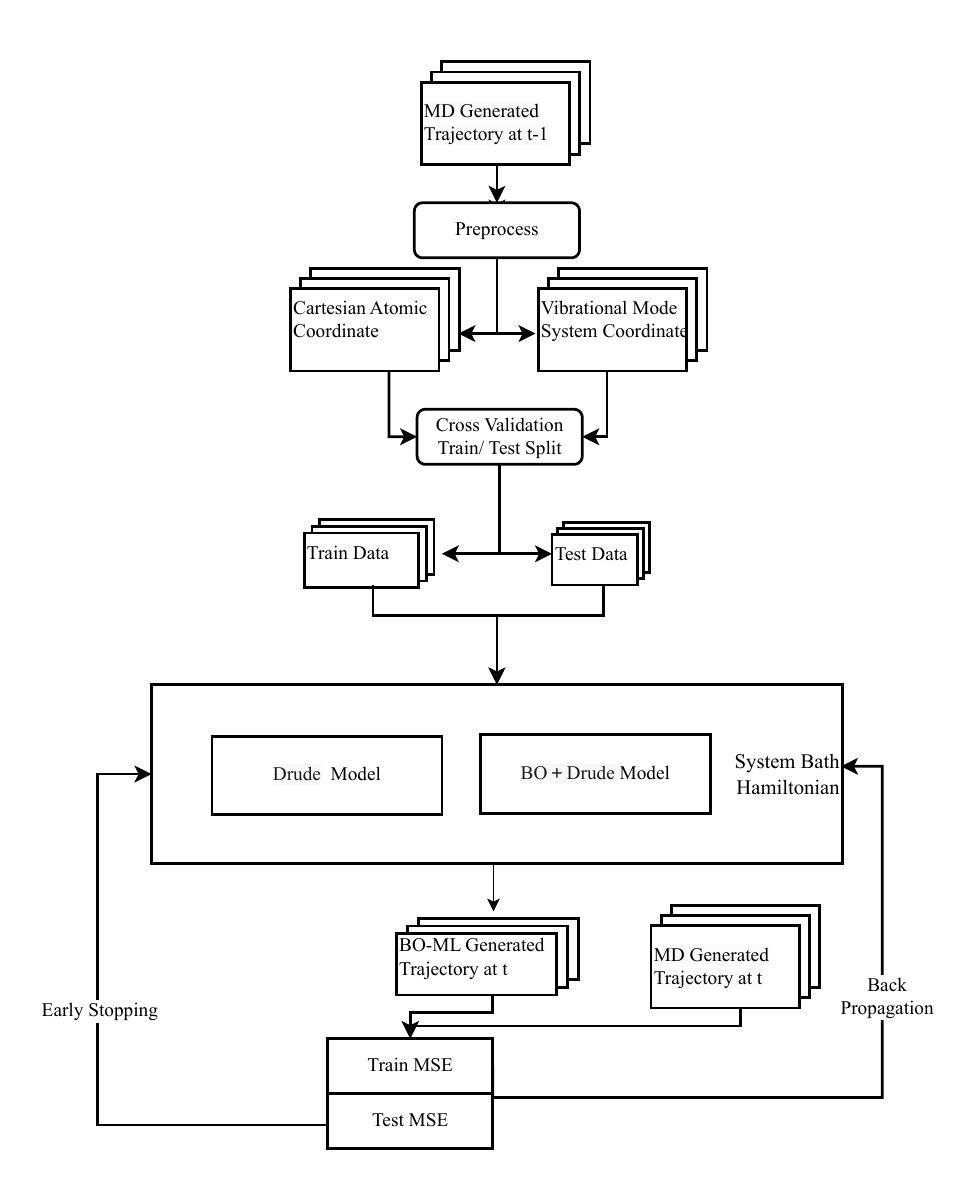}
  \caption{\label{fgr:algorithmflow}Flowchart of the algorithm used to optimize the parameters of the MAB model based on atomic trajectories obtained from MD simulations.}
\end{figure*}

\begin{figure*}[!t]
  \centering
  \includegraphics[scale=0.24]{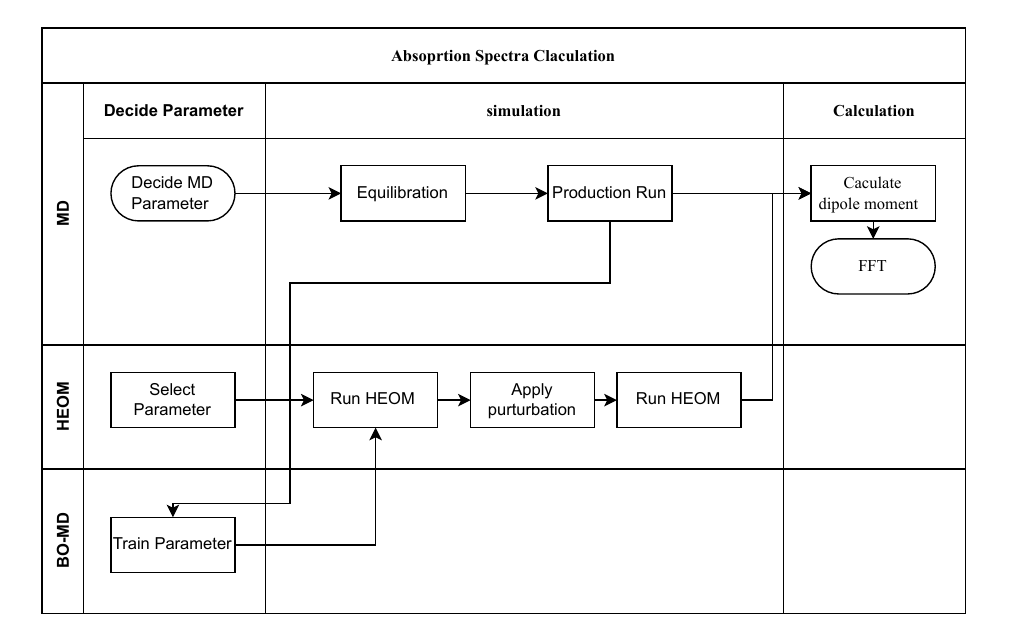}
  \caption{\label{fig:absorption-workflow}Schematic workflow for spectral calculation. Starting from MD trajectories, we first train the bath parameters and mode–mode coupling strength. The trained parameters are then passed to HEOM propagation, and the Fourier transform yields the final IR absorption spectrum.}
\end{figure*}

The methodology developed herein is broadly applicable to molecular systems embedded in diverse environments, such as biomolecular assemblies,\cite{FujihashiIshizaki2015} solid-state matrices,\cite{CBT22JCP} and solutions.\cite{UT21JCTC} To provide a clear and quantitative demonstration of its performance, we focus on liquid water as a representative system.
\cite{Ohmine_ChemRev93,OCSACR1999,Nibbering2004UltrafastVD,bagchi_2013}

Water and aqueous solutions have been extensively characterized using a range of advanced spectroscopic techniques, including 2D IR,\cite{ElsaesserDwaynePNAS2008,Tokmakoff2016H2O,VothTokmakoff_St-BendJCP2017,Tokmakoff2022,Kuroda_BendPCCP2014,Cho2009,Hamm2011ConceptsAM} 2D IR-Raman,\cite{grechko2018,Bonn2DTZIFvis2021,Begusic2023} and 2D THz-Raman spectroscopy.\cite{HammTHz2012,HSOT12JCP,Hamm2013PNAS,hamm2014,HammPerspH2O2017} These experimental approaches have been complemented by MD simulations,\cite{YagasakiSaitoJCP20082DIR,YagasakiSaitoJCP2011Relax,Yagasaki_ARPC64,ImotXanteasSaitoJCP2013H2O,Imotobend-lib2015} which yield detailed insights essential for the development and validation of theoretical models.

In addition, the HEOM framework incorporating the MAB models has been successfully employed to predict 2D Raman\cite{TS20JPSJ} and 2D IR-Raman\cite{IT16JCP,TT23JCP1} signals prior to their experimental realization. The framework also exhibits sufficient flexibility to reproduce experimentally measured 2D IR-Raman\cite{TT23JCP2} and 2DTHz-Raman spectra.\cite{IIT15JCP,HT25JCP2} Furthermore, computational tools integrating quantum and classical methodologies for simulating 2D spectra have been developed and made publicly available,\cite{TT23JCP2,HT25JCP2} thereby enabling rigorous validation of the theoretical approach.

The ML methodology employed in this study builds upon {a previous study.\cite{UT20JCTC} The main advance here is the adoption of a fixed functional form for the SDF, tailored for compatibility with the HEOM formalism.} In contrast to earlier models constructed using atomic coordinates, the present framework utilizes normal mode coordinates within an optimization scheme that is inherently compatible with the HEOM formalism. This coordinate choice facilitates efficient optimization, eliminates rotational and librational contributions, and enables the treatment of each vibrational mode with independent anharmonic potentials.

We consider the intramolecular vibrational modes of a water molecule extracted from MD simulations. These modes are described in terms of the two O–H bond lengths and the H–O–H bond angle of the $k$th water molecule, defined as
\begin{eqnarray}
  r_1^k= \left| \mathbf{x} _\mathrm{O}^k -  \mathbf{x} _\mathrm{H_1}^k \right|, \label{eq:def_molcoord1}
    \end{eqnarray}
  \begin{eqnarray}
  r_2^k = \left| \mathbf{x} _\mathrm{O}^k -  \mathbf{x} _\mathrm{H_2}^k\right|, \label{eq:def_molcoord2}
  \end{eqnarray}
and
  \begin{eqnarray}
 \theta^k = \arccos \left( \frac{\left(  \mathbf{x} _\mathrm{O}^k -  \mathbf{x}_\mathrm{H_1}^k \right)
  \cdot\left( \mathbf{x} _\mathrm{O} -  \mathbf{x} _\mathrm{H_2}\right)}{r_1r_2}\right),
  \label{eq:def_molcoord3}
\end{eqnarray}
where $ \mathbf{x}_\mathrm{O}$, $ \mathbf{x}_{\mathrm{H}_1}$, and  $ \mathbf{x}_{\mathrm{H}_e}$are the positions of the oxygen, the 1st, and 2nd hydrogen atoms, respectively, describing the intramolecular motion of the $k$th molecule.

The MAB model presented in Eqs.~\eqref{eqn:H_total}–\eqref{eqn:LLSL} comprises three intramolecular vibrational modes of the water molecule: (1) symmetric O–H stretching, ($1'$) asymmetric O–H stretching, and (2) H–O–H bending. These modes are respectively expressed as
\begin{eqnarray}
(1) \quad q_{1}^k = \frac{1}{2} \left( r_1^k + r_2^k - r_0 \right), \nonumber
\end{eqnarray}
\begin{eqnarray}
(1') \quad q_{2}^k = \frac{1}{2} \left( r_1^k - r_2^k \right), \nonumber
\end{eqnarray}
and
\begin{eqnarray}
(2) \quad q_{3}^k = \theta^k - \theta_0, \nonumber
\end{eqnarray}
where $r_0$ is the equilibrium length of the OH bond and $\theta_0$ is the
equilibrium bending angle. The learning MAB model includes the anharmonic interactions between the modes. The thermal effects, including vibrational dephasing, are described as interactions between each mode and its harmonic bath. 

To optimize the parameter set of the MAB model using MD trajectories, we employ a generative ML approach comprising the following steps. (i) MD trajectories $q_{j}^k$ are generated for water molecules. (ii) For the $k$th molecule, the trajectory of $q_{j}^k$ from time $t-1$ to $t$ is simulated using the MAB model with a trial parameter set. (iii) A loss function at time $t$ is evaluated to quantify the discrepancy between the reference MD trajectory from step (i) and the ML-generated trajectory from step (ii).
(iv) The loss is backpropagated to update the parameters of the MAB model, thereby iteratively enhancing its predictive accuracy. This procedure was {applied to} both case (a) in Eq.~\eqref{eq:drude} and case (b) in Eq.~\eqref{eq:drudePlusBO}.

More specifically, we analyze the trajectory set for the $k$th water molecule, represented as $\left(\mathbf{q}^k(t), \mathbf{p}^k(t)\right) \equiv \left( \{ q_s^k(t) \}, \{ p_s^k(t) \} \right)$, where $s$ indexes the vibrational modes. From MD simulations, we obtain a sequence of phase-space trajectories $\left(\mathbf{q}^k(t_0 + i\Delta t), \mathbf{p}^k(t_0 + i\Delta t)\right)$ sampled at time intervals $\Delta t$, with $i$ satisfying $0 \leq i \leq N - 1$, and $N$ denoting the total number of time steps. Using the MAB model, we generate a corresponding sequence of predicted trajectories, denoted as $\left(\bar{\mathbf{q}}^k(t_0 + i\Delta t), \bar{\mathbf{p}}^k(t_0 + i\Delta t)\right)$. Within the ML framework, we optimize the parameters in Eqs.~\eqref{eqn:H_total}–\eqref{eqn:LLSL}, along with the SDF parameters in Eq.~\eqref{eq:drude} or Eq.~\eqref{eq:drudePlusBO}, to reproduce the reference MD trajectories.

The thermal bath {associated with} the $s$th vibrational mode of the $k$th water molecule is modeled as a finite set of harmonic oscillators, each described by a coordinate $x_{j_s}$. The trajectory of this composite system is assumed to take the form\cite{UT20JCTC}
\begin{equation}
\tilde{x}_{j_s} (t) = A_{j_s} \sin(\omega_{j_s} t + \phi_{j_s}),
\end{equation}
where $A_{j_s}$ and $\phi_{j_s}$ is the amplitude and phase of the $j_s$th bath oscillator.
While $\phi_{j_s}$ is chosen randomly to prevent recursive motion,  $A_{j_s}$ are evaluated from  Eq. \eqref{eq:drude} or \eqref{eq:drudePlusBO} as learning parameters. 
For LL coupling, the bath parameters and the S-B interactions are expressed as a set of latent variables:
\begin{equation}
z_k = \{c_{j_s}^k\},
\end{equation}
where $c_{j_s}^k$ is defined as:
\begin{equation}
c_{j_s}^k = \alpha_{js} V_{\rm LL}(q_s) A_{j_s}.
\end{equation}
Additionally, $V_{SL}$ is learned as the ratio of these latent variables.

The trajectory at time $t_0 + i\Delta t$ can then be calculated using the MAB model:
\begin{eqnarray}
&&\left(\tilde{\mathbf{q}}^k(t_0 + i\Delta t), \tilde{\mathbf{p}}^k(t_0 + i\Delta t)\right) 
 =  \hat{L}(\Delta t; z_k, \Sigma) \nonumber \\ 
 &&~~~~~~~~~~~\times
 (\tilde{\mathbf{q}}^k(t_0 + (i-1)\Delta t), \tilde{\mathbf{p}}^k(t_0 + (i-1)\Delta t)),\nonumber \\ 
\end{eqnarray}
where $\left(\tilde{\mathbf{q}}^k(t), \tilde{\mathbf{p}}^k(t)\right)$ is the momentum and coordinate of the $k$th molecule, and $\hat{L}(\Delta t; z_k, \Sigma)$ is the Liouvillian for  Eqs. \eqref{eqn:H_total}-\eqref{eqn:LLSL} with the discretized heat bath, and $\Sigma$ represents the set of system and bath parameters.

We define the loss function as the Mean Squared Error (MSE) between the predicted and actual MD trajectories for the $s$th mode:
\begin{align}
\text{MSE}_{q_s} &\equiv \frac{1}{N} \sum_{i=1}^{N} \left[ \tilde{q}_s^k(t_i) -  q_s^k(t_i) \right]^2.
\end{align}

Minimization of the loss functions corresponds to the optimization of the learning model parameters. These include the anharmonicity of the potential energy surfaces, intermode anharmonic couplings, coupling strengths for LL and SL interactions, and the SDF parameters associated with each vibrational mode. 
We further evaluated the descriptive efficiency of atomic versus normal mode coordinates by computing the MSE for each representation (see Appendix~\ref{sec:coordinatemapping}). A schematic overview of the learning algorithm is provided in Fig.~\ref{fgr:algorithmflow}.

\subsection{HEOM with BO + Drude SDF}

The phase-space formulation of the HEOM, originally developed for the MAB system with the Drude SDF [case (a)], has been extended to quantum two-mode systems via the quantum hierarchical Fokker–Planck equations (QHFPE),\cite{TT23JCP1,TT23JCP2} and to classical three-mode systems via the classical hierarchical Fokker–Planck equations (CHFPE).\cite{IT16JCP,HT25JCP1} Source codes for both implementations are publicly available.\cite{TT23JCP2,HT25JCP2}

{For case (b), which involves the BO + Drude SDF, the hierarchy space must be further extended. This extended framework has previously been employed to simulate two-dimensional electronic spectra (2DES) of electron transfer systems.\cite{T12JCP} In the present study, we apply this formalism to the reduced density operator of the MAB system, $\hat \rho_A (t)$, thereby enabling explicit treatment of intramolecular vibrational modes. In contrast, for case (a), calculations can be performed without such extension by deactivating the BO bath within the same BO+Drude HEOM framework.}

{Note that the HEOM is introduced for a single-molecule picture, where the bath and S-B interaction are described by Eq.~\eqref{eq:h_bathint}. Accordingly, the counter term is not explicitly considered.}
 
{For the BO+Drude spectral density function (SDF), Eq.~\eqref{eq:drudePlusBO}, he antisymmetric and symmetric correlation functions are now evaluated as\cite{T12JCP}
\begin{align}
\label{eq:L1}
iL_1^{(s)}(t) &= \frac{im_s \zeta_s^{\rm D}\gamma_s^{\rm D}}{2} {{\rm{e}}^{ - {\gamma_s^{\rm D}} 
t}} \nonumber \\
 &- \frac{{i m_s \zeta_s^{\rm B} \gamma_s^{\rm B} (\omega_s^{\rm B})^2}}{4 \delta_s^{\rm B}}\left[ {{\rm{e}}^{ - \left( {\frac{\gamma_s^{\rm B}}{2} - i{{\delta_s^{\rm B}}}} \right)t}} - {\rm{e}}^{ - \left( \frac{\gamma_s^{\rm B}}{2} + i\delta_s^{\rm B} \right)t} \right],
\end{align}
and
\begin{align}
\label{eq:L2}
L_2^{(s)}(t) &=  \frac{m_s \zeta_s^{\rm D} \hbar ( \gamma_s^{\rm D} )^2}{4}\cot \left( \frac{\beta \hbar \gamma_s^{\rm D}}{2} \right) \rm{e}^{ - \gamma_s^{\rm D} t}  \nonumber \\
&
+ \frac{\gamma_s^{\rm B} ( \omega_s^{\rm B} )^2 m_s \zeta_s^{\rm B} \hbar}{8 \delta_s^{\rm B}} \nonumber \\
&\times \left[ A_s^- \rm{e}^{ - \left( \frac{\gamma_s^{\rm B}}{2} - i \delta_s^{\rm B} \right ) t} 
- A_s^+ \rm{e}^{ - \left( \frac{ \gamma_s^{\rm B} }{2} + i\delta_s^{\rm B} \right) t } \right] 
\nonumber \\
& - \sum\limits_{k = 1}^\infty B_s^k  {e^{ - {\nu _k}t}},
\end{align}
where  $\delta_s^{\rm B} = \sqrt {(\omega_s^{\rm B})^2 - {({\gamma_s^{\rm B}})^2}/4} $, $A_s^ \pm  = \coth \left( {\beta \hbar i\left( \gamma_s^{\rm B} \pm 2 i \delta_s^{\rm B} \right)/4} \right)$, and
\begin{align}
B_s^k &= \left\{ \frac{m_s \zeta_s^{\rm D} (\gamma_s^{\rm D})^2 }{\beta \hbar }\frac{\nu _k}{ {(\gamma_s^{\rm D})^2} - \nu _k^2} \right. \nonumber \\
&~~~~+ \left. \frac{m_s \zeta_s^{\rm B} ( \gamma_s^{\rm B} )^2 (\omega_s^{\rm B})^2}{{\beta \hbar }}\frac{\nu _k}{{\left[ {(\omega_s^{\rm B})^2 + \nu _k^2} \right]}^2 - (\gamma_s^{\rm B})^2\nu _k^2} \right\},
\end{align}
In this work, we employ the [$K_s - 1 / K_s$] Padé decomposition to incorporate temperature effects into the fluctuation and dissipation operators,\cite{hu2010communication} where $K_s$ is an integer associated with the $s$th bath mode. The HEOM is then formulated using the Padé frequencies $\nu_k^{s}$, with 
$k = \{-2, -1, \cdots, K_s\}$, defined as $\nu_{-2}^s \equiv \gamma_{s}^{B}+i\delta_s^{B}$, $\nu_{-1}^s \equiv \gamma_{s}^{B}-i\delta_s^{B}$, and $\nu_0^s \equiv \gamma_{s}^{D}$, are then expressed as
\begin{eqnarray}
\label{HEOM}
\frac{d}{dt} \hat{\rho} _{\{{\bf n}_{s}\}}=&&-\left[ \frac{i}{\hbar}\hat{H}_{S}^{\times}+\sum_{s}\sum_{k=-2}^{K_s} \left( n_k^s \nu_k^s \right) 
 \right]\hat{\rho} _{\{{\bf n}_{s}\}} \nonumber \\
&&-i\sum_{s}\sum_{k=-2}^{K_s} n_k^s\hat{\Theta}_k^s\hat{\rho}_{\{{\bf n}_{s} - {\bf e}^{k}_{s} \}}  \nonumber \\
&&-i\sum_{s}\sum_{k=-2}^{K_s} \hat{V}_s^\times\hat{\rho}
_{\{{\bf n}_{s} + {\bf e}^{k}_{s} \}} , 
\end{eqnarray}
The hierarchy elements are indexed by the set ${\{{\bf n}_{s}\}} \equiv ({\bf n}_{1}, {\bf n}_{2}, {\bf n}_{3})$, where each ${\bf n}_{s}$ is a multi-index defined as ${\bf n}_{s}=(n_{-2}^{s}, n_{-1}^{s},n_0^{s}, n_1^{s},  \cdots,   n_{K_{s}}^{s}$ for the three-mode case ($s = 1, 2, 3$). The notation ${\{{\bf n}_{s} \pm {\bf e}^{k}_{s}\}}$ indicates an increment or decrement of the $k$th component of ${\bf n}_{s}$, where ${\bf e}^{k}_{s}$ is the unit vector corresponding to the $k$th frequency component in the $s$th bath. The operators are defined as 
\begin{equation}
\hat \Theta _{-2}^{(s)} = \frac{m_s \zeta_s^{\rm B} \gamma_s^{\rm B} (\omega_s^{\rm B})^2}{8\hbar  \delta_s^{\rm B}} 
\left\{  -i {\hat V}_s^\circ  + \bar{A}_s^+  {\hat V}_s^ \times  \right\},
\end{equation}
\begin{equation}
\hat \Theta _{-1}^{(s)} = \frac{m_s \zeta_s^{\rm B} \gamma_s^{\rm B} ( \omega_s^{\rm B} )^2}{8\hbar \delta_s^{\rm B}}\left\{ i {\hat V}_s^\circ  + \bar{A}_s^- {\hat V}_s^ \times  \right\} ,
\end{equation}
\begin{align}
\label{Pade1}
\hat{\Theta}_0^{(s)} &= \frac{m_s \zeta_s^{\rm D}\gamma_s^{\rm D}}{4\hbar \beta} 
\left(1+\sum_{k=1}^{K_s} \frac{2\eta_k^s \gamma_{s}^2 }{({\gamma_s^{\rm D}})^2 
-{\nu_k^s}^2}\right)\hat{V}_s^{\times},
\end{align}
and
\begin{equation}
\label{Pade2}
\Theta_{k> 0}^{(s)}=-\frac{B_s^k}{\hbar} \hat{V}_s^\times,
\end{equation}
where we introduce the hyperoperators \(\hat{A}^{\times} \hat{B} \equiv \hat{A} \hat{B} - \hat{B} \hat{A}\) and \(\hat{A}^{\circ} \hat{B} \equiv \hat{A} \hat{B} + \hat{B} \hat{A}\), defined for arbitrary operators \(\hat{A}\) and \(\hat{B}\). The parameters \(\eta_k^s\) and \(\nu_k^s\) denote the Padé-approximated coupling intensity and frequency, respectively. }

\section{Application to water}
\label{sec:result}

\subsection{Collective Coordinates versus Single Molecule Coordinates}

{In previous studies, nonlinear spectra were calculated using the HEOM formalism based on the MAB model, with model parameters tuned to} reproduce the peak positions and spectral features of  the 2D spectrum obtained from MD simulations.\cite{IIT15JCP,HT25JCP2,IT16JCP,TT23JCP1,ST11JPCA,TT23JCP2,HT25JCP1} 
{Within this framework, the coordinates $q_s(t)$ 
assigned to} each mode are interpreted as representing collective motions. The bath parameters associated with these modes are not directly extracted from $q_s(t)$, but are instead inferred from its time correlation function through the dipole response.

In contrast, the ML approach presented in this study builds upon {previous work analyzing} single-molecule trajectories,\cite{UT20JCTC} and {offers a fundamentally different physical interpretation of $q_s(t)$. 
For example, earlier studies have treated the stretching and bending motions of surrounding molecules as bath components, leading to spectral peaks at the corresponding frequencies in the SDF. In this study, we constrain the SDF to the Drude or BO + Drude form, thereby guiding the learning process to interpret surrounding intramolecular vibrational modes not as bath components, but rather as contributors to mode–mode coupling.}

{It should also be noted that, although here we adopt the HEOM framework, the parameter values obtained in this study may differ from those previously derived from 2D IR–Raman spectral profiles.\cite{IIT15JCP,HT25JCP2,IT16JCP,TT23JCP1,ST11JPCA,TT23JCP2,HT25JCP1} This is because, while earlier HEOM datasets were typically constructed to capture bulk (or collective-mode) spectral characteristics, the parameters obtained in this study are trained to reproduce single-molecule dynamics.} 

\subsection{Details of the ML Approach}

We demonstrate our approach by optimizing the parameters of the MAB model for water, thereby providing a parameter set suitable for computing a wide range of spectra. 
MD trajectories for machine learning were generated using a system consisting of 392 water molecules confined within a cubic box measuring 2.3 nanometers per side. The system was maintained at a temperature of 300 K. {Simulations were performed for 50 picoseconds using GROMACS\cite{ABRAHAM201519}, with water molecules represented by the flexible SPC/E model\cite{doi:10.1021/j100308a038,GROMACS2025Manual} and Ferguson potential model\cite{1995JCoCh..16..501F} with Amber03 force field.}More detailed explanation about potential function of each MD simulation can be found in Appendix~\ref{sec:mdpotential}

The resulting trajectories were transformed into normal mode coordinates corresponding to intramolecular vibrations (see Appendix \ref{sec:coordinatemapping}). We then trained models for two cases: (a) the Drude SDF case and (b) the BO+Drude SDF case. For each training and testing split, early stopping was applied with a patience threshold of 300 epochs. Optimization was terminated when the test loss did not improve for 300 consecutive epochs (see Appendix \ref{sec:earlystopping}). This strategy reduced overfitting while ensuring a consistent stopping criterion.

In the evaluation of mode coupling, each coupling coefficient was calculated twice because the trajectories of the two modes were optimized independently. Specifically, the coefficient was computed once during the optimization of mode $s$ and once during the optimization of mode $s'$. Although these two estimates were obtained from separate optimization processes, they correspond to the same physical interaction. Therefore, we adopt the averaged values for three types of mode coupling: linear–linear, square–linear, and linear–squared. These are defined respectively as $\overline{g}_{s's}=(g_{ss'}+g_{s's})/2$, $\overline{ g}_{s^2 s'}=( g_{s^2 s'}+ g_{s' s^2})/2$, and $\overline{g}_{s s'^2}=( g_{s s'^2}+ g_{s'^2 s})/2$.

To implement early stopping, we employed Time-Step Cross-Validation (TSCV), a method designed to preserve the temporal continuity of the system's dynamics. 
This strategy enables systematic evaluation of how various optimization conditions influence model accuracy, including molecular sampling schemes derived from MD trajectories, time step resolutions, and model flexibility or adjustability
In the TSCV, each fold was constructed by training on the initial 4000~fs of the trajectory and testing on the subsequent 1000~fs (i.e., the next 1000~time steps). 
This procedure was repeated to generate four non-overlapping train–test splits. Within this framework, the same molecules interact in the same bath environment up to 4~ps, aligning well with the single-molecule perspective. Alternative cross-validation strategies were assessed in Appendix \ref{sec:cv}, yet TSCV offered a more consistent and physically grounded basis for spectral simulation.

During the development of the learning framework, the choice of initial values emerged as a critical factor shaping optimization outcomes. In non-convex landscapes, poor initialization can trap algorithms near local optima or saddle points. This directly impacts the S–B trade-off. For example, prior ML research\cite{UT20JCTC} reported substantially weaker S–B coupling compared to results obtained using the collective coordinate framework. Such a discrepancy is anticipated: both that study and the present work rely on single-molecule trajectories for ML, which inherently reflect more localized and weaker bath environments than those revealed through spectral analysis of collective coordinates. Furthermore, when the system model possesses sufficient flexibility, the optimization process may allocate residual variance to the system rather than the bath, further diminishing the apparent S–B coupling strength.
The discrepancy was traced to initial parameter values that constrained the system to a local minimum associated with an elevated harmonic potential. To mitigate this, initial values for the system potential parameters were selected to correspond to infrared stretch and bend peaks, providing physically motivated starting points for optimization.

Guided by this rationale, the training procedure was structured into two sequential stages. In the first stage, the system potential parameters and bath terms were jointly optimized to establish a consistent baseline representation of the vibrational modes and their surrounding environment. Upon convergence, these parameters were held fixed.
In the second stage, we refined the higher-order interactions, focusing specifically on the mode–mode coupling terms and anharmonicity of potential. This staged protocol reflects the logic of the initialization: by constraining the baseline potential and bath response, the subsequent estimation of mode–mode couplings is less susceptible to spurious minima and yields parameters that more faithfully capture the intrinsic physical correlations among vibrational modes.

For ML, model training was performed using Python 3.9.18 in conjunction with TensorFlow 2.15 and CUDA 12.2. 
All computations were executed on a system equipped with an Intel Core i9-13900H CPU and an NVIDIA GeForce RTX 4070 GPU. Each training fold required approximately 2–4 hours per mode case; with four folds, the total wall-time per model ranged from approximately 24 to 48 hours.

\subsection{Optimized parameter set}

\begin{table*}[!htb]
\caption{\label{tab:drude_bath_without_coupling_without_vll_bath}
Optimized parameters of the MAB model trained from {Ferguson potential} incorporating the Drude SDF and SL interaction are presented for the vibrational modes: 
(1) asymmetric stretch, ($1'$) symmetric stretch, and (2) bending.
Here,  $\tilde{\zeta_s^{\rm D}}$ denotes the normalized S-B coupling strength, and $\gamma_s^{\rm D}$ denotes the inverse correlation time of the bath fluctuations,  $V_{\mathrm{SL}}^{(s)}$ and $V_{\mathrm{LL}}^{(s)}$ denote the SL and LL interactions, 
$\tilde{g}_{s^3}$ is the cubic anharmonicity for the $s$ vibrational mode, respectively.}
\begin{tabular}{ccccccc}
  \hline \hline
\toprule
  \hline
 & $\omega_s$ (cm$^{-1}$)  & $\gamma_s^{\rm D}/\omega_0$ & $\tilde{\zeta}_s^{\rm D}$ & $\tilde{V}_{LL}^{(s)}$  &  $\tilde{V}_{SL}^{(s)}$  & $\tilde{g}_{s^3}$ \\
\midrule
1 & 3202 & $8.90\times10^{-3}$ & $1.39$ & $0$ & $1.00$ & $9.35\times10^{-9}$ \\
$1'$ & 3123& $2.66\times10^{-2}$ & $3.19\times10^{-2}$ & $0$ & $1.00$ & $1.12\times10^{-8}$ \\
2 & 1648 & $2.31\times10^{-2}$ & $3.85\times10^{4}$ & $0$ & $1.00$ & $-1.75\times10^{-4}$ \\
\bottomrule
  \hline \hline
\end{tabular}
\end{table*}

\begin{table*}[!htb]
\caption{\label{tab:DBCoupleSqcoupleAnh_coupling_mean_combo_wihtout_vll}
Optimized mode–mode coupling parameters of the MAB model trained from {Ferguson potential}  with the Drude SDF and SL interaction for (1) asymmetric stretch, ($1'$) symmetric stretch, and (2) bending modes.}
\begin{tabular}{cccc}
  \hline \hline
\toprule
  \hline
 $\mathrm{s-s'}$ & $\tilde{g}_{ss'}$  & $\tilde{g}_{s^2s'}$ & $\tilde{g}_{s{s'}^2}$  \\
\midrule
  $\mathrm{1-1'}$ & $1.32\times10^{-5}$ & $1.05\times10^{-8}$ & $9.96\times10^{-9}$ \\
  $\mathrm{1-2}$ & $-8.45\times10^{-5}$ & $-5.93\times10^{-9}$ & $-4.42\times10^{-7}$ \\
  $\mathrm{1'-2}$  & $-1.77\times10^{-4}$ & $1.20\times10^{-8}$ & $8.37\times10^{-8}$ \\
\bottomrule
\hline \hline
\end{tabular}
\end{table*}

\begin{table*}[!htb]
\caption{\label{tab:drude_bath_without_coupling_without_vll_bath_ferguson}
Optimized parameters of the MAB model trained from Ferguson potential with more sensitive anharmonicity setting incorporating the Drude SDF and SL interaction are presented for the vibrational modes: 
(1) asymmetric stretch, ($1'$) symmetric stretch, and (2) bending.
Here,  $\tilde{\zeta_s^{\rm D}}$ denotes the normalized S-B coupling strength, and $\gamma_s^{\rm D}$ denotes the inverse correlation time of the bath fluctuations,  $V_{\mathrm{SL}}^{(s)}$ and $V_{\mathrm{LL}}^{(s)}$ denote the SL and LL interactions, 
$\tilde{g}_{s^3}$ is the cubic anharmonicity for the $s$ vibrational mode, respectively.}
\begin{tabular}{ccccccc}
  \hline \hline
\toprule
  \hline
 & $\omega_s$ (cm$^{-1}$)  & $\gamma_s^{\rm D}/\omega_0$ & $\tilde{\zeta}_s^{\rm D}$ & $\tilde{V}_{LL}^{(s)}$  &  $\tilde{V}_{SL}^{(s)}$  & $\tilde{g}_{s^3}$ \\
\midrule
1 & 3513 & $2.42\times10^{-2}$ & $3.42\times10^{-2}$ & $0$ & $1.00$& $3.64\times10^{-3}$ \\
$1'$ & 3413 & $2.42\times10^{-2}$ & $3.43\times10^{-2}$ & $0$ & $1.00$ & $0.125$\\
2 & 1636 & $1.73\times10^{-3}$ & $1.79\times10^{4}$ & $0$ & $1.00$  & $2.895$\\
\bottomrule
  \hline \hline
\end{tabular}
\end{table*}

\begin{table*}[!htb]
\caption{\label{tab:DBCoupleSqcoupleAnh_coupling_mean_combo_wihtout_vll_ferguson}
Optimized mode–mode coupling parameters of the MAB model trained from Ferguson potential with more sensitive anharmonicity setting with the Drude SDF and SL interaction for (1) asymmetric stretch, ($1'$) symmetric stretch, and (2) bending modes.}
\begin{tabular}{cccc}
  \hline \hline
\toprule
  \hline
 $\mathrm{s-s'}$ & $\tilde{g}_{ss'}$  & $\tilde{g}_{s^2s'}$ & $\tilde{g}_{s{s'}^2}$  \\
\midrule
  $\mathrm{1-1'}$ &$3.78\times10^{-2}$ & $1.11\times10^{-2}$ & $0.192$ \\
  $\mathrm{1-2}$ & $5.15\times10^{-2}$ & $7.74\times10^{-2}$ & $5.13\times10^{-2}$ \\
  $\mathrm{1'-2}$ & $-2.282$ & $9.15\times10^{-2}$ & $0.172$\\
\bottomrule
\hline \hline
\end{tabular}
\end{table*}

\begin{table*}[!htb]
\caption{\label{tab:drude_bath_without_coupling_without_vll_bath_big_anharmonicity}
Optimized parameters of the MAB model from {Ferguson potential} with more sensitive anharmonicity setting incorporating the Drude SDF and SL interaction are presented for the vibrational modes: 
(1) asymmetric stretch, ($1'$) symmetric stretch, and (2) bending.
Here,  $\tilde{\zeta_s^{\rm D}}$ denotes the normalized S-B coupling strength, and $\gamma_s^{\rm D}$ denotes the inverse correlation time of the bath fluctuations,  $V_{\mathrm{SL}}^{(s)}$ and $V_{\mathrm{LL}}^{(s)}$ denote the SL and LL interactions, 
$\tilde{g}_{s^3}$ is the cubic anharmonicity for the $s$ vibrational mode, respectively.}
\begin{tabular}{ccccccc}
  \hline \hline
\toprule
  \hline
 & $\omega_s$ (cm$^{-1}$)  & $\gamma_s^{\rm D}/\omega_0$ & $\tilde{\zeta}_s^{\rm D}$ & $\tilde{V}_{LL}^{(s)}$  &  $\tilde{V}_{SL}^{(s)}$  & $\tilde{g}_{s^3}$ \\
\midrule
1 & 3202 & $2.41\times10^{-2}$ & $2.80\times10^{-2}$ & $0$ & $1.00$ &  $1.71\times10^{-2}$  \\
$1'$ & 3123 & $2.41\times10^{-2}$ & $5.45\times10^{-2}$ & $0$ & $1.00$ &  $2.51\times10^{-2}$ \\
2 & 1592 & $1.85\times10^{-3}$ & $1.84\times10^{4}$ & $0$ & $1.00$ & $3.357$  \\
\bottomrule
  \hline \hline
\end{tabular}
\end{table*}

\begin{table*}[!htb]
\caption{\label{tab:DBCoupleSqcoupleAnh_coupling_mean_combo_wihtout_vll_big_anharmonicity}
Optimized mode–mode coupling parameters of the MAB model trained from {Ferguson potential} with more sensitive anharmonicity setting with the Drude SDF and SL interaction for (1) asymmetric stretch, ($1'$) symmetric stretch, and (2) bending modes.}
\begin{tabular}{cccc}
  \hline \hline
\toprule
  \hline
 $\mathrm{s-s'}$ & $\tilde{g}_{ss'}$  & $\tilde{g}_{s^2s'}$ & $\tilde{g}_{s{s'}^2}$  \\
\midrule
  $\mathrm{1-1'}$ & $-1.01\times10^{-2}$ & $5.02\times10^{-3}$ & $3.43\times10^{-2}$ \\
  $\mathrm{1-2}$ &  $-6.12\times10^{-2}$ & $0.128$ & $3.53\times10^{-2}$ \\
  $\mathrm{1'-2}$  & $-2.965$ & $0.136$ & $9.77\times10^{-2}$ \\
\bottomrule
\hline \hline
\end{tabular}
\end{table*}

\begin{table*}[!htb]
\caption{\label{tab:drude_bath_without_coupling_bath}Optimized parameters of the MAB model  trained from {Ferguson potential}  with the Drude SDF and LL+SL interaction for (1) asymmetric stretching, (1$'$) symmetric stretching, and (2) bending modes. 
Here,  $\tilde{\zeta_s^{\rm D}}$ denotes the normalized S-B coupling strength, and $\gamma_s^{\rm D}$ denotes the inverse correlation time of the bath fluctuations,  $V_{\mathrm{SL}}^{(s)}$ and $V_{\mathrm{LL}}^{(s)}$ denote the SL and LL interactions, 
$\tilde{g}_{s^3}$ is the cubic {anharmonicity} for the $s$ vibrational mode, respectively.}
\begin{tabular}{ccccccc}
  \hline \hline
\toprule
  \hline
 & $\omega_s$ (cm$^{-1}$)  & $\gamma_s^{\rm D}/\omega_0$ & $\tilde{\zeta}_s^{\rm D}$ & $\tilde{V}_{LL}^{(s)}$  &  $\tilde{V}_{SL}^{(s)}$  & $\tilde{g}_{s^3}$ \\
\midrule
1 & 3202 & $9.50\times10^{-3}$ & $1.19$ & $2.15\times10^{-1}$ & $1.00$ & $9.34\times10^{-9}$ \\
$1'$ & 3123 & $2.67\times10^{-2}$ & $1.71\times10^{-2}$ & $1.37\times10^{-1}$ & $1.00$ & $1.12\times10^{-8}$\\
2 &1622 & $2.17\times10^{-2}$ & $1.47\times10^{4}$ & $5.64\times10^{-2}$ & $1.00$ & $-1.54\times10^{-4}$\\
\bottomrule
  \hline \hline
\end{tabular}
\end{table*}

\begin{table*}[!htb]
\caption{\label{tab:DBCoupleSqcoupleAnh_coupling}Optimized mode-mode coupling strength of the MAB model  trained from {Ferguson potential}  with the Drude SDF and LL + SL interaction for (1) asymmetric stretch, ($1'$) symmetric stretch, and (2) bending modes.}
\begin{tabular}{cccc}
  \hline \hline
\toprule
  \hline
 $\mathrm{s-s'}$ & $\tilde{g}_{ss'}$  & $\tilde{g}_{s^2s'}$ & $\tilde{g}_{s{s'}^2}$  \\
\midrule
  $\mathrm{1-1'}$ & $1.31\times10^{-5}$ & $1.05\times10^{-8}$ & $9.96\times10^{-9}$ \\
  $\mathrm{1-2}$ & $-7.45\times10^{-5}$ & $1.70\times10^{-9}$ & $-5.14\times10^{-7}$ \\
  $\mathrm{1'-2}$ & $-1.42\times10^{-4}$ & $1.61\times10^{-8}$ & $7.25\times10^{-8}$ \\
\bottomrule
\hline \hline
\end{tabular}
\end{table*}

\begin{table*}[!htb]
\caption{\label{tab:drude_bath_without_coupling_bath}Optimized parameters of the MAB model  trained from {Ferguson potential}  with the Drude SDF and LL+SL interaction for (1) asymmetric stretching, (1$'$) symmetric stretching, and (2) bending modes. 
Here,  $\tilde{\zeta_s^{\rm D}}$ denotes the normalized S-B coupling strength, and $\gamma_s^{\rm D}$ denotes the inverse correlation time of the bath fluctuations,  $V_{\mathrm{SL}}^{(s)}$ and $V_{\mathrm{LL}}^{(s)}$ denote the SL and LL interactions, 
$\tilde{g}_{s^3}$ is the cubic {anharmonicity} for the $s$ vibrational mode, respectively.}
\begin{tabular}{ccccccc}
  \hline \hline
\toprule
  \hline
 & $\omega_s$ (cm$^{-1}$)  & $\gamma_s^{\rm D}/\omega_0$ & $\tilde{\zeta}_s^{\rm D}$ & $\tilde{V}_{LL}^{(s)}$  &  $\tilde{V}_{SL}^{(s)}$  & $\tilde{g}_{s^3}$ \\
\midrule
1 & 3202 & $9.50\times10^{-3}$ & $1.19$ & $2.15\times10^{-1}$ & $1.00$ & $9.34\times10^{-9}$ \\
$1'$ & 3123 & $2.67\times10^{-2}$ & $1.71\times10^{-2}$ & $1.37\times10^{-1}$ & $1.00$ & $1.12\times10^{-8}$\\
2 &1622 & $2.17\times10^{-2}$ & $1.47\times10^{4}$ & $5.64\times10^{-2}$ & $1.00$ & $-1.54\times10^{-4}$\\
\bottomrule
  \hline \hline
\end{tabular}
\end{table*}

\begin{table*}[!htb]
\caption{\label{tab:DBCoupleSqcoupleAnh_coupling}Optimized mode-mode coupling strength of the MAB model  trained from {Ferguson potential}  with the Drude SDF and LL + SL interaction for (1) asymmetric stretch, ($1'$) symmetric stretch, and (2) bending modes.}
\begin{tabular}{cccc}
  \hline \hline
\toprule
  \hline
 $\mathrm{s-s'}$ & $\tilde{g}_{ss'}$  & $\tilde{g}_{s^2s'}$ & $\tilde{g}_{s{s'}^2}$  \\
\midrule
  $\mathrm{1-1'}$ & $1.31\times10^{-5}$ & $1.05\times10^{-8}$ & $9.96\times10^{-9}$ \\
  $\mathrm{1-2}$ & $-7.45\times10^{-5}$ & $1.70\times10^{-9}$ & $-5.14\times10^{-7}$ \\
  $\mathrm{1'-2}$ & $-1.42\times10^{-4}$ & $1.61\times10^{-8}$ & $7.25\times10^{-8}$ \\
\bottomrule
\hline \hline
\end{tabular}
\end{table*}

\begin{table*}[!htb]
\caption{\label{tab:drude_bath_without_coupling_bath}Optimized parameters of the MAB model  trained from {Ferguson potential} with more sensitive anharmonicity setting  with the Drude SDF and LL+SL interaction for (1) asymmetric stretching, (1$'$) symmetric stretching, and (2) bending modes. 
Here,  $\tilde{\zeta_s^{\rm D}}$ denotes the normalized S-B coupling strength, and $\gamma_s^{\rm D}$ denotes the inverse correlation time of the bath fluctuations,  $V_{\mathrm{SL}}^{(s)}$ and $V_{\mathrm{LL}}^{(s)}$ denote the SL and LL interactions, 
$\tilde{g}_{s^3}$ is the cubic {anharmonicity} for the $s$ vibrational mode, respectively.}
\begin{tabular}{ccccccc}
  \hline \hline
\toprule
  \hline
 & $\omega_s$ (cm$^{-1}$)  & $\gamma_s^{\rm D}/\omega_0$ & $\tilde{\zeta}_s^{\rm D}$ & $\tilde{V}_{LL}^{(s)}$  &  $\tilde{V}_{SL}^{(s)}$  & $\tilde{g}_{s^3}$ \\
\midrule
1 & 3202 & $2.41\times10^{-2}$ & $2.75\times10^{-2}$ & $3.16\times10^{-1}$ & $1.00$ &$1.46\times10^{-2}$\\
$1'$ & 3123 &  $2.41\times10^{-2}$ & $3.46\times10^{-2}$ & $3.12\times10^{-1}$ & $1.00$ & $3.61\times10^{-2}$\\
2 & 1592 & $1.82\times10^{-3}$ & $1.89\times10^{4}$ & $2.58\times10^{-1}$ & $1.00$ & $3.469$ \\
\bottomrule
  \hline \hline
\end{tabular}
\end{table*}

\begin{table*}[!htb]
\caption{\label{tab:DBCoupleSqcoupleAnh_coupling}Optimized mode-mode coupling strength of the MAB model  trained from {Ferguson potential} with more sensitive anharmonicity setting with the Drude SDF and LL + SL interaction for (1) asymmetric stretch, ($1'$) symmetric stretch, and (2) bending modes.}
\begin{tabular}{cccc}
  \hline \hline
\toprule
  \hline
 $\mathrm{s-s'}$ & $\tilde{g}_{ss'}$  & $\tilde{g}_{s^2s'}$ & $\tilde{g}_{s{s'}^2}$  \\
\midrule
  $\mathrm{1-1'}$ & $-1.46\times10^{-2}$ & $1.01\times10^{-2}$ & $3.72\times10^{-2}$ \\
  $\mathrm{1-2}$ & $-4.17\times10^{-2}$ & $0.134$ & $0.182$  \\
  $\mathrm{1'-2}$ & $-3.375$ & $0.153$ & $0.279$  \\
\bottomrule
\hline \hline
\end{tabular}
\end{table*}

\begin{table*}[!htb]
\caption{\label{tab:drude_bath_without_coupling_bath_ferguson}Optimized parameters of the MAB model  trained from Ferguson potential with more sensitive anharmonicity setting  with the Drude SDF and LL+SL interaction for (1) asymmetric stretching, (1$'$) symmetric stretching, and (2) bending modes. 
Here,  $\tilde{\zeta_s^{\rm D}}$ denotes the normalized S-B coupling strength, and $\gamma_s^{\rm D}$ denotes the inverse correlation time of the bath fluctuations,  $V_{\mathrm{SL}}^{(s)}$ and $V_{\mathrm{LL}}^{(s)}$ denote the SL and LL interactions, 
$\tilde{g}_{s^3}$ is the cubic {anharmonicity} for the $s$ vibrational mode, respectively.}
\begin{tabular}{ccccccc}
  \hline \hline
\toprule
  \hline
 & $\omega_s$ (cm$^{-1}$)  & $\gamma_s^{\rm D}/\omega_0$ & $\tilde{\zeta}_s^{\rm D}$ & $\tilde{V}_{LL}^{(s)}$  &  $\tilde{V}_{SL}^{(s)}$  & $\tilde{g}_{s^3}$ \\
\midrule
1 & 3513 & $2.41\times10^{-2}$ & $1.97\times10^{-2}$ & $3.31\times10^{-1}$ & $1.00$ & $1.08\times10^{-2}$\\
$1'$ & 3413 & $2.42\times10^{-2}$ & $3.31\times10^{-2}$ & $3.26\times10^{-1}$ & $1.00$ & $0.115$\\
2 & 1636 & $1.80\times10^{-3}$ & $1.71\times10^{4}$ & $2.58\times10^{-1}$ & $1.00$ & $2.927$\\
\bottomrule
  \hline \hline
\end{tabular}
\end{table*}

\begin{table*}[!htb]
\caption{\label{tab:DBCoupleSqcoupleAnh_coupling_ferguson}Optimized mode-mode coupling strength of the MAB model  trained from Ferguson potential with more sensitive anharmonicity setting with the Drude SDF and LL + SL interaction for (1) asymmetric stretch, ($1'$) symmetric stretch, and (2) bending modes.}
\begin{tabular}{cccc}
  \hline \hline
\toprule
  \hline
 $\mathrm{s-s'}$ & $\tilde{g}_{ss'}$  & $\tilde{g}_{s^2s'}$ & $\tilde{g}_{s{s'}^2}$  \\
\midrule
  $\mathrm{1-1'}$ &$-1.21\times10^{-2}$ & $1.09\times10^{-2}$ & $0.199$  \\
  $\mathrm{1-2}$ & $-2.10\times10^{-2}$ & $7.96\times10^{-2}$ & $8.17\times10^{-2}$  \\
  $\mathrm{1'-2}$ &$-2.295$ & $7.77\times10^{-2}$ & $0.224$ \\
\bottomrule
\hline \hline
\end{tabular}
\end{table*}

We optimized the parameters {associated with} the intramolecular vibrational modes under {two distinct frameworks:} (a) the Drude SDF, and (b) BO+ Drude SDF. The computational {protocol employed to generate} the absorption spectra is schematically depicted in Fig.~\ref{fig:absorption-workflow}.

{Previous studies\cite{UT20JCTC} have demonstrated} that the SDF obtained via ML in the single-molecule framework is significantly more intricate than the Drude SDF, displaying distinct peaks corresponding to each intermolecular and intramolecular vibrational mode. 
{Notably,} the contribution of each intramolecular three-mode component, when modeled as a bath, can be effectively captured by the Drude representation, provided that strong mode–mode coupling is adequately incorporated. 
In contrast, low-frequency intermolecular vibrations are {not sufficiently described by the Drude bath alone. To remedy this limitation,} we incorporated a BO mode to account for relaxation pathways into the intermolecular vibrational manifold. {This hybrid BO + Drude bath model was found to yield a reduced} learning loss relative to the Drude-only counterpart, thereby offering a more accurate representation of the S-B interaction.

Several studies have been conducted using the MAB model for the collective mode of water in the Drude framework.\cite{IT16JCP,TT23JCP1,TT23JCP2,ST11JPCA,HT25JCP1,HT25JCP2} 
In these investigations, the model parameters for water were selected to reproduce the peak positions and spectral profiles observed in 1D and 2D spectra obtained directly from MD simulations. In the current study, {these} parameters were determined using ML based on atomic trajectories obtained from the MD. 

Accordingly, the parameters of the MAB model introduced herein are defined with respect to single-molecule coordinates and are therefore not directly comparable to prior results formulated in terms of collective coordinates. It is also noteworthy that earlier modeling efforts have predominantly targeted optical observables—such as polarization and its temporal derivative—without explicitly resolving the underlying molecular coordinates. {In contrast, the present approach affords a more granular depiction of MD} at the microscopic level. Importantly, it enables the identification and characterization of spectroscopically inactive (dark) states that are inaccessible via conventional optical probes.

Despite the methodological {disparities, we endeavor to} compare the present results with model parameters reported in prior studies. Such a comparison facilitates a critical examination of the differences between collective and molecular coordinate representations, with particular emphasis on the respective roles of vibrational mode coupling and environmental bath interactions.
{The} non-Markovian nature of the bath is essential, as the correlation time of the bath noise determines the vibrational dephasing time.\cite{IT06JCP,ST11JPCA,TI09ACR} 
In the high-frequency  {regime—typified by intramolecular vibrations—the impact} of LL coupling on the spectral profile is negligible relative to that of SL coupling. Accordingly, we begin by considering a simplified scenario in which LL system–bath coupling is omitted.

\subsubsection{Drude bath with SL interaction}

The parameters of the MAB model, derived via the ML approach, are compiled in Tables \ref{tab:drude_bath_without_coupling_without_vll_bath} and \ref{tab:DBCoupleSqcoupleAnh_coupling_mean_combo_wihtout_vll}, expressed as functions of molecular coordinates. 
{To facilitate direct comparison with previous studies} employing collective modes\cite{IT16JCP,TT23JCP1,TT23JCP2}, and to streamline the computation of the 2D spectrum using the source code currently under development for the quantum three-mode system, we adopted the same formatting conventions as those used in Refs. \onlinecite{HT25JCP1,HT25JCP2}. Accordingly, the scaling of quantities follows $\tilde{\zeta_s^{\rm D}} \equiv \zeta_s^{\rm D} (\omega_0/\omega_s)^2$, with $\omega_0 = 4000\,\mathrm{cm}^{-1}$, and $\gamma_s^{\rm D}$ is reported as $\gamma_s^{\rm D}/\omega_0$. For mode mode coupling, the parameters are normalized as $\tilde g_{s^3}=\overline{g}_{s^3}(\omega_0/\omega_s)^3$, $\tilde g_{s's}=\overline{ g}_{s's}(\omega_0/\omega_s)(\omega_0/\omega{s'})$, $\tilde g_{s^2 s'}=\overline{g}_{s^2 s'}(\omega_0/\omega_s)^2(\omega_0/\omega{s'})$, and $\tilde g_{s s'^2}=\overline{g}_{s s'^2}(\omega_0/\omega_s)(\omega_0/\omega{s'})^2$.

Tables \ref{tab:drude_bath_without_coupling_without_vll_bath} and \ref{tab:DBCoupleSqcoupleAnh_coupling_mean_combo_wihtout_vll} correspond to Tables IV and II–III, respectively, in Ref. \onlinecite{HT25JCP1}.
Table \ref{tab:drude_bath_without_coupling_without_vll_bath} reveals notably low anharmonicity, which is primarily attributed to the use of the {Ferguson potential} for the ML algorithm’s development. These models do not incorporate anharmonicity in OH stretching vibrations, so the anharmonic component only weakly appears in the models. 
For linear absorption spectra, where the impact of anharmonicity and mode coupling is minimal, such force fields are sufficiently accurate. However, the simulation of 2D vibrational spectra demands a more refined representation, such as that provided by POLI2VS.\cite{HT11JPCB} {and MB-POL}\cite{Babin2013_MBpol_I,Babin2014_MBpol_II,Medders2014_MBpol_III}
 Indeed, to evaluate anharmonicity and mode coupling within the collective coordinate framework, 2D IR-Raman spectra computed using POLI2VS were utilized.\cite{IT16JCP}

Although the absolute values differ, the inverse correlation times of the bath noise exhibit similar trends. The primary distinction arises from the coupling strength between the bath and the symmetric stretching and bending vibrational modes. For the bending mode, the limited angular variation allowed by the MD force field may introduce an artifact, potentially contributing to the observed discrepancy. In contrast, the deviation in the stretching mode may be linked to hydrogen bonding and could reflect intrinsic features of the microscopic water structure. It should be noted that the collective coordinate approach does not differentiate between symmetric and asymmetric stretching modes; hence, the observed behavior should be regarded as suggestive rather than definitive.

Table \ref{tab:DBCoupleSqcoupleAnh_coupling_mean_combo_wihtout_vll} presents the results for anharmonic coupling. Compared to the collective coordinate results (Tables II and III in Ref. \onlinecite{HT25JCP1}), the magnitude of the anharmonic coupling is estimated to be small, on the same order as $\tilde g_{s^3}$ in Table \ref{tab:drude_bath_without_coupling_without_vll_bath}. This reduction is primarily attributed to the use of an OH stretching potential that lacks intrinsic anharmonicity.

Table \ref{tab:DBCoupleSqcoupleAnh_coupling_mean_combo_wihtout_vll} shows the results for anharmonic coupling. Compared to the results of the collective coordinate (Tables II and III in Ref. \onlinecite{HT25JCP1}), the magnitude of the anharmonic coupling is estimated to be small, on the same order as $\tilde g_{s^3}$ in Table  \ref{tab:drude_bath_without_coupling_without_vll_bath}. This may also be considered a consequence of using the potential for OH stretching without anharmonicity.

\subsubsection{Drude bath with LL+SL interaction}

{The results incorporating LL interactions are summarized} in Tables \ref{tab:drude_bath_without_coupling_bath} and \ref{tab:DBCoupleSqcoupleAnh_coupling}.
A comparison between Tables \ref{tab:drude_bath_without_coupling_without_vll_bath} and \ref{tab:drude_bath_without_coupling_bath} reveals that, {even with the inclusion of $V_{\rm LL}$ in the optimization,} its overall contribution remains relatively minor. {Accordingly,} variations in other bath parameters are also limited.

Next, we examine the differences in mode coupling parameters listed in Tables \ref{tab:DBCoupleSqcoupleAnh_coupling_mean_combo_wihtout_vll} and \ref{tab:DBCoupleSqcoupleAnh_coupling}. 
{As with the bath parameters,} the distinction between SL and LL+SL coupling remains marginal.
In both cases, the strongest coupling is observed for $\tilde{g}_{s{s'}}$ with $s = 2$ and $s' = 1$, followed by the coupling between $s = 2$ and $s' = 1'$. The former is more pronounced due to the shared symmetry between the bending mode and the symmetric stretch mode, which facilitates vibrational interaction.

Regarding anharmonic coupling, the parameter $\tilde{g}_{s{s'}^2}$ is notably large when $s = 1$ or $1'$ and $s' = 2$, reflecting the near-resonance between the overtone of the bending mode and the fundamental frequency of the stretch mode. However, the difference between the $1$–$2$ and $1'$–$2$ couplings is less pronounced than in previous studies.\cite{UT20JCTC} 
{This may be attributed to the relatively} weak anharmonicity of the potential employed in this study or differences in the model training protocol.

\subsubsection{BO + Drude bath with SL interaction}

\begin{table*}[!htb]
\caption{\label{tab:brownian_drude_bath_without_coupling_without_vll_bath}Optimized parameters of the MAB model  trained from {Ferguson potential} with BO + Drude SDF and SL interaction for (1) asymmetric stretching, (1$'$) symmetric stretching, and (2) bending modes. 
Here,  $\tilde{\zeta_s^{\rm D}}$ denotes the normalized S-B coupling strength, and $\gamma_s^{\rm D}$ denotes the inverse correlation time of the bath fluctuations for the Drude mode and $\tilde{\zeta_s^{\rm B}}$ and $\gamma_s^{\rm B}$ are those for the BO mode.  The central frequency of the BO mode is expressed as $\omega_s^{\rm B}$, and  $V_{\mathrm{SL}}^{(s)}$ and $V_{\mathrm{LL}}^{(s)}$ denote the SL and LL interactions, 
$\tilde{g}_{s^3}$ is the cubic {anharmonicity} for the $s$ vibrational mode, respectively.
}
\begin{tabular}{cccccccccc}
  \hline \hline
\toprule
  \hline
 & $\omega_s$ (cm$^{-1}$) & $\gamma_s^{\rm D}/\omega_0$ & $\tilde{\zeta}_s^{\rm D}$  & $\gamma_s^{\rm B}/\omega_0$ & $\tilde{\zeta}_s^{\rm B}$ & $\omega_s^{\rm B}/\omega_0$   & $\tilde{V}_{\rm LL}^{(s)}$ &  $\tilde{V}_{\rm SL}^{(s)}$  & $\tilde{g}_{s^3}$  \\
\midrule
1 & 3202 & $7.27\times10^{-1}$ & $1.95\times10^{-2}$ & $56.69$ & $1.31\times10^{-2}$ & $8.60\times10^{-3}$ & $0$ & $1.00$  & $9.58\times10^{-9}$  \\
$1'$ & 3123 & $6.22$ & $1.54\times10^{-2}$ & $54.85$ & $1.69\times10^{-2}$ & $1.13\times10^{-2}$ & $0$ & $1.00$  & $1.12\times10^{-8}$ \\
2 & 1596 & $19.88$ & $1.42\times10^{-1}$ & $8.12\times10^{2}$ & $9.22\times10^{-2}$ & $1.16\times10^{-1}$ & $0$ & $1.00$  & $-2.05\times10^{-4}$ \\
\bottomrule
  \hline \hline
\end{tabular}
\end{table*}

\begin{table*}[!htb]
\caption{\label{tab:BBDBCwoLL}
Optimized mode–mode coupling parameters of the MAB model  trained from {Ferguson potential} with the BO+Drude SDF and SL interaction for (1) asymmetric stretch, ($1'$) symmetric stretch, and (2) bending modes.}
\begin{tabular}{cccc}
  \hline \hline
\toprule
  \hline
 $\mathrm{s-s'}$ & $\tilde{g}_{ss'}$  & $\tilde{g}_{s^2s'}$ & $\tilde{g}_{s{s'}^2}$  \\
\midrule
  $\mathrm{1-1'}$ & $6.84\times10^{-6}$ & $1.06\times10^{-8}$ & $1.01\times10^{-8}$ \\
  $\mathrm{1-2}$ & $-8.61\times10^{-5}$ & $-6.32\times10^{-9}$ & $6.91\times10^{-8}$ \\
  $\mathrm{1'-2}$  & $-1.71\times10^{-4}$ & $1.28\times10^{-8}$ & $1.69\times10^{-7}$ \\
\bottomrule
  \hline \hline
\end{tabular}
\end{table*}

\begin{table*}[!htb]
\caption{\label{tab:brownian_drude_bath_without_coupling_without_vll_bath_bigger_anharmonicity}Optimized parameters of the MAB model trained from {Ferguson potential} with more sensitive anharmonicity setting with BO + Drude SDF and SL interaction for (1) asymmetric stretching, (1$'$) symmetric stretching, and (2) bending modes. 
Here,  $\tilde{\zeta_s^{\rm D}}$ denotes the normalized S-B coupling strength, and $\gamma_s^{\rm D}$ denotes the inverse correlation time of the bath fluctuations for the Drude mode and $\tilde{\zeta_s^{\rm B}}$ and $\gamma_s^{\rm B}$ are those for the BO mode.  The central frequency of the BO mode is expressed as $\omega_s^{\rm B}$, and  $V_{\mathrm{SL}}^{(s)}$ and $V_{\mathrm{LL}}^{(s)}$ denote the SL and LL interactions, 
$\tilde{g}_{s^3}$ is the cubic {anharmonicity} for the $s$ vibrational mode, respectively.
}
\begin{tabular}{cccccccccc}
  \hline \hline
\toprule
  \hline
 & $\omega_s$ (cm$^{-1}$) & $\gamma_s^{\rm D}/\omega_0$ & $\tilde{\zeta}_s^{\rm D}$  & $\gamma_s^{\rm B}/\omega_0$ & $\tilde{\zeta}_s^{\rm B}$ & $\omega_s^{\rm B}/\omega_0$   & $\tilde{V}_{\rm LL}^{(s)}$ &  $\tilde{V}_{\rm SL}^{(s)}$  & $\tilde{g}_{s^3}$  \\
\midrule
1 & 3202 & $2.41\times10^{-2}$ & $3.58\times10^{-2}$ & $5.49\times10^{-5}$ & $2.31\times10^{3}$ & $1.12\times10^{-2}$ & $0$ & $1.00$ &  $1.28\times10^{-2}$  \\
$1'$ & 3123 & $2.41\times10^{-2}$ & $2.90\times10^{-2}$ & $5.73\times10^{-5}$ & $2.38\times10^{3}$ & $1.15\times10^{-2}$ & $0$ & $1.00$&  $3.70\times10^{-2}$ \\
2 & 1592 & $2.43\times10^{-3}$ & $7.61\times10^{3}$ & $8.36\times10^{-3}$ & $1.35\times10^{3}$ & $5.00\times10^{-2}$ & $0$ & $1.00$& $2.398$ \\
\bottomrule
  \hline \hline
\end{tabular}
\end{table*}

\begin{table*}[!htb]
\caption{\label{tab:BBDBCwoLL_bigger_anharmonicity}
Optimized mode–mode coupling parameters of the MAB model trained from {Ferguson potential} with more sensitive anharmonicity setting with the BO+Drude SDF and SL interaction for (1) asymmetric stretch, ($1'$) symmetric stretch, and (2) bending modes.}
\begin{tabular}{cccc}
  \hline \hline
\toprule
  \hline
 $\mathrm{s-s'}$ & $\tilde{g}_{ss'}$  & $\tilde{g}_{s^2s'}$ & $\tilde{g}_{s{s'}^2}$  \\
\midrule
  $\mathrm{1-1'}$ &$2.36\times10^{-2}$ & $9.35\times10^{-3}$ & $3.44\times10^{-2}$ \\
  $\mathrm{1-2}$ &$-6.75\times10^{-2}$ & $0.103$ & $-2.65\times10^{-2}$  \\
  $\mathrm{1'-2}$  &$-3.390$ & $0.109$ & $2.93\times10^{-2}$ \\
\bottomrule
  \hline \hline
\end{tabular}
\end{table*}

\begin{table*}[!htb]
\caption{\label{tab:brownian_drude_bath_without_coupling_without_vll_bath_bigger_anharmonicity_ferguson}Optimized parameters of the MAB model trained from {Ferguson potential} with more sensitive anharmonicity setting with BO + Drude SDF and SL interaction for (1) asymmetric stretching, (1$'$) symmetric stretching, and (2) bending modes. 
Here,  $\tilde{\zeta_s^{\rm D}}$ denotes the normalized S-B coupling strength, and $\gamma_s^{\rm D}$ denotes the inverse correlation time of the bath fluctuations for the Drude mode and $\tilde{\zeta_s^{\rm B}}$ and $\gamma_s^{\rm B}$ are those for the BO mode.  The central frequency of the BO mode is expressed as $\omega_s^{\rm B}$, and  $V_{\mathrm{SL}}^{(s)}$ and $V_{\mathrm{LL}}^{(s)}$ denote the SL and LL interactions, 
$\tilde{g}_{s^3}$ is the cubic {anharmonicity} for the $s$ vibrational mode, respectively.
}
\begin{tabular}{cccccccccc}
  \hline \hline
\toprule
  \hline
 & $\omega_s$ (cm$^{-1}$) & $\gamma_s^{\rm D}/\omega_0$ & $\tilde{\zeta}_s^{\rm D}$  & $\gamma_s^{\rm B}/\omega_0$ & $\tilde{\zeta}_s^{\rm B}$ & $\omega_s^{\rm B}/\omega_0$   & $\tilde{V}_{\rm LL}^{(s)}$ &  $\tilde{V}_{\rm SL}^{(s)}$  & $\tilde{g}_{s^3}$  \\
\midrule
1 & 3513 & $2.42\times10^{-2}$ & $2.01\times10^{-2}$ & $6.99\times10^{-5}$ & $1.37\times10^{3}$ & $8.95\times10^{-3}$ & $0$ & $1.00$ & $9.44\times10^{-3}$  \\
$1'$ & 3413 & $2.42\times10^{-2}$ & $2.10\times10^{-2}$ & $9.21\times10^{-5}$ & $1.22\times10^{3}$ & $9.52\times10^{-3}$ & $0$ & $1.00$ & $0.127$ \\
2 & 1636 & $2.47\times10^{-3}$ & $6.65\times10^{3}$ & $8.21\times10^{-3}$ & $1.22\times10^{3}$ & $5.35\times10^{-2}$ & $0$ & $1.00$  & $1.662$ \\
\bottomrule
  \hline \hline
\end{tabular}
\end{table*}

\begin{table*}[!htb]
\caption{\label{tab:BBDBCwoLL_bigger_anharmonicity_ferguson}
Optimized mode–mode coupling parameters of the MAB model trained from {Ferguson potential} with more sensitive anharmonicity setting with the BO+Drude SDF and SL interaction for (1) asymmetric stretch, ($1'$) symmetric stretch, and (2) bending modes.}
\begin{tabular}{cccc}
  \hline \hline
\toprule
  \hline
 $\mathrm{s-s'}$ & $\tilde{g}_{ss'}$  & $\tilde{g}_{s^2s'}$ & $\tilde{g}_{s{s'}^2}$  \\
\midrule
  $\mathrm{1-1'}$ & $3.03\times10^{-2}$ & $1.35\times10^{-2}$ & $0.204$ \\
  $\mathrm{1-2}$ &$-3.32\times10^{-2}$ & $5.66\times10^{-2}$ & $-6.02\times10^{-2}$ \\
  $\mathrm{1'-2}$  & $-2.259$ & $5.36\times10^{-2}$ & $0.115$\\
\bottomrule
  \hline \hline
\end{tabular}
\end{table*}

\begin{table*}[!htb]
\caption{\label{tab:brownian_drude_bath_without_coupling_bath}Optimized parameters of the MAB model trained from {Ferguson potential} with BO + Drude SDF and LL+SL interaction for (1) asymmetric stretching, (1$'$) symmetric stretching, and (2) bending modes. Here,  $\tilde{\zeta_s^{\rm D}}$ denotes the normalized S-B coupling strength, and $\gamma_s^{\rm D}$ denotes the inverse correlation time of the bath fluctuations for the Drude mode and $\tilde{\zeta_s^{\rm B}}$ and $\gamma_s^{\rm B}$ are those for the BO mode.  The central frequency of the BO mode is expressed as $\omega_s^{\rm B}$, and  $V_{\mathrm{SL}}^{(s)}$ and $V_{\mathrm{LL}}^{(s)}$ denote the SL and LL interactions, 
$\tilde{g}_{s^3}$ is the qubic {anharmonicity} for the $s$ vibrational mode, respectively..}
\begin{tabular}{cccccccccc}
  \hline \hline
\toprule
 \hline
 & $\omega_s$ (cm$^{-1}$) & $\gamma_s^{\rm D}/\omega_0$ & $\tilde{\zeta}_s^{\rm D}$  & $\gamma_s^{\rm B}/\omega_0$ & $\tilde{\zeta}_s^{\rm B}$ & $\omega_s^{\rm B}/\omega_0$   & $\tilde{V}_{\rm LL}^{(s)}$ &  $\tilde{V}_{\rm SL}^{(s)}$  & $\tilde{g}_{s^3}$  \\
\midrule
1 & 3202 & $1.98\times10^{-2}$ & $1.24$ & $1.26\times10^{-2}$ & $61.57$ & $8.27\times10^{-3}$ & $6.13\times10^{-2}$ & $1.00$ & $9.58\times10^{-9}$ \\
$1'$ & 3123 & $1.51\times10^{-2}$ & $5.14$ & $1.77\times10^{-2}$ & $46.64$ & $1.19\times10^{-2}$ & $5.02\times10^{-2}$ & $1.00$ & $1.12\times10^{-8}$ \\
2 & 1596 & $1.38\times10^{-1}$ & $17.38$ & $9.77\times10^{-2}$ & $4.92\times10^{2}$ & $1.23\times10^{-1}$ & $1.51\times10^{-1}$ & $1.00$& $-2.06\times10^{-4}$\\
\bottomrule
  \hline \hline
\end{tabular}
\end{table*}

\begin{table*}[!htb]
\caption{\label{tab:DBCoupleSqcoupleAnh_coupling_mean_combo}
Optimized mode–mode coupling parameters of the MAB model  trained from {Ferguson potential}  with the BO+Drude SDF andLL+SLL interaction for (1) asymmetric stretch, ($1'$) symmetric stretch, and (2) bending modes.}
\begin{tabular}{cccc}
  \hline \hline
\toprule
  \hline
 $\mathrm{s-s'}$ & $\tilde{g}_{ss'}$  & $\tilde{g}_{s^2s'}$ & $\tilde{g}_{s{s'}^2}$  \\
\midrule
  $\mathrm{1-1'}$  & $6.80\times10^{-6}$ & $1.06\times10^{-8}$ & $1.01\times10^{-8}$ \\
  $\mathrm{1-2}$ & $-7.15\times10^{-5}$ & $-6.11\times10^{-9}$ & $3.86\times10^{-8}$ \\
  $\mathrm{1'-2}$& $-1.78\times10^{-4}$ & $1.29\times10^{-8}$ & $1.91\times10^{-7}$ \\
\bottomrule
  \hline \hline
\end{tabular}
\end{table*}

\begin{table*}[!htb]
\caption{\label{tab:brownian_drude_bath_without_coupling_bath_big_anharmonicity}Optimized parameters of the MAB model trained from {Ferguson potential} with more sensitive anharmonicity setting with BO + Drude SDF and LL+SL interaction for (1) asymmetric stretching, (1$'$) symmetric stretching, and (2) bending modes. Here,  $\tilde{\zeta_s^{\rm D}}$ denotes the normalized S-B coupling strength, and $\gamma_s^{\rm D}$ denotes the inverse correlation time of the bath fluctuations for the Drude mode and $\tilde{\zeta_s^{\rm B}}$ and $\gamma_s^{\rm B}$ are those for the BO mode.  The central frequency of the BO mode is expressed as $\omega_s^{\rm B}$, and  $V_{\mathrm{SL}}^{(s)}$ and $V_{\mathrm{LL}}^{(s)}$ denote the SL and LL interactions, 
$\tilde{g}_{s^3}$ is the qubic {anharmonicity} for the $s$ vibrational mode, respectively..}
\begin{tabular}{cccccccccc}
  \hline \hline
\toprule
 \hline
 & $\omega_s$ (cm$^{-1}$) & $\gamma_s^{\rm D}/\omega_0$ & $\tilde{\zeta}_s^{\rm D}$  & $\gamma_s^{\rm B}/\omega_0$ & $\tilde{\zeta}_s^{\rm B}$ & $\omega_s^{\rm B}/\omega_0$   & $\tilde{V}_{\rm LL}^{(s)}$ &  $\tilde{V}_{\rm SL}^{(s)}$  & $\tilde{g}_{s^3}$  \\
\midrule
1 & 3202 & $2.41\times10^{-2}$ & $4.09\times10^{-2}$ & $4.72\times10^{-5}$ & $2.67\times10^{3}$ & $1.07\times10^{-2}$ & $3.16\times10^{-1}$ & $1.00$&$1.72\times10^{-2}$ \\
$1'$ & 3123 & $2.42\times10^{-2}$ & $7.16\times10^{-2}$ & $5.32\times10^{-5}$ & $2.57\times10^{3}$ & $1.15\times10^{-2}$ & $3.12\times10^{-1}$ & $1.00$  & $4.28\times10^{-2}$\\
2 & 1592 & $2.55\times10^{-3}$ & $7.08\times10^{3}$ & $8.52\times10^{-3}$ & $1.37\times10^{3}$ & $5.17\times10^{-2}$ & $2.94\times10^{-1}$ & $1.00$  & $3.016$\\
\bottomrule
  \hline \hline
\end{tabular}
\end{table*}

\begin{table*}[!htb]
\caption{\label{tab:BBDBCoupleSqcoupleAnh_coupling_mean_combo_big_anharmonicity}
Optimized mode–mode coupling parameters of the MAB model  trained from {Ferguson potential} with more sensitive anharmonicity setting with the BO+Drude SDF andLL+SLL interaction for (1) asymmetric stretch, ($1'$) symmetric stretch, and (2) bending modes.}
\begin{tabular}{cccc}
  \hline \hline
\toprule
  \hline
 $\mathrm{s-s'}$ & $\tilde{g}_{ss'}$  & $\tilde{g}_{s^2s'}$ & $\tilde{g}_{s{s'}^2}$  \\
\midrule
  $\mathrm{1-1'}$ & $-4.97\times10^{-2}$ & $1.43\times10^{-3}$ & $4.04\times10^{-2}$  \\
  $\mathrm{1-2}$& $0.181$ & $0.101$ & $0.133$ \\
  $\mathrm{1'-2}$ & $-3.305$ & $0.101$ & $0.116$\\
\bottomrule
  \hline \hline
\end{tabular}
\end{table*}

\begin{table*}[!htb]
\caption{\label{tab:brownian_drude_bath_without_coupling_bath_ferguson_big_anharmonicity}Optimized parameters of the MAB model trained from {Ferguson potential} with more sensitive anharmonicity setting with BO + Drude SDF and LL+SL interaction for (1) asymmetric stretching, (1$'$) symmetric stretching, and (2) bending modes. Here,  $\tilde{\zeta_s^{\rm D}}$ denotes the normalized S-B coupling strength, and $\gamma_s^{\rm D}$ denotes the inverse correlation time of the bath fluctuations for the Drude mode and $\tilde{\zeta_s^{\rm B}}$ and $\gamma_s^{\rm B}$ are those for the BO mode.  The central frequency of the BO mode is expressed as $\omega_s^{\rm B}$, and  $V_{\mathrm{SL}}^{(s)}$ and $V_{\mathrm{LL}}^{(s)}$ denote the SL and LL interactions, 
$\tilde{g}_{s^3}$ is the qubic {anharmonicity} for the $s$ vibrational mode, respectively..}
\begin{tabular}{cccccccccc}
  \hline \hline
\toprule
 \hline
 & $\omega_s$ (cm$^{-1}$) & $\gamma_s^{\rm D}/\omega_0$ & $\tilde{\zeta}_s^{\rm D}$  & $\gamma_s^{\rm B}/\omega_0$ & $\tilde{\zeta}_s^{\rm B}$ & $\omega_s^{\rm B}/\omega_0$   & $\tilde{V}_{\rm LL}^{(s)}$ &  $\tilde{V}_{\rm SL}^{(s)}$  & $\tilde{g}_{s^3}$  \\
\midrule
1 & 3513 & $2.41\times10^{-2}$ & $2.24\times10^{-2}$ & $5.96\times10^{-5}$ & $1.57\times10^{3}$ & $9.06\times10^{-3}$ & $3.31\times10^{-1}$ & $1.00$ & $1.76\times10^{-2}$\\
$1'$ & 3413 & $2.42\times10^{-2}$ & $3.39\times10^{-2}$ & $8.61\times10^{-5}$ & $1.29\times10^{3}$ & $9.75\times10^{-3}$ & $3.26\times10^{-1}$ & $1.00$ &$0.132$  \\
2 & 1636 & $2.02\times10^{-3}$ & $8.32\times10^{3}$ & $8.90\times10^{-3}$ & $1.10\times10^{3}$ & $5.19\times10^{-2}$ & $2.97\times10^{-1}$ & $1.00$ & $1.731$\\
\bottomrule
  \hline \hline
\end{tabular}
\end{table*}

\begin{table*}[!htb]
\caption{\label{tab:BBDBCoupleSqcoupleAnh_coupling_mean_combo_big_anharmonicity}
Optimized mode–mode coupling parameters of the MAB model  trained from {Ferguson potential} with more sensitive anharmonicity setting with the BO+Drude SDF andLL+SLL interaction for (1) asymmetric stretch, ($1'$) symmetric stretch, and (2) bending modes.}
\begin{tabular}{cccc}
  \hline \hline
\toprule
  \hline
 $\mathrm{s-s'}$ & $\tilde{g}_{ss'}$  & $\tilde{g}_{s^2s'}$ & $\tilde{g}_{s{s'}^2}$  \\
\midrule
  $\mathrm{1-1'}$ & $5.23\times10^{-3}$ & $1.05\times10^{-2}$ & $0.204$\\
  $\mathrm{1-2}$ & $-5.71\times10^{-2}$ & $7.21\times10^{-2}$ & $-4.84\times10^{-2}$ \\
  $\mathrm{1'-2}$& $-2.322$ & $7.05\times10^{-2}$ & $4.56\times10^{-2}$ \\
\bottomrule
  \hline \hline
\end{tabular}
\end{table*}
{The results} {for} BO + Drude SDF with SL interaction are presented below.
In this framework, the overdamped Drude SDF {describes} the relaxation of intramolecular modes, while the underdamped BO SDF {represents interactions with low-frequency intermolecular modes, including libration and hydrogen-bond translation. The inclusion of BO baths was guided by} insights from previous studies.\cite{UT20JCTC} 
{Incorporating BO modes was found to enhance learning efficiency relative to the Drude SDF alone.}

{For consistency with} the collective coordinate representation, the BO parameters were normalized {according to} $\tilde{\zeta}_s^{\rm B} \equiv \zeta_s^{\rm B} (\omega_0/\omega_s)^2$, with $\gamma_s^{\rm B}$ and $\omega_s^{\rm B}$ reported as $\gamma_s^{\rm B}/\omega_0$ and $\omega_s^{\rm B}/\omega_0$, respectively.

Table \ref{tab:brownian_drude_bath_without_coupling_without_vll_bath} summarizes the bath parameters and the potential anharmonicity as evaluated by ML. Although the inclusion of BO modes does not {alter} the correlation time of the Drude mode, it leads to a reduction in its coupling strength. In contrast, the coupling strength associated with the BO modes is {substantially larger, indicating pronounced} coupling between intramolecular and intermolecular vibrational modes.

Table \ref{tab:BBDBCwoLL} presents the evaluated mode coupling parameters. These values remain largely unchanged despite {changes in the bath configuration, indicating that mode coupling, as a mechanical interaction, may} be treated independently of the thermal bath configuration.

\subsubsection{BO + Drude bath with LL+SL interaction}

We finally present the BO + Drude SDF results  incorporating both LL and SL interactions.
The results are shown in Tables \ref{tab:brownian_drude_bath_without_coupling_bath} and \ref{tab:DBCoupleSqcoupleAnh_coupling_mean_combo}. 
{As in the Drude-only case, inclusion of LL coupling results in minimal changes to bath parameters and mode coupling strengths relative to the SL-only optimization.}

Within the MAB framework, the addition of LL coupling does not modify the optical profile of high-frequency intramolecular modes relative to the SL-only case, {indicating that its exclusion during model parameterization may be justified.}
Nonetheless, {improved ML efficiency would enable seamless incorporation of $V_{\rm LL}$ into the training process.}

The ML results described above faithfully reproduce the behavior of the original MD trajectories. Consequently, when constructing models for 2D spectroscopic simulations, it is essential to utilize MD trajectories generated with a highly descriptive force field. {Leveraging both quantum-dynamic and classical-dynamic trajectories facilitates more effective identification of quantum effects.}

\subsection{Linear absorption spectra}
\begin{figure}[!t]
  \centering
  \includegraphics[scale=0.35]{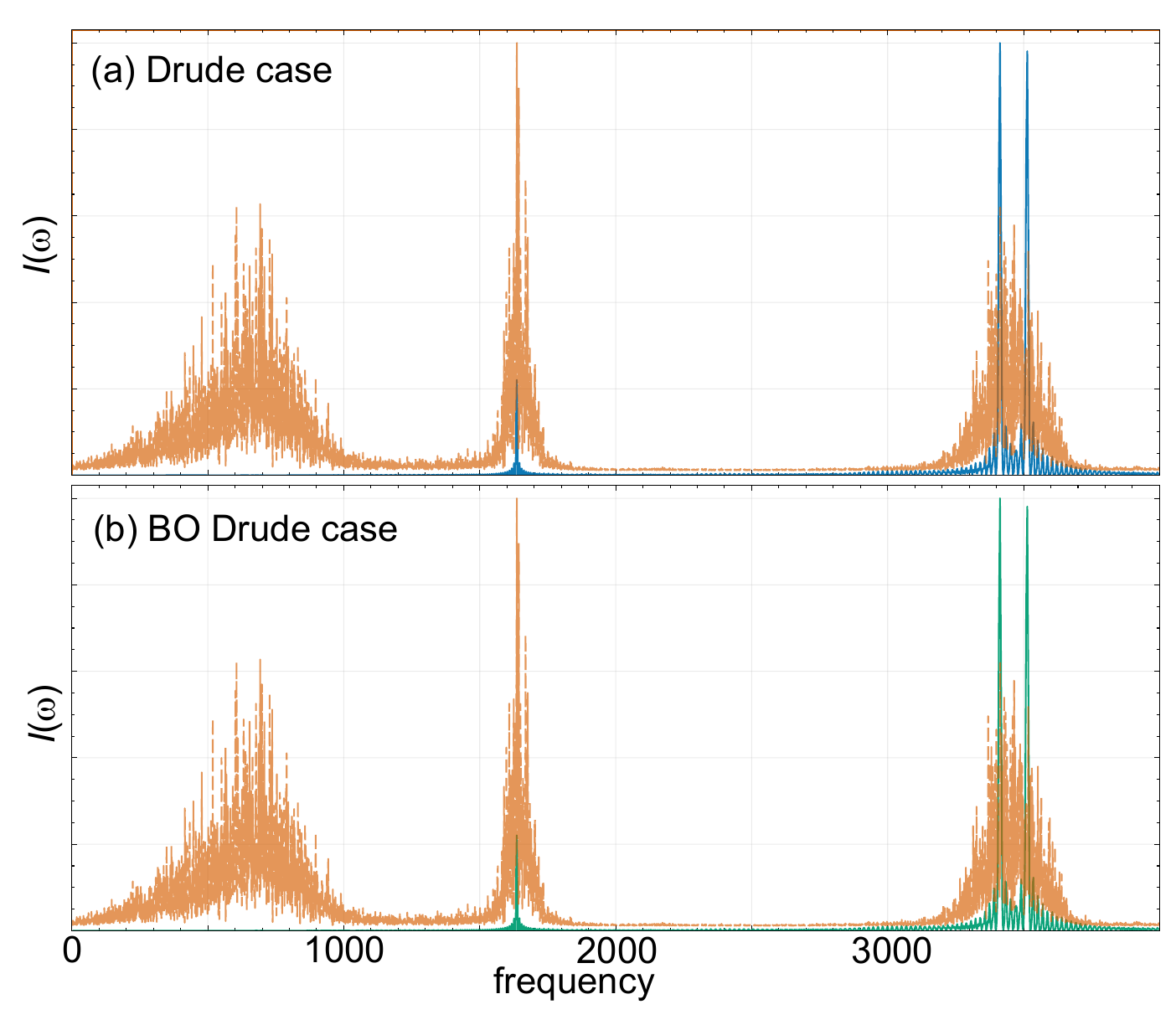}
    \caption{\label{fgr:db_ir}Infrared absorption spectra obtained from HEOM calculations using the optimized MAB model parameters for (a) Drude SDF case (blue curves) and (b) BO+Drude case (green curves). For comparison, each figure also includes results from MD simulations (orange lines) and experimental data.(black dotted curve).\cite{IRexp2011}
 }
\end{figure}
The infrared absorption spectrum is calculated from\cite{IT16JCP,TT23JCP1}
\begin{equation}
I(\omega) = \omega\,\Im\!\int_{0}^{\infty}\!dt\,e^{i\omega t}\,R^{(1)}(t),
\end{equation}
where the first-order response function of dipole moment is defined as
$R^{(1)}(t)=i\big\langle[\hat\mu(t),\hat\mu(0)]\big\rangle/\hbar$.
We can rewrite the response function as\cite{T06JPSJ,T20JCP}
\begin{eqnarray}
\label{R_1t}
R^{(1)}(t) = \frac{i}{\hbar}{\rm Tr}\left\{ \hat\mu \hat{\mathcal{G}} (t) 
\hat\mu^{\times} \hat \rho^{eq} \right\},
\end{eqnarray}
where $\hat{\mathcal{G}}(t)$ represents the Green's function associated with Eq. \eqref{eq:HEOM_DB}, while $\hat{\rho}^{eq}$ represents the equilibrium density operator, derived from the steady-state solution of the HEOM.
To evaluate $R^{(1)}(t)$, we solve the HEOM starting from the initial state at $t=0$, given as $\hat{\mu}^{\times} \hat{\rho}^{eq}$. The solution obtained at time $t$ is referred to as $\hat \rho'(t)$. The response function is then calculated as the expectation value: $R^{(1)}(t) = i {\rm tr}\{ \hat\mu \hat \rho'(t) \}/\hbar $.
HEOM calculations were implemented in Python using NumPy 1.26.3 and Numba 0.60.0. Time-dependent HEOM equations were solved via the fourth-order Runge–Kutta method.
By employing the learned parameters of the MAB model, we integrate the HEOM formalism to derive linear-response spectra without relying on ad hoc fitting approaches, either to experimental data or simulated spectra. The training process, grounded in MD trajectories, ensures that the resultant parameters remain consistent with actual microscopic dynamics, thereby reducing ambiguities inherent in purely spectral-based fitting.
For the (a) Drude case, both the quantum hierarchical Fokker–Planck equation (QHFPE) \cite{TT23JCP1,TT23JCP2} and the classical hierarchical Fokker–Planck equation (CHFPE) \cite{IIT15JCP,IT16JCP,HT25JCP1,HT25JCP2} have been developed to compute 2D vibrational spectra.

The HEOM code capable of simulating 2D vibrational spectrosctra in BO+Drude case is currently under development. Here, as a demonstration, we present the results of simulating the linear absorption spectrum using HEOM (Eq.\eqref{HEOM}) for each mode solbed independently, ignoring coupling between modes in both (a) Drude and (b) BO+Drude cases on the basis of the program developed for BO+Drude 2D electronic spectroscopy.\cite{T12JCP}
Notably, in the context of linear absorption spectroscopy, the influence of both potential anharmonicity and anharmonic mode coupling is typically limited. This is especially true in the present case, where their contributions are vanishingly small.

To apply the HEOM formalism, the eigenenergies of the Hamiltonian for each vibrational mode were calculated and quantized, incorporating the effects of zero-point vibrations. The linear response function in Eq. \eqref{R_1t} was evaluated by numerically integrating the HEOM in Eq. \eqref{HEOM} for two cases: (a) the Drude model, with parameters listed in Table \ref{tab:drude_bath_without_coupling_without_vll_bath}, and (b) the Brownian oscillator (BO) + Drude model, with parameters given in Table \ref{tab:brownian_drude_bath_without_coupling_without_vll_bath}.

{Figure \ref{fgr:db_ir} presents the calculated absorption spectra. For comparison, each panel also includes spectra obtained from MD simulations. In the MD analysis, the Cartesian components of the dipole moment were extracted from the trajectory, followed by computation of the autocorrelation functions, averaging, and Fourier transformation to yield the absorption spectra.}

{In the MD results, the symmetric and antisymmetric stretching peaks appear broadened and overlapping, whereas the HEOM spectra resolve these peaks distinctly. This discrepancy arises because the MD spectra reflect dipole fluctuations influenced by many-body interactions, while the HEOM spectra are derived from a single-molecule model. Notably, both the Drude and BO+Drude cases yield similar spectral profiles, indicating that the linear absorption spectrum is governed primarily by simple excitation processes. The influence of the thermal bath—crucial for relaxation dynamics—is relatively minor in this context. Therefore, further investigation using 2D spectroscopy is essential to elucidate the underlying mechanisms in greater detail.}

\section{Conclusion}
\label{sec:conclusion}

We developed {a} ML algorithm that optimizes the parametric variables of the MAB model{---}a framework that captures intramolecular vibrational modes in concert with their surrounding environments{---}using MD  trajectories as the foundational data source.

Cross-validation across diverse time windows and molecular systems confirms the model's {generalizability} beyond the training subsets. Key physical parameters remain stable under data resampling, underscoring the robustness of the proposed approach. The results enable precise determination of essential quantities, including anharmonic mode coupling and the characteristics of Drude and BO + Drude baths. 
{These parameters provide the foundation}
for constructing a reliable model capable of calculating 2D vibrational spectra, including 
2D IR spectra via the various forms of HEOM. This extension facilitates systematic evaluation of mode couplings and bath architectures within 2D spectra, while maintaining consistency with the underlying MD.

While the present study focuses on water and specific bath couplings, the methodology is broadly applicable to other molecular systems. Its accuracy is expected to improve with trajectories incorporating quantum nuclear effects or refined interaction potentials.{\cite{JianLiu2018H2OMP,Imoto_JCP135,medders2015irraman,Paesan2018H2OCMD} }

{For both the Drude model and the BO+Drude model, we calculated and compared the linear absorption spectra. However, in this spectrum, which examines only the excitation process from the ground state, no difference between the two models was observed. Although 2D spectral calculations are not presented in this paper, they will be reported separately.}

\section*{Acknowledgments}
Y. T. was supported by JST (Grant No. CREST 1002405000170). K. P. acknowledges a fellowship supported by JST SPRING, the establishment of university fellowships toward the creation of science technology innovation (Grant No.~JPMJSP2110). J. J. was supported by JSPS KAKENHI (Grant No. 24K23103)

\section*{Author declarations}
\subsection*{Conflict of Interest}
The authors have no conflicts to disclose.

\section*{Data availability}
The data that support the findings of this study are available from the corresponding author upon reasonable request.

\appendix
\section{Coordiniate Mapping}
\label{sec:coordinatemapping}

To evaluate training efficiency, we compared two distinct {representations:} one based on (a) Cartesian atomic coordinates and the other on (b) normal-mode vibrational coordinates.
In the (a) Cartesian {representation,} the potential profile of the learnable system is optimized using internal coordinates, specifically, the two OH bond lengths, the HOH bending angle, and the SDF. {Following optimization, the MAB model parameters are evaluated in the normal-mode coordinate space,} denoted as $q_s$.

In contrast, under the framework referred to as case (b), each vibrational mode and its associated Liouvillian are constructed directly from Cartesian atomic coordinates. The effective potential $U_s(q_s)$ and the {$J_s(\omega)$ for each mode are subsequently} optimized as functions of $q_s$. This approach enables more direct optimization with respect to the MAB variable $q_s$, rather than relying on atomic coordinates as in case (a).

\begin{figure}[!htb]
  \centering
\includegraphics[scale=0.5]{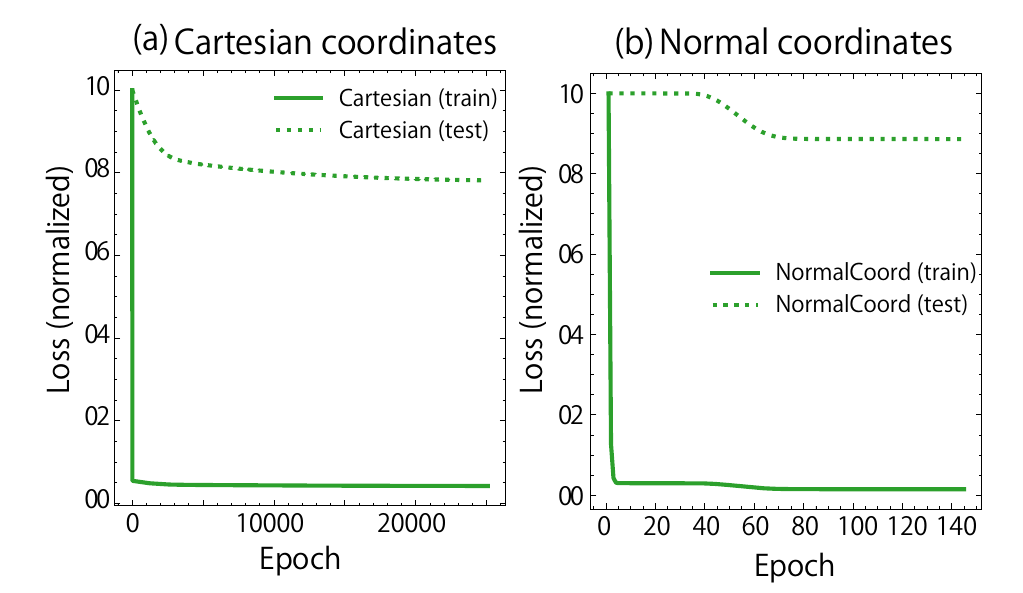}
    \caption{\label{fgr:atomicVSmolecular}Training and testing losses were evaluated by comparing predicted model trajectories with actual MD trajectories, using two coordinate systems: atomic coordinates in Cartesian space and normal-mode coordinates. In both cases, the system's time evolution was governed by the corresponding MD Liouvillian.}
\end{figure}

\begin{figure}[!htb]
  \centering
  \includegraphics[scale=0.55]{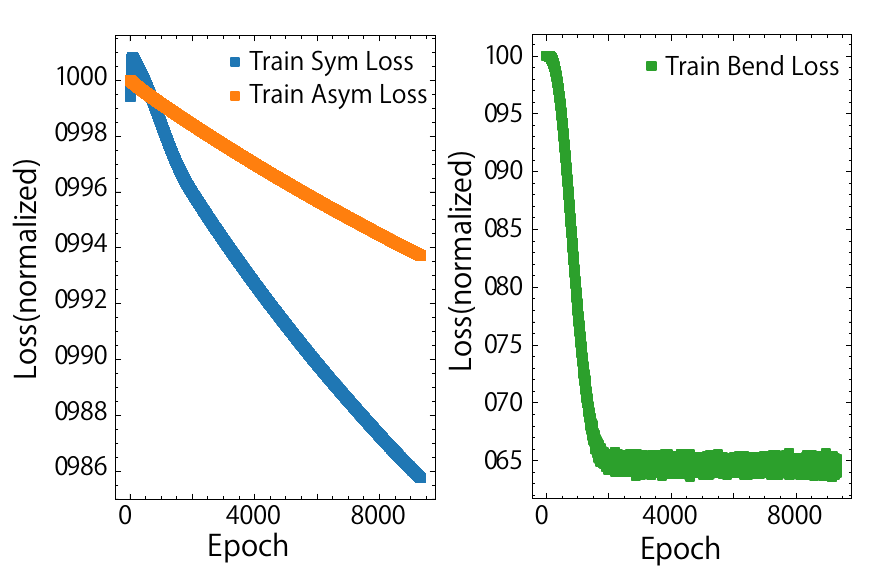}
    \caption{\label{fgr:sym_asym_decompose}The left panel shows training losses for the OH symmetric and OH asymmetric stretch modes, while the right panel displays the loss for the HOH bending mode. The bending mode demonstrates a more rapid learning process compared to the stretch modes.}
\end{figure}

Figure~\ref{fgr:atomicVSmolecular} presents a comparison of training and test losses for cases (a) and (b). The normal mode framework exhibits faster convergence than the atomic framework, although its generality across {molecular systems is reduced by} the molecule-specific nature of normal modes. A further advantage of the normal mode formulation is that {the objective function naturally decomposes by mode, facilitating} mode-resolved diagnostics. Figure~\ref{fgr:sym_asym_decompose} illustrates the training losses for the OH symmetric stretch, OH asymmetric stretch, and HOH bending modes. Notably, the bending mode converges substantially faster than the stretching modes.

\section{Early Stopping in Model Training}
\label{sec:earlystopping}

{To mitigate overfitting and enhance} generalization, early stopping\cite{NIPS1993_43fa7f58} was applied {to each model}. Validation 
{loss was monitored continuously, and training was halted upon stagnation or degradation of validation loss.} Specifically, a patience threshold of 300 epochs was employed. If the best validation loss remained unchanged for 300 consecutive epochs, early stopping was triggered. Upon activation, training resumed once with a reduced learning rate for a further 300 epochs. If no improvement occurred during this second phase, training was subsequently terminated. This protocol effectively mitigated overfitting {while conservatively enabling escape from} shallow plateaus via adaptive learning rate adjustment.

\section{Cross-Validation}
\label{sec:cv}

Model performance was assessed via cross-validation (CV). For each {fold,} the fitted physical parameters were recorded, and their {inter-fold} variation across folds was analyzed. This dispersion {serves as a quantitative indicator of parameter stability and reflects} the model's sensitivity to the choice of time windows and molecular subsets employed during training. 	The present analysis focuses on two key aspects: the representation of water molecules extracted from MD trajectories and the influence of the selected time window.

Accordingly, two distinct CV strategies were evaluated: molecule-level cross-validation (MOLVC) and time-step cross-validation (TSCV). The MOLVC approach {represents a scenario in which each molecule interacts with multiple baths,} while the TSCV reflects a situation in which a single molecule interacts with a single bath over an extended duration. Notably, the molecule-level strategy 
{systematically underestimates the strength of system–bath coupling. Therefore, the time-step approach 
was therefore adopted for subsequent evaluation.}

\subsection{Molecule-Level Cross-Validation (MOLVC)}

For molecule-level assessment, four-fold cross-validation was conducted over individual molecules. In each {fold, three subset} were used for training and one for testing, with roles rotated such that each subset served as the test set once. This protocol ensured rigorous separation between training and testing data, {enabling evaluation of the model’s ability to generalize across distinct molecular configurations exhibiting} potentially diverse dynamical behavior.

\subsection{Time-Step Cross-Validation (TSCV)}

{Within the time-step framework, data partitions were constructed to retain the temporal ordering of the dynamics.
This approach is specifically intended to evaluate} temporal consistency, requiring the model to {produce forward predictions from earlier to later configurations. By preserving chronological order,this strategy enables a more realistic assessment of the model’s predictive performance over time.}

\subsection{Ferguson potential(flexible SPC water)}
\label{sec:mdpotential}
Molecules are indexed by $i,j$; atomic sites within a molecule by
$a,b\in\{\mathrm O,\mathrm H_1,\mathrm H_2\}$.
$r_{OO}^{ij}$ is the O--O distance between molecules $i$ and $j$.
$r_{ab}$ is the distance between site $a\in i$ and $b\in j$ (intermolecular).
$q_a$ are partial charges; $k_e=1/(4\pi\varepsilon_0)$.
Lennard--Jones parameters are $\sigma$ (size) and $\varepsilon$ (well depth).
Intramolecular geometry uses two O--H bond lengths
$r_{i,\mathrm{OH}_1}, r_{i,\mathrm{OH}_2}$ and the H--O--H angle $\theta_i$.
Stretch/bend parameters are $(k_b,r_0)$ and $(k_\theta,\theta_0)$, respectively.
Unless noted, intramolecular nonbonded interactions are excluded by the topology.

This model keeps SPC--style nonbonded interactions and introduces
an anharmonic(cubic) O--H stretch plus a harmonic bend:
\begin{eqnarray}
V_{\mathrm{Ferguson}}
&&= \sum_{i<j}\Bigg[
4\varepsilon\!\left(\Big(\tfrac{\sigma}{r_{OO}^{ij}}\Big)^{12}
                 - \Big(\tfrac{\sigma}{r_{OO}^{ij}}\Big)^6\right)
+ \sum_{a\in i}\sum_{b\in j} \frac{k_e\,q_a q_b}{r_{ab}}
\Bigg] \nonumber\\
&&\quad + \sum_i \Big[
k_b(r_{i,\mathrm{OH}_1}-r_0)^2
+ k_b k_{\mathrm{cub}}(r_{i,\mathrm{OH}_1}-r_0)^3 \nonumber\\
&&\qquad\ \ + k_b(r_{i,\mathrm{OH}_2}-r_0)^2
+ k_b k_{\mathrm{cub}}(r_{i,\mathrm{OH}_2}-r_0)^3  \nonumber\\
&&\qquad\ \ + \tfrac12 k_\theta(\theta_i-\theta_0)^2
\Big].
\label{eq:Ferguson}
\end{eqnarray}
\noindent The cubic coefficient $k_{\mathrm{cub}}$ renders the stretch
asymmetric about $r_0$, improving vibrational behavior versus purely
harmonic stretches; the bend remains harmonic about $\theta_0$.

\bibliography{tanimura_publist.bib,TT23.bib,HT24.bib,PJT25.bib}

\begin{thebibliography}{85}%
\makeatletter
\providecommand \@ifxundefined [1]{%
 \@ifx{#1\undefined}
}%
\providecommand \@ifnum [1]{%
 \ifnum #1\expandafter \@firstoftwo
 \else \expandafter \@secondoftwo
 \fi
}%
\providecommand \@ifx [1]{%
 \ifx #1\expandafter \@firstoftwo
 \else \expandafter \@secondoftwo
 \fi
}%
\providecommand \natexlab [1]{#1}%
\providecommand \enquote  [1]{``#1''}%
\providecommand \bibnamefont  [1]{#1}%
\providecommand \bibfnamefont [1]{#1}%
\providecommand \citenamefont [1]{#1}%
\providecommand \href@noop [0]{\@secondoftwo}%
\providecommand \href [0]{\begingroup \@sanitize@url \@href}%
\providecommand \@href[1]{\@@startlink{#1}\@@href}%
\providecommand \@@href[1]{\endgroup#1\@@endlink}%
\providecommand \@sanitize@url [0]{\catcode `\\12\catcode `\$12\catcode
  `\&12\catcode `\#12\catcode `\^12\catcode `\_12\catcode `\%12\relax}%
\providecommand \@@startlink[1]{}%
\providecommand \@@endlink[0]{}%
\providecommand \url  [0]{\begingroup\@sanitize@url \@url }%
\providecommand \@url [1]{\endgroup\@href {#1}{\urlprefix }}%
\providecommand \urlprefix  [0]{URL }%
\providecommand \Eprint [0]{\href }%
\providecommand \doibase [0]{https://doi.org/}%
\providecommand \selectlanguage [0]{\@gobble}%
\providecommand \bibinfo  [0]{\@secondoftwo}%
\providecommand \bibfield  [0]{\@secondoftwo}%
\providecommand \translation [1]{[#1]}%
\providecommand \BibitemOpen [0]{}%
\providecommand \bibitemStop [0]{}%
\providecommand \bibitemNoStop [0]{.\EOS\space}%
\providecommand \EOS [0]{\spacefactor3000\relax}%
\providecommand \BibitemShut  [1]{\csname bibitem#1\endcsname}%
\let\auto@bib@innerbib\@empty
\bibitem [{\citenamefont {Mukamel}(1999)}]{mukamel1999principles}%
  \BibitemOpen
  \bibfield  {author} {\bibinfo {author} {\bibfnamefont {S.}~\bibnamefont
  {Mukamel}},\ }\href@noop {} {\emph {\bibinfo {title} {Principles of nonlinear
  optical spectroscopy}}},\ \bibinfo {number} {6}\ (\bibinfo  {publisher}
  {Oxford University Press on Demand},\ \bibinfo {year} {1999})\BibitemShut
  {NoStop}%
\bibitem [{\citenamefont {Cho}(2009)}]{Cho2009}%
  \BibitemOpen
  \bibfield  {author} {\bibinfo {author} {\bibfnamefont {M.}~\bibnamefont
  {Cho}},\ }\href {https://doi.org/https://doi.org/10.1201/9781420084306}
  {\emph {\bibinfo {title} {Two-Dimensional Optical Spectroscopy}}}\ (\bibinfo
  {publisher} {CRC Press},\ \bibinfo {year} {2009})\BibitemShut {NoStop}%
\bibitem [{\citenamefont {Hamm}\ and\ \citenamefont
  {Zanni}(2011)}]{Hamm2011ConceptsAM}%
  \BibitemOpen
  \bibfield  {author} {\bibinfo {author} {\bibfnamefont {P.}~\bibnamefont
  {Hamm}}\ and\ \bibinfo {author} {\bibfnamefont {M.~T.}\ \bibnamefont
  {Zanni}},\ }\href {https://doi.org/https://doi.org/10.1017/CBO9780511675935}
  {\emph {\bibinfo {title} {Concepts and Methods of 2D Infrared
  Spectroscopy}}}\ (\bibinfo  {publisher} {Cambridge University Press},\
  \bibinfo {year} {2011})\BibitemShut {NoStop}%
\bibitem [{\citenamefont {Hamm}\ and\ \citenamefont
  {Shalit}(2017)}]{HammPerspH2O2017}%
  \BibitemOpen
  \bibfield  {author} {\bibinfo {author} {\bibfnamefont {P.}~\bibnamefont
  {Hamm}}\ and\ \bibinfo {author} {\bibfnamefont {A.}~\bibnamefont {Shalit}},\
  }\bibfield  {title} {\enquote {\bibinfo {title} {Perspective: Echoes in
  2d-\uppercase{R}aman-\uppercase{TH}z spectroscopy},}\ }\href
  {https://doi.org/10.1063/1.4979288} {\bibfield  {journal} {\bibinfo
  {journal} {The Journal of Chemical Physics}\ }\textbf {\bibinfo {volume}
  {146}},\ \bibinfo {pages} {130901} (\bibinfo {year} {2017})},\ \Eprint
  {https://arxiv.org/abs/https://doi.org/10.1063/1.4979288}
  {https://doi.org/10.1063/1.4979288} \BibitemShut {NoStop}%
\bibitem [{\citenamefont {Jansen}\ \emph {et~al.}(2019)\citenamefont {Jansen},
  \citenamefont {Saito}, \citenamefont {Jeon},\ and\ \citenamefont
  {Cho}}]{JansenChoShinji2DVPerspe2019}%
  \BibitemOpen
  \bibfield  {author} {\bibinfo {author} {\bibfnamefont {T.~l.~C.}\
  \bibnamefont {Jansen}}, \bibinfo {author} {\bibfnamefont {S.}~\bibnamefont
  {Saito}}, \bibinfo {author} {\bibfnamefont {J.}~\bibnamefont {Jeon}},\ and\
  \bibinfo {author} {\bibfnamefont {M.}~\bibnamefont {Cho}},\ }\bibfield
  {title} {\enquote {\bibinfo {title} {Theory of coherent two-dimensional
  vibrational spectroscopy},}\ }\href {https://doi.org/10.1063/1.5083966}
  {\bibfield  {journal} {\bibinfo  {journal} {The Journal of Chemical Physics}\
  }\textbf {\bibinfo {volume} {150}},\ \bibinfo {pages} {100901} (\bibinfo
  {year} {2019})},\ \Eprint
  {https://arxiv.org/abs/https://doi.org/10.1063/1.5083966}
  {https://doi.org/10.1063/1.5083966} \BibitemShut {NoStop}%
\bibitem [{\citenamefont {Saito}\ and\ \citenamefont
  {Ohmine}(2006)}]{Shinji2DRaman2006}%
  \BibitemOpen
  \bibfield  {author} {\bibinfo {author} {\bibfnamefont {S.}~\bibnamefont
  {Saito}}\ and\ \bibinfo {author} {\bibfnamefont {I.}~\bibnamefont {Ohmine}},\
  }\bibfield  {title} {\enquote {\bibinfo {title} {Fifth-order two-dimensional
  \uppercase{R}aman spectroscopy of liquid water, crystalline ice ih and
  amorphous ices: Sensitivity to anharmonic dynamics and local hydrogen bond
  network structure},}\ }\href {https://doi.org/10.1063/1.2232254} {\bibfield
  {journal} {\bibinfo  {journal} {The Journal of Chemical Physics}\ }\textbf
  {\bibinfo {volume} {125}},\ \bibinfo {pages} {084506} (\bibinfo {year}
  {2006})},\ \Eprint {https://arxiv.org/abs/https://doi.org/10.1063/1.2232254}
  {https://doi.org/10.1063/1.2232254} \BibitemShut {NoStop}%
\bibitem [{\citenamefont {Hasegawa}\ and\ \citenamefont
  {Tanimura}(2006)}]{HT06JCP}%
  \BibitemOpen
  \bibfield  {author} {\bibinfo {author} {\bibfnamefont {T.}~\bibnamefont
  {Hasegawa}}\ and\ \bibinfo {author} {\bibfnamefont {Y.}~\bibnamefont
  {Tanimura}},\ }\bibfield  {title} {\enquote {\bibinfo {title} {Calculating
  fifth-order \uppercase{R}aman signals for various molecular liquids by
  equilibrium and nonequilibrium hybrid molecular dynamics simulation
  algorithms},}\ }\href {https://doi.org/10.1063/1.2217947} {\bibfield
  {journal} {\bibinfo  {journal} {The Journal of Chemical Physics}\ }\textbf
  {\bibinfo {volume} {125}},\ \bibinfo {pages} {074512} (\bibinfo {year}
  {2006})}\BibitemShut {NoStop}%
\bibitem [{\citenamefont {Li}\ \emph {et~al.}(2008)\citenamefont {Li},
  \citenamefont {Huang}, \citenamefont {Dwayne~Miller}, \citenamefont
  {Hasegawa},\ and\ \citenamefont {Tanimura}}]{LHDHT08JCP}%
  \BibitemOpen
  \bibfield  {author} {\bibinfo {author} {\bibfnamefont {Y.~L.}\ \bibnamefont
  {Li}}, \bibinfo {author} {\bibfnamefont {L.}~\bibnamefont {Huang}}, \bibinfo
  {author} {\bibfnamefont {R.~J.}\ \bibnamefont {Dwayne~Miller}}, \bibinfo
  {author} {\bibfnamefont {T.}~\bibnamefont {Hasegawa}},\ and\ \bibinfo
  {author} {\bibfnamefont {Y.}~\bibnamefont {Tanimura}},\ }\bibfield  {title}
  {\enquote {\bibinfo {title} {Two-dimensional fifth-order \uppercase{R}aman
  spectroscopy of liquid formamide: Experiment and theory},}\ }\href
  {https://doi.org/10.1063/1.2927311} {\bibfield  {journal} {\bibinfo
  {journal} {The Journal of Chemical Physics}\ }\textbf {\bibinfo {volume}
  {128}},\ \bibinfo {pages} {234507} (\bibinfo {year} {2008})}\BibitemShut
  {NoStop}%
\bibitem [{\citenamefont {Hasegawa}\ and\ \citenamefont
  {Tanimura}(2008)}]{HT08JCP}%
  \BibitemOpen
  \bibfield  {author} {\bibinfo {author} {\bibfnamefont {T.}~\bibnamefont
  {Hasegawa}}\ and\ \bibinfo {author} {\bibfnamefont {Y.}~\bibnamefont
  {Tanimura}},\ }\bibfield  {title} {\enquote {\bibinfo {title} {Nonequilibrium
  molecular dynamics simulations with a backward-forward trajectories sampling
  for multidimensional infrared spectroscopy of molecular vibrational modes},}\
  }\href {https://doi.org/10.1063/1.2828189} {\bibfield  {journal} {\bibinfo
  {journal} {The Journal of Chemical Physics}\ }\textbf {\bibinfo {volume}
  {128}},\ \bibinfo {pages} {064511} (\bibinfo {year} {2008})}\BibitemShut
  {NoStop}%
\bibitem [{\citenamefont {Pan}\ \emph {et~al.}(2015)\citenamefont {Pan},
  \citenamefont {Wu}, \citenamefont {Jin}, \citenamefont {Liu}, \citenamefont
  {Nagata}, \citenamefont {Zhang},\ and\ \citenamefont
  {Zhuang}}]{Wei2015Nagata2DRamanTHz}%
  \BibitemOpen
  \bibfield  {author} {\bibinfo {author} {\bibfnamefont {Z.}~\bibnamefont
  {Pan}}, \bibinfo {author} {\bibfnamefont {T.}~\bibnamefont {Wu}}, \bibinfo
  {author} {\bibfnamefont {T.}~\bibnamefont {Jin}}, \bibinfo {author}
  {\bibfnamefont {Y.}~\bibnamefont {Liu}}, \bibinfo {author} {\bibfnamefont
  {Y.}~\bibnamefont {Nagata}}, \bibinfo {author} {\bibfnamefont
  {R.}~\bibnamefont {Zhang}},\ and\ \bibinfo {author} {\bibfnamefont
  {W.}~\bibnamefont {Zhuang}},\ }\bibfield  {title} {\enquote {\bibinfo {title}
  {Low frequency 2\uppercase{D} \uppercase{R}aman-\uppercase{TH}z spectroscopy
  of ionic solution: A simulation study},}\ }\href
  {https://doi.org/10.1063/1.4917260} {\bibfield  {journal} {\bibinfo
  {journal} {The Journal of Chemical Physics}\ }\textbf {\bibinfo {volume}
  {142}},\ \bibinfo {pages} {212419} (\bibinfo {year} {2015})},\ \Eprint
  {https://arxiv.org/abs/https://pubs.aip.org/aip/jcp/article-pdf/doi/10.1063/1.4917260/13242549/212419\_1\_online.pdf}
  {https://pubs.aip.org/aip/jcp/article-pdf/doi/10.1063/1.4917260/13242549/212419\_1\_online.pdf}
  \BibitemShut {NoStop}%
\bibitem [{\citenamefont {Ito}, \citenamefont {Hasegawa},\ and\ \citenamefont
  {Tanimura}(2014)}]{IHT14JCP}%
  \BibitemOpen
  \bibfield  {author} {\bibinfo {author} {\bibfnamefont {H.}~\bibnamefont
  {Ito}}, \bibinfo {author} {\bibfnamefont {T.}~\bibnamefont {Hasegawa}},\ and\
  \bibinfo {author} {\bibfnamefont {Y.}~\bibnamefont {Tanimura}},\ }\bibfield
  {title} {\enquote {\bibinfo {title} {Calculating two-dimensional
  \uppercase{TH}z-\uppercase{R}aman-\uppercase{TH}z and
  \uppercase{R}aman-\uppercase{TH}z-\uppercase{TH}z signals for various
  molecular liquids: The samplers},}\ }\href
  {https://doi.org/10.1063/1.4895908} {\bibfield  {journal} {\bibinfo
  {journal} {The Journal of Chemical Physics}\ }\textbf {\bibinfo {volume}
  {141}},\ \bibinfo {pages} {124503} (\bibinfo {year} {2014})}\BibitemShut
  {NoStop}%
\bibitem [{\citenamefont {Ito}, \citenamefont {Hasegawa},\ and\ \citenamefont
  {Tanimura}(2016)}]{IHT16JPCL}%
  \BibitemOpen
  \bibfield  {author} {\bibinfo {author} {\bibfnamefont {H.}~\bibnamefont
  {Ito}}, \bibinfo {author} {\bibfnamefont {T.}~\bibnamefont {Hasegawa}},\ and\
  \bibinfo {author} {\bibfnamefont {Y.}~\bibnamefont {Tanimura}},\ }\bibfield
  {title} {\enquote {\bibinfo {title} {Effects of intermolecular charge
  transfer in liquid water on \uppercase{R}aman spectra},}\ }\href
  {https://doi.org/10.1021/acs.jpclett.6b01766} {\bibfield  {journal} {\bibinfo
   {journal} {The Journal of Physical Chemistry Letters}\ }\textbf {\bibinfo
  {volume} {7}},\ \bibinfo {pages} {4147--4151} (\bibinfo {year}
  {2016})}\BibitemShut {NoStop}%
\bibitem [{\citenamefont {Steinel}\ \emph {et~al.}(2004)\citenamefont
  {Steinel}, \citenamefont {Asbury}, \citenamefont {Corcelli}, \citenamefont
  {Lawrence}, \citenamefont {Skinner},\ and\ \citenamefont
  {Fayer}}]{SkinnerCPL2004}%
  \BibitemOpen
  \bibfield  {author} {\bibinfo {author} {\bibfnamefont {T.}~\bibnamefont
  {Steinel}}, \bibinfo {author} {\bibfnamefont {J.~B.}\ \bibnamefont {Asbury}},
  \bibinfo {author} {\bibfnamefont {S.}~\bibnamefont {Corcelli}}, \bibinfo
  {author} {\bibfnamefont {C.}~\bibnamefont {Lawrence}}, \bibinfo {author}
  {\bibfnamefont {J.}~\bibnamefont {Skinner}},\ and\ \bibinfo {author}
  {\bibfnamefont {M.}~\bibnamefont {Fayer}},\ }\bibfield  {title} {\enquote
  {\bibinfo {title} {Water dynamics: dependence on local structure probed with
  vibrational echo correlation spectroscopy},}\ }\href
  {https://doi.org/https://doi.org/10.1016/j.cplett.2004.01.042} {\bibfield
  {journal} {\bibinfo  {journal} {Chemical Physics Letters}\ }\textbf {\bibinfo
  {volume} {386}},\ \bibinfo {pages} {295--300} (\bibinfo {year}
  {2004})}\BibitemShut {NoStop}%
\bibitem [{\citenamefont {Ishizaki}\ and\ \citenamefont
  {Tanimura}(2008)}]{IT08CP}%
  \BibitemOpen
  \bibfield  {author} {\bibinfo {author} {\bibfnamefont {A.}~\bibnamefont
  {Ishizaki}}\ and\ \bibinfo {author} {\bibfnamefont {Y.}~\bibnamefont
  {Tanimura}},\ }\bibfield  {title} {\enquote {\bibinfo {title}
  {Nonperturbative non-\uppercase{M}arkovian quantum master equation:
  \uppercase{V}alidity and limitation to calculate nonlinear response
  functions},}\ }\href
  {https://doi.org/https://doi.org/10.1016/j.chemphys.2007.10.037} {\bibfield
  {journal} {\bibinfo  {journal} {Chemical Physics}\ }\textbf {\bibinfo
  {volume} {347}},\ \bibinfo {pages} {185--193} (\bibinfo {year} {2008})},\
  \bibinfo {note} {ultrafast Photoinduced Processes in Polyatomic
  Molecules}\BibitemShut {NoStop}%
\bibitem [{\citenamefont {Tanimura}\ and\ \citenamefont
  {Ishizaki}(2009)}]{TI09ACR}%
  \BibitemOpen
  \bibfield  {author} {\bibinfo {author} {\bibfnamefont {Y.}~\bibnamefont
  {Tanimura}}\ and\ \bibinfo {author} {\bibfnamefont {A.}~\bibnamefont
  {Ishizaki}},\ }\bibfield  {title} {\enquote {\bibinfo {title} {Modeling,
  calculating, and analyzing multidimensional vibrational spectroscopies},}\
  }\href {https://doi.org/10.1021/ar9000444} {\bibfield  {journal} {\bibinfo
  {journal} {Accounts of Chemical Research}\ }\textbf {\bibinfo {volume}
  {42}},\ \bibinfo {pages} {1270--1279} (\bibinfo {year} {2009})}\BibitemShut
  {NoStop}%
\bibitem [{\citenamefont {Tanimura}(2020)}]{T20JCP}%
  \BibitemOpen
  \bibfield  {author} {\bibinfo {author} {\bibfnamefont {Y.}~\bibnamefont
  {Tanimura}},\ }\bibfield  {title} {\enquote {\bibinfo {title} {Numerically
  "exact" approach to open quantum dynamics: The hierarchical equations of
  motion (\uppercase{HEOM})},}\ }\href {https://doi.org/10.1063/5.0011599}
  {\bibfield  {journal} {\bibinfo  {journal} {The Journal of Chemical Physics}\
  }\textbf {\bibinfo {volume} {153}},\ \bibinfo {pages} {020901} (\bibinfo
  {year} {2020})}\BibitemShut {NoStop}%
\bibitem [{\citenamefont {Liu}\ and\ \citenamefont
  {Liu}(2018)}]{JianLiu2018H2OMP}%
  \BibitemOpen
  \bibfield  {author} {\bibinfo {author} {\bibfnamefont {X.}~\bibnamefont
  {Liu}}\ and\ \bibinfo {author} {\bibfnamefont {J.}~\bibnamefont {Liu}},\
  }\bibfield  {title} {\enquote {\bibinfo {title} {Critical role of quantum
  dynamical effects in the \uppercase{R}aman spectroscopy of liquid water},}\
  }\href {https://doi.org/10.1080/00268976.2018.1434907} {\bibfield  {journal}
  {\bibinfo  {journal} {Molecular Physics}\ }\textbf {\bibinfo {volume}
  {116}},\ \bibinfo {pages} {755--779} (\bibinfo {year} {2018})},\ \Eprint
  {https://arxiv.org/abs/https://doi.org/10.1080/00268976.2018.1434907}
  {https://doi.org/10.1080/00268976.2018.1434907} \BibitemShut {NoStop}%
\bibitem [{\citenamefont {Liu}\ \emph {et~al.}(2011)\citenamefont {Liu},
  \citenamefont {Miller}, \citenamefont {Fanourgakis}, \citenamefont
  {Xantheas}, \citenamefont {Imoto},\ and\ \citenamefont
  {Saito}}]{Imoto_JCP135}%
  \BibitemOpen
  \bibfield  {author} {\bibinfo {author} {\bibfnamefont {J.}~\bibnamefont
  {Liu}}, \bibinfo {author} {\bibfnamefont {W.~H.}\ \bibnamefont {Miller}},
  \bibinfo {author} {\bibfnamefont {G.~S.}\ \bibnamefont {Fanourgakis}},
  \bibinfo {author} {\bibfnamefont {S.~S.}\ \bibnamefont {Xantheas}}, \bibinfo
  {author} {\bibfnamefont {S.}~\bibnamefont {Imoto}},\ and\ \bibinfo {author}
  {\bibfnamefont {S.}~\bibnamefont {Saito}},\ }\bibfield  {title} {\enquote
  {\bibinfo {title} {Insights in quantum dynamical effects in the infrared
  spectroscopy of liquid water from a semiclassical study with an ab
  initio-based flexible and polarizable force field},}\ }\href
  {https://doi.org/10.1063/1.3670960} {\bibfield  {journal} {\bibinfo
  {journal} {The Journal of Chemical Physics}\ }\textbf {\bibinfo {volume}
  {135}},\ \bibinfo {pages} {244503} (\bibinfo {year} {2011})},\ \Eprint
  {https://arxiv.org/abs/https://doi.org/10.1063/1.3670960}
  {https://doi.org/10.1063/1.3670960} \BibitemShut {NoStop}%
\bibitem [{\citenamefont {Hunter}, \citenamefont {Shakib},\ and\ \citenamefont
  {Paesani}(2018)}]{Paesan2018H2OCMD}%
  \BibitemOpen
  \bibfield  {author} {\bibinfo {author} {\bibfnamefont {K.~M.}\ \bibnamefont
  {Hunter}}, \bibinfo {author} {\bibfnamefont {F.~A.}\ \bibnamefont {Shakib}},\
  and\ \bibinfo {author} {\bibfnamefont {F.}~\bibnamefont {Paesani}},\
  }\bibfield  {title} {\enquote {\bibinfo {title} {Disentangling coupling
  effects in the infrared spectra of liquid water},}\ }\href
  {https://doi.org/10.1021/acs.jpcb.8b09910} {\bibfield  {journal} {\bibinfo
  {journal} {The Journal of Physical Chemistry B}\ }\textbf {\bibinfo {volume}
  {122}},\ \bibinfo {pages} {10754--10761} (\bibinfo {year} {2018})},\ \bibinfo
  {note} {pMID: 30403350},\ \Eprint
  {https://arxiv.org/abs/https://doi.org/10.1021/acs.jpcb.8b09910}
  {https://doi.org/10.1021/acs.jpcb.8b09910} \BibitemShut {NoStop}%
\bibitem [{\citenamefont {Cho}\ \emph {et~al.}(1994)\citenamefont {Cho},
  \citenamefont {Fleming}, \citenamefont {Saito}, \citenamefont {Ohmine},\ and\
  \citenamefont {Stratt}}]{ChoOhmine1994}%
  \BibitemOpen
  \bibfield  {author} {\bibinfo {author} {\bibfnamefont {M.}~\bibnamefont
  {Cho}}, \bibinfo {author} {\bibfnamefont {G.~R.}\ \bibnamefont {Fleming}},
  \bibinfo {author} {\bibfnamefont {S.}~\bibnamefont {Saito}}, \bibinfo
  {author} {\bibfnamefont {I.}~\bibnamefont {Ohmine}},\ and\ \bibinfo {author}
  {\bibfnamefont {R.~M.}\ \bibnamefont {Stratt}},\ }\bibfield  {title}
  {\enquote {\bibinfo {title} {Instantaneous normal mode analysis of liquid
  water},}\ }\href {https://doi.org/10.1063/1.467027} {\bibfield  {journal}
  {\bibinfo  {journal} {The Journal of Chemical Physics}\ }\textbf {\bibinfo
  {volume} {100}},\ \bibinfo {pages} {6672--6683} (\bibinfo {year} {1994})},\
  \Eprint
  {https://arxiv.org/abs/https://pubs.aip.org/aip/jcp/article-pdf/100/9/6672/19219400/6672\_1\_online.pdf}
  {https://pubs.aip.org/aip/jcp/article-pdf/100/9/6672/19219400/6672\_1\_online.pdf}
  \BibitemShut {NoStop}%
\bibitem [{\citenamefont {Paarmann}\ \emph {et~al.}(2009)\citenamefont
  {Paarmann}, \citenamefont {Hayashi}, \citenamefont {Mukamel},\ and\
  \citenamefont {Miller}}]{Mukamel2009}%
  \BibitemOpen
  \bibfield  {author} {\bibinfo {author} {\bibfnamefont {A.}~\bibnamefont
  {Paarmann}}, \bibinfo {author} {\bibfnamefont {T.}~\bibnamefont {Hayashi}},
  \bibinfo {author} {\bibfnamefont {S.}~\bibnamefont {Mukamel}},\ and\ \bibinfo
  {author} {\bibfnamefont {R.~J.~D.}\ \bibnamefont {Miller}},\ }\bibfield
  {title} {\enquote {\bibinfo {title} {Nonlinear response of vibrational
  excitons: Simulating the two-dimensional infrared spectrum of liquid
  water},}\ }\href {https://doi.org/10.1063/1.3139003} {\bibfield  {journal}
  {\bibinfo  {journal} {The Journal of Chemical Physics}\ }\textbf {\bibinfo
  {volume} {130}},\ \bibinfo {pages} {204110} (\bibinfo {year} {2009})},\
  \Eprint {https://arxiv.org/abs/https://doi.org/10.1063/1.3139003}
  {https://doi.org/10.1063/1.3139003} \BibitemShut {NoStop}%
\bibitem [{\citenamefont {Piryatinski}, \citenamefont {Lawrence},\ and\
  \citenamefont {Skinner}(2003{\natexlab{a}})}]{SkinnerStochs2003}%
  \BibitemOpen
  \bibfield  {author} {\bibinfo {author} {\bibfnamefont {A.}~\bibnamefont
  {Piryatinski}}, \bibinfo {author} {\bibfnamefont {C.~P.}\ \bibnamefont
  {Lawrence}},\ and\ \bibinfo {author} {\bibfnamefont {J.~L.}\ \bibnamefont
  {Skinner}},\ }\bibfield  {title} {\enquote {\bibinfo {title} {Vibrational
  spectroscopy of \uppercase{HOD} in liquid \uppercase{D}$_2$\uppercase{O}. v.
  infrared three-pulse photon echoes},}\ }\href
  {https://doi.org/10.1063/1.1569474} {\bibfield  {journal} {\bibinfo
  {journal} {The Journal of Chemical Physics}\ }\textbf {\bibinfo {volume}
  {118}},\ \bibinfo {pages} {9672--9679} (\bibinfo {year}
  {2003}{\natexlab{a}})},\ \Eprint
  {https://arxiv.org/abs/https://doi.org/10.1063/1.1569474}
  {https://doi.org/10.1063/1.1569474} \BibitemShut {NoStop}%
\bibitem [{\citenamefont {Schmidt}, \citenamefont {Corcelli},\ and\
  \citenamefont {Skinner}(2005)}]{Skinner2005}%
  \BibitemOpen
  \bibfield  {author} {\bibinfo {author} {\bibfnamefont {J.~R.}\ \bibnamefont
  {Schmidt}}, \bibinfo {author} {\bibfnamefont {S.~A.}\ \bibnamefont
  {Corcelli}},\ and\ \bibinfo {author} {\bibfnamefont {J.~L.}\ \bibnamefont
  {Skinner}},\ }\bibfield  {title} {\enquote {\bibinfo {title} {Pronounced
  non-condon effects in the ultrafast infrared spectroscopy of water},}\ }\href
  {https://doi.org/10.1063/1.1961472} {\bibfield  {journal} {\bibinfo
  {journal} {The Journal of Chemical Physics}\ }\textbf {\bibinfo {volume}
  {123}},\ \bibinfo {pages} {044513} (\bibinfo {year} {2005})},\ \Eprint
  {https://arxiv.org/abs/https://pubs.aip.org/aip/jcp/article-pdf/doi/10.1063/1.1961472/13644860/044513\_1\_online.pdf}
  {https://pubs.aip.org/aip/jcp/article-pdf/doi/10.1063/1.1961472/13644860/044513\_1\_online.pdf}
  \BibitemShut {NoStop}%
\bibitem [{\citenamefont {Jansen}\ \emph {et~al.}(2010)\citenamefont {Jansen},
  \citenamefont {Auer}, \citenamefont {Yang},\ and\ \citenamefont
  {Skinner}}]{JansenSkinnerJCP2010}%
  \BibitemOpen
  \bibfield  {author} {\bibinfo {author} {\bibfnamefont {T.~l.~C.}\
  \bibnamefont {Jansen}}, \bibinfo {author} {\bibfnamefont {B.~M.}\
  \bibnamefont {Auer}}, \bibinfo {author} {\bibfnamefont {M.}~\bibnamefont
  {Yang}},\ and\ \bibinfo {author} {\bibfnamefont {J.~L.}\ \bibnamefont
  {Skinner}},\ }\bibfield  {title} {\enquote {\bibinfo {title} {Two-dimensional
  infrared spectroscopy and ultrafast anisotropy decay of water},}\ }\href
  {https://doi.org/10.1063/1.3454733} {\bibfield  {journal} {\bibinfo
  {journal} {The Journal of Chemical Physics}\ }\textbf {\bibinfo {volume}
  {132}},\ \bibinfo {pages} {224503} (\bibinfo {year} {2010})},\ \Eprint
  {https://arxiv.org/abs/https://pubs.aip.org/aip/jcp/article-pdf/doi/10.1063/1.3454733/15953822/224503\_1\_online.pdf}
  {https://pubs.aip.org/aip/jcp/article-pdf/doi/10.1063/1.3454733/15953822/224503\_1\_online.pdf}
  \BibitemShut {NoStop}%
\bibitem [{\citenamefont {Sakurai}\ and\ \citenamefont
  {Tanimura}(2011)}]{ST11JPCA}%
  \BibitemOpen
  \bibfield  {author} {\bibinfo {author} {\bibfnamefont {A.}~\bibnamefont
  {Sakurai}}\ and\ \bibinfo {author} {\bibfnamefont {Y.}~\bibnamefont
  {Tanimura}},\ }\bibfield  {title} {\enquote {\bibinfo {title} {Does $\hbar$
  play a role in multidimensional spectroscopy? reduced hierarchy equations of
  motion approach to molecular vibrations},}\ }\href
  {https://doi.org/10.1021/jp1095618} {\bibfield  {journal} {\bibinfo
  {journal} {The Journal of Physical Chemistry A}\ }\textbf {\bibinfo {volume}
  {115}},\ \bibinfo {pages} {4009--4022} (\bibinfo {year} {2011})}\BibitemShut
  {NoStop}%
\bibitem [{\citenamefont {Ikeda}, \citenamefont {Ito},\ and\ \citenamefont
  {Tanimura}(2015)}]{IIT15JCP}%
  \BibitemOpen
  \bibfield  {author} {\bibinfo {author} {\bibfnamefont {T.}~\bibnamefont
  {Ikeda}}, \bibinfo {author} {\bibfnamefont {H.}~\bibnamefont {Ito}},\ and\
  \bibinfo {author} {\bibfnamefont {Y.}~\bibnamefont {Tanimura}},\ }\bibfield
  {title} {\enquote {\bibinfo {title} {Analysis of 2\uppercase{D}
  \uppercase{TH}z-\uppercase{R}aman spectroscopy using a
  non-\uppercase{M}arkovian \uppercase{B}rownian oscillator model with
  nonlinear system-bath interactions},}\ }\href
  {https://doi.org/10.1063/1.4917033} {\bibfield  {journal} {\bibinfo
  {journal} {The Journal of Chemical Physics}\ }\textbf {\bibinfo {volume}
  {142}},\ \bibinfo {pages} {212421} (\bibinfo {year} {2015})}\BibitemShut
  {NoStop}%
\bibitem [{\citenamefont {Ito}\ and\ \citenamefont {Tanimura}(2016)}]{IT16JCP}%
  \BibitemOpen
  \bibfield  {author} {\bibinfo {author} {\bibfnamefont {H.}~\bibnamefont
  {Ito}}\ and\ \bibinfo {author} {\bibfnamefont {Y.}~\bibnamefont {Tanimura}},\
  }\bibfield  {title} {\enquote {\bibinfo {title} {Simulating two-dimensional
  infrared-\uppercase{R}aman and \uppercase{R}aman spectroscopies for
  intermolecular and intramolecular modes of liquid water},}\ }\href
  {https://doi.org/10.1063/1.4941842} {\bibfield  {journal} {\bibinfo
  {journal} {The Journal of Chemical Physics}\ }\textbf {\bibinfo {volume}
  {144}},\ \bibinfo {pages} {074201} (\bibinfo {year} {2016})}\BibitemShut
  {NoStop}%
\bibitem [{\citenamefont {Takahashi}\ and\ \citenamefont
  {Tanimura}(2023{\natexlab{a}})}]{TT23JCP1}%
  \BibitemOpen
  \bibfield  {author} {\bibinfo {author} {\bibfnamefont {H.}~\bibnamefont
  {Takahashi}}\ and\ \bibinfo {author} {\bibfnamefont {Y.}~\bibnamefont
  {Tanimura}},\ }\bibfield  {title} {\enquote {\bibinfo {title} {Discretized
  hierarchal equations of motion in mixed
  \uppercase{L}iouville--\uppercase{W}igner space for two-dimensional
  vibrational spectroscopies of water},}\ }\href
  {https://doi.org/10.1063/5.0135725} {\bibfield  {journal} {\bibinfo
  {journal} {The Journal of Chemical Physics}\ }\textbf {\bibinfo {volume}
  {158}},\ \bibinfo {pages} {044115} (\bibinfo {year} {2023}{\natexlab{a}})},\
  \Eprint {https://arxiv.org/abs/2302.09799} {arXiv:2302.09799} \BibitemShut
  {NoStop}%
\bibitem [{\citenamefont {Takahashi}\ and\ \citenamefont
  {Tanimura}(2023{\natexlab{b}})}]{TT23JCP2}%
  \BibitemOpen
  \bibfield  {author} {\bibinfo {author} {\bibfnamefont {H.}~\bibnamefont
  {Takahashi}}\ and\ \bibinfo {author} {\bibfnamefont {Y.}~\bibnamefont
  {Tanimura}},\ }\bibfield  {title} {\enquote {\bibinfo {title} {{Simulating
  two-dimensional correlation spectroscopies with third-order infrared and
  fifth-order infrared–\uppercase{R}aman processes of liquid water}},}\
  }\href {https://doi.org/10.1063/5.0141181} {\bibfield  {journal} {\bibinfo
  {journal} {The Journal of Chemical Physics}\ }\textbf {\bibinfo {volume}
  {158}},\ \bibinfo {pages} {124108} (\bibinfo {year} {2023}{\natexlab{b}})},\
  \Eprint {https://arxiv.org/abs/2302.09760} {arXiv:2302.09760} \BibitemShut
  {NoStop}%
\bibitem [{\citenamefont {Hoshino}\ and\ \citenamefont
  {Tanimura}(2025{\natexlab{a}})}]{HT25JCP1}%
  \BibitemOpen
  \bibfield  {author} {\bibinfo {author} {\bibfnamefont {R.}~\bibnamefont
  {Hoshino}}\ and\ \bibinfo {author} {\bibfnamefont {Y.}~\bibnamefont
  {Tanimura}},\ }\bibfield  {title} {\enquote {\bibinfo {title} {Analysis of
  intramolecular modes of liquid water in two-dimensional spectroscopy: A
  classical hierarchical equations of motion approach},}\ }\href
  {https://doi.org/10.1063/5.0245564} {\bibfield  {journal} {\bibinfo
  {journal} {The Journal of Chemical Physics}\ }\textbf {\bibinfo {volume}
  {162}},\ \bibinfo {pages} {044105} (\bibinfo {year} {2025}{\natexlab{a}})},\
  \Eprint
  {https://arxiv.org/abs/https://pubs.aip.org/aip/jcp/article-pdf/doi/10.1063/5.0245564/20360227/044105\_1\_5.0245564.pdf}
  {https://pubs.aip.org/aip/jcp/article-pdf/doi/10.1063/5.0245564/20360227/044105\_1\_5.0245564.pdf}
  \BibitemShut {NoStop}%
\bibitem [{\citenamefont {Hoshino}\ and\ \citenamefont
  {Tanimura}(2025{\natexlab{b}})}]{HT25JCP2}%
  \BibitemOpen
  \bibfield  {author} {\bibinfo {author} {\bibfnamefont {R.}~\bibnamefont
  {Hoshino}}\ and\ \bibinfo {author} {\bibfnamefont {Y.}~\bibnamefont
  {Tanimura}},\ }\bibfield  {title} {\enquote {\bibinfo {title} {A multimode
  classical hierarchical \uppercase{F}okker--\uppercase{P}lanck equations
  approach to molecular vibrations: Simulating two-dimensional spectra},}\
  }\href {https://doi.org/xxxx} {\bibfield  {journal} {\bibinfo  {journal} {The
  Journal of Chemical Physics}\ }\textbf {\bibinfo {volume} {163}},\ \bibinfo
  {pages} {xxxxx} (\bibinfo {year} {2025}{\natexlab{b}})}\BibitemShut {NoStop}%
\bibitem [{\citenamefont {Tanimura}(2006)}]{T06JPSJ}%
  \BibitemOpen
  \bibfield  {author} {\bibinfo {author} {\bibfnamefont {Y.}~\bibnamefont
  {Tanimura}},\ }\bibfield  {title} {\enquote {\bibinfo {title} {Stochastic
  \uppercase{L}iouville, \uppercase{L}angevin,
  \uppercase{F}okker-\uppercase{P}lanck, and master equation qpproaches to
  quantum dissipative systems},}\ }\href
  {https://doi.org/10.1143/JPSJ.75.082001} {\bibfield  {journal} {\bibinfo
  {journal} {Journal of the Physical Society of Japan}\ }\textbf {\bibinfo
  {volume} {75}},\ \bibinfo {pages} {082001} (\bibinfo {year}
  {2006})}\BibitemShut {NoStop}%
\bibitem [{\citenamefont {Steffen}\ and\ \citenamefont
  {Tanimura}(2000)}]{ST20JPSJ}%
  \BibitemOpen
  \bibfield  {author} {\bibinfo {author} {\bibfnamefont {T.}~\bibnamefont
  {Steffen}}\ and\ \bibinfo {author} {\bibfnamefont {Y.}~\bibnamefont
  {Tanimura}},\ }\bibfield  {title} {\enquote {\bibinfo {title}
  {Two-dimensional spectroscopy for harmonic vibrational modes with nonlinear
  system-bath interactions. i. \uppercase{G}aussian-white case},}\ }\href
  {https://doi.org/10.1143/JPSJ.69.3115} {\bibfield  {journal} {\bibinfo
  {journal} {Journal of the Physical Society of Japan}\ }\textbf {\bibinfo
  {volume} {69}},\ \bibinfo {pages} {3115--3132} (\bibinfo {year}
  {2000})}\BibitemShut {NoStop}%
\bibitem [{\citenamefont {Tanimura}\ and\ \citenamefont
  {Steffen}(2000)}]{TS20JPSJ}%
  \BibitemOpen
  \bibfield  {author} {\bibinfo {author} {\bibfnamefont {Y.}~\bibnamefont
  {Tanimura}}\ and\ \bibinfo {author} {\bibfnamefont {T.}~\bibnamefont
  {Steffen}},\ }\bibfield  {title} {\enquote {\bibinfo {title} {Two-dimensional
  spectroscopy for harmonic vibrational modes with nonlinear system-bath
  interactions.ii. \uppercase{G}aussian-\uppercase{M}arkovian case},}\ }\href
  {https://doi.org/10.1143/JPSJ.69.4095} {\bibfield  {journal} {\bibinfo
  {journal} {Journal of the Physical Society of Japan}\ }\textbf {\bibinfo
  {volume} {69}},\ \bibinfo {pages} {4095--4106} (\bibinfo {year}
  {2000})}\BibitemShut {NoStop}%
\bibitem [{\citenamefont {Ueno}\ and\ \citenamefont
  {Tanimura}(2020)}]{UT20JCTC}%
  \BibitemOpen
  \bibfield  {author} {\bibinfo {author} {\bibfnamefont {S.}~\bibnamefont
  {Ueno}}\ and\ \bibinfo {author} {\bibfnamefont {Y.}~\bibnamefont
  {Tanimura}},\ }\bibfield  {title} {\enquote {\bibinfo {title} {Modeling
  intermolecular and intramolecular modes of liquid water using multiple heat
  baths: Machine learning approach},}\ }\href
  {https://doi.org/10.1021/acs.jctc.9b01288} {\bibfield  {journal} {\bibinfo
  {journal} {Journal of Chemical Theory and Computation}\ }\textbf {\bibinfo
  {volume} {16}},\ \bibinfo {pages} {2099--2108} (\bibinfo {year}
  {2020})}\BibitemShut {NoStop}%
\bibitem [{\citenamefont {Ueno}\ and\ \citenamefont
  {Tanimura}(2021)}]{UT21JCTC}%
  \BibitemOpen
  \bibfield  {author} {\bibinfo {author} {\bibfnamefont {S.}~\bibnamefont
  {Ueno}}\ and\ \bibinfo {author} {\bibfnamefont {Y.}~\bibnamefont
  {Tanimura}},\ }\bibfield  {title} {\enquote {\bibinfo {title} {Modeling and
  simulating the excited-state dynamics of a system with condensed phases:
  \uppercase{A} machine learning approach},}\ }\href
  {https://doi.org/10.1021/acs.jctc.1c00104} {\bibfield  {journal} {\bibinfo
  {journal} {Journal of Chemical Theory and Computation}\ }\textbf {\bibinfo
  {volume} {17}},\ \bibinfo {pages} {3618--3628} (\bibinfo {year}
  {2021})}\BibitemShut {NoStop}%
\bibitem [{\citenamefont {Kato}\ and\ \citenamefont
  {Tanimura}(2002)}]{KT02JCP1}%
  \BibitemOpen
  \bibfield  {author} {\bibinfo {author} {\bibfnamefont {T.}~\bibnamefont
  {Kato}}\ and\ \bibinfo {author} {\bibfnamefont {Y.}~\bibnamefont
  {Tanimura}},\ }\bibfield  {title} {\enquote {\bibinfo {title} {Vibrational
  spectroscopy of a harmonic oscillator system nonlinearly coupled to a heat
  bath},}\ }\href {https://doi.org/10.1063/1.1503778} {\bibfield  {journal}
  {\bibinfo  {journal} {The Journal of Chemical Physics}\ }\textbf {\bibinfo
  {volume} {117}},\ \bibinfo {pages} {6221--6234} (\bibinfo {year}
  {2002})}\BibitemShut {NoStop}%
\bibitem [{\citenamefont {Kato}\ and\ \citenamefont
  {Tanimura}(2004)}]{KT04JCP}%
  \BibitemOpen
  \bibfield  {author} {\bibinfo {author} {\bibfnamefont {T.}~\bibnamefont
  {Kato}}\ and\ \bibinfo {author} {\bibfnamefont {Y.}~\bibnamefont
  {Tanimura}},\ }\bibfield  {title} {\enquote {\bibinfo {title}
  {Two-dimensional \uppercase{R}aman and infrared vibrational spectroscopy for
  a harmonic oscillator system nonlinearly coupled with a colored noise
  bath},}\ }\href {https://doi.org/10.1063/1.1629272} {\bibfield  {journal}
  {\bibinfo  {journal} {The Journal of Chemical Physics}\ }\textbf {\bibinfo
  {volume} {120}},\ \bibinfo {pages} {260--271} (\bibinfo {year}
  {2004})}\BibitemShut {NoStop}%
\bibitem [{\citenamefont {Ishizaki}\ and\ \citenamefont
  {Tanimura}(2006)}]{IT06JCP}%
  \BibitemOpen
  \bibfield  {author} {\bibinfo {author} {\bibfnamefont {A.}~\bibnamefont
  {Ishizaki}}\ and\ \bibinfo {author} {\bibfnamefont {Y.}~\bibnamefont
  {Tanimura}},\ }\bibfield  {title} {\enquote {\bibinfo {title} {Modeling
  vibrational dephasing and energy relaxation of intramolecular anharmonic
  modes for multidimensional infrared spectroscopies},}\ }\href
  {https://doi.org/10.1063/1.2244558} {\bibfield  {journal} {\bibinfo
  {journal} {The Journal of Chemical Physics}\ }\textbf {\bibinfo {volume}
  {125}},\ \bibinfo {pages} {084501} (\bibinfo {year} {2006})}\BibitemShut
  {NoStop}%
\bibitem [{\citenamefont {Okumura}\ and\ \citenamefont
  {Tanimura}(1997)}]{OT97PRE}%
  \BibitemOpen
  \bibfield  {author} {\bibinfo {author} {\bibfnamefont {K.}~\bibnamefont
  {Okumura}}\ and\ \bibinfo {author} {\bibfnamefont {Y.}~\bibnamefont
  {Tanimura}},\ }\bibfield  {title} {\enquote {\bibinfo {title} {Two-time
  correlation functions of a harmonic system nonbilinearly coupled to a heat
  bath: Spontaneous \uppercase{R}aman spectroscopy},}\ }\href
  {https://doi.org/10.1103/PhysRevE.56.2747} {\bibfield  {journal} {\bibinfo
  {journal} {Phys. Rev. E}\ }\textbf {\bibinfo {volume} {56}},\ \bibinfo
  {pages} {2747--2750} (\bibinfo {year} {1997})}\BibitemShut {NoStop}%
\bibitem [{\citenamefont {Tanimura}\ and\ \citenamefont
  {Wolynes}(1991)}]{TW91PRA}%
  \BibitemOpen
  \bibfield  {author} {\bibinfo {author} {\bibfnamefont {Y.}~\bibnamefont
  {Tanimura}}\ and\ \bibinfo {author} {\bibfnamefont {P.~G.}\ \bibnamefont
  {Wolynes}},\ }\bibfield  {title} {\enquote {\bibinfo {title} {Quantum and
  classical \uppercase{F}okker-\uppercase{P}lanck equations for a
  \uppercase{G}aussian-\uppercase{M}arkovian noise bath},}\ }\href
  {https://doi.org/10.1103/PhysRevA.43.4131} {\bibfield  {journal} {\bibinfo
  {journal} {Phys. Rev. A}\ }\textbf {\bibinfo {volume} {43}},\ \bibinfo
  {pages} {4131--4142} (\bibinfo {year} {1991})}\BibitemShut {NoStop}%
\bibitem [{\citenamefont {Lawrence}\ and\ \citenamefont
  {Skinner}(2002{\natexlab{a}})}]{SkinnerI}%
  \BibitemOpen
  \bibfield  {author} {\bibinfo {author} {\bibfnamefont {C.~P.}\ \bibnamefont
  {Lawrence}}\ and\ \bibinfo {author} {\bibfnamefont {J.~L.}\ \bibnamefont
  {Skinner}},\ }\bibfield  {title} {\enquote {\bibinfo {title} {Vibrational
  spectroscopy of \uppercase{HOD} in liquid \uppercase{D}$_2$\uppercase{O}.
  \uppercase{I}. vibrational energy relaxation},}\ }\href
  {https://doi.org/10.1063/1.1502248} {\bibfield  {journal} {\bibinfo
  {journal} {The Journal of Chemical Physics}\ }\textbf {\bibinfo {volume}
  {117}},\ \bibinfo {pages} {5827--5838} (\bibinfo {year}
  {2002}{\natexlab{a}})},\ \Eprint
  {https://arxiv.org/abs/https://pubs.aip.org/aip/jcp/article-pdf/117/12/5827/19037401/5827\_1\_online.pdf}
  {https://pubs.aip.org/aip/jcp/article-pdf/117/12/5827/19037401/5827\_1\_online.pdf}
  \BibitemShut {NoStop}%
\bibitem [{\citenamefont {Lawrence}\ and\ \citenamefont
  {Skinner}(2002{\natexlab{b}})}]{SkinnerII}%
  \BibitemOpen
  \bibfield  {author} {\bibinfo {author} {\bibfnamefont {C.~P.}\ \bibnamefont
  {Lawrence}}\ and\ \bibinfo {author} {\bibfnamefont {J.~L.}\ \bibnamefont
  {Skinner}},\ }\bibfield  {title} {\enquote {\bibinfo {title} {Vibrational
  spectroscopy of \uppercase{HOD} in liquid \uppercase{D}$_2$\uppercase{O}.
  \uppercase{II}. infrared line shapes and vibrational stokes shift},}\ }\href
  {https://doi.org/10.1063/1.1514652} {\bibfield  {journal} {\bibinfo
  {journal} {The Journal of Chemical Physics}\ }\textbf {\bibinfo {volume}
  {117}},\ \bibinfo {pages} {8847--8854} (\bibinfo {year}
  {2002}{\natexlab{b}})},\ \Eprint
  {https://arxiv.org/abs/https://pubs.aip.org/aip/jcp/article-pdf/117/19/8847/19017834/8847\_1\_online.pdf}
  {https://pubs.aip.org/aip/jcp/article-pdf/117/19/8847/19017834/8847\_1\_online.pdf}
  \BibitemShut {NoStop}%
\bibitem [{\citenamefont {Lawrence}\ and\ \citenamefont
  {Skinner}(2003{\natexlab{a}})}]{SkinnerIII}%
  \BibitemOpen
  \bibfield  {author} {\bibinfo {author} {\bibfnamefont {C.~P.}\ \bibnamefont
  {Lawrence}}\ and\ \bibinfo {author} {\bibfnamefont {J.~L.}\ \bibnamefont
  {Skinner}},\ }\bibfield  {title} {\enquote {\bibinfo {title} {Vibrational
  spectroscopy of \uppercase{HOD} in liquid \uppercase{D}$_2$\uppercase{O}.
  \uppercase{III}. spectral diffusion, and hydrogen-bonding and rotational
  dynamics},}\ }\href {https://doi.org/10.1063/1.1525802} {\bibfield  {journal}
  {\bibinfo  {journal} {The Journal of Chemical Physics}\ }\textbf {\bibinfo
  {volume} {118}},\ \bibinfo {pages} {264--272} (\bibinfo {year}
  {2003}{\natexlab{a}})},\ \Eprint
  {https://arxiv.org/abs/https://pubs.aip.org/aip/jcp/article-pdf/118/1/264/19152490/264\_1\_online.pdf}
  {https://pubs.aip.org/aip/jcp/article-pdf/118/1/264/19152490/264\_1\_online.pdf}
  \BibitemShut {NoStop}%
\bibitem [{\citenamefont {Piryatinski}, \citenamefont {Lawrence},\ and\
  \citenamefont {Skinner}(2003{\natexlab{b}})}]{SkinnerIV}%
  \BibitemOpen
  \bibfield  {author} {\bibinfo {author} {\bibfnamefont {A.}~\bibnamefont
  {Piryatinski}}, \bibinfo {author} {\bibfnamefont {C.~P.}\ \bibnamefont
  {Lawrence}},\ and\ \bibinfo {author} {\bibfnamefont {J.~L.}\ \bibnamefont
  {Skinner}},\ }\bibfield  {title} {\enquote {\bibinfo {title} {Vibrational
  spectroscopy of hod in liquid \uppercase{D}$_2$\uppercase{O}. \uppercase{IV}.
  infrared two-pulse photon echoes},}\ }\href
  {https://doi.org/10.1063/1.1566434} {\bibfield  {journal} {\bibinfo
  {journal} {The Journal of Chemical Physics}\ }\textbf {\bibinfo {volume}
  {118}},\ \bibinfo {pages} {9664--9671} (\bibinfo {year}
  {2003}{\natexlab{b}})},\ \Eprint
  {https://arxiv.org/abs/https://pubs.aip.org/aip/jcp/article-pdf/118/21/9664/19025048/9664\_1\_online.pdf}
  {https://pubs.aip.org/aip/jcp/article-pdf/118/21/9664/19025048/9664\_1\_online.pdf}
  \BibitemShut {NoStop}%
\bibitem [{\citenamefont {Piryatinski}, \citenamefont {Lawrence},\ and\
  \citenamefont {Skinner}(2003{\natexlab{c}})}]{SkinnerV}%
  \BibitemOpen
  \bibfield  {author} {\bibinfo {author} {\bibfnamefont {A.}~\bibnamefont
  {Piryatinski}}, \bibinfo {author} {\bibfnamefont {C.~P.}\ \bibnamefont
  {Lawrence}},\ and\ \bibinfo {author} {\bibfnamefont {J.~L.}\ \bibnamefont
  {Skinner}},\ }\bibfield  {title} {\enquote {\bibinfo {title} {Vibrational
  spectroscopy of \uppercase{HOD} in liquid \uppercase{D}$_2$\uppercase{O}.
  \uppercase{V}. infrared three-pulse photon echoes},}\ }\href
  {https://doi.org/10.1063/1.1569474} {\bibfield  {journal} {\bibinfo
  {journal} {The Journal of Chemical Physics}\ }\textbf {\bibinfo {volume}
  {118}},\ \bibinfo {pages} {9672--9679} (\bibinfo {year}
  {2003}{\natexlab{c}})},\ \Eprint
  {https://arxiv.org/abs/https://pubs.aip.org/aip/jcp/article-pdf/118/21/9672/19025656/9672\_1\_online.pdf}
  {https://pubs.aip.org/aip/jcp/article-pdf/118/21/9672/19025656/9672\_1\_online.pdf}
  \BibitemShut {NoStop}%
\bibitem [{\citenamefont {Lawrence}\ and\ \citenamefont
  {Skinner}(2003{\natexlab{b}})}]{SkinnerVI}%
  \BibitemOpen
  \bibfield  {author} {\bibinfo {author} {\bibfnamefont {C.~P.}\ \bibnamefont
  {Lawrence}}\ and\ \bibinfo {author} {\bibfnamefont {J.~L.}\ \bibnamefont
  {Skinner}},\ }\bibfield  {title} {\enquote {\bibinfo {title} {Vibrational
  spectroscopy of \uppercase{HOD} in liquid \uppercase{D}$_2$\uppercase{O}.
  \uppercase{VI}. intramolecular and intermolecular vibrational energy flow},}\
  }\href {https://doi.org/10.1063/1.1582173} {\bibfield  {journal} {\bibinfo
  {journal} {The Journal of Chemical Physics}\ }\textbf {\bibinfo {volume}
  {119}},\ \bibinfo {pages} {1623--1633} (\bibinfo {year}
  {2003}{\natexlab{b}})},\ \Eprint
  {https://arxiv.org/abs/https://pubs.aip.org/aip/jcp/article-pdf/119/3/1623/19007336/1623\_1\_online.pdf}
  {https://pubs.aip.org/aip/jcp/article-pdf/119/3/1623/19007336/1623\_1\_online.pdf}
  \BibitemShut {NoStop}%
\bibitem [{\citenamefont {Lawrence}\ and\ \citenamefont
  {Skinner}(2003{\natexlab{c}})}]{SkinnerVII}%
  \BibitemOpen
  \bibfield  {author} {\bibinfo {author} {\bibfnamefont {C.~P.}\ \bibnamefont
  {Lawrence}}\ and\ \bibinfo {author} {\bibfnamefont {J.~L.}\ \bibnamefont
  {Skinner}},\ }\bibfield  {title} {\enquote {\bibinfo {title} {Vibrational
  spectroscopy of \uppercase{HOD} in liquid \uppercase{D}$_2$\uppercase{O}.
  \uppercase{VII}. temperature and frequency dependence of the \uppercase{OH}
  stretch lifetime},}\ }\href {https://doi.org/10.1063/1.1591178} {\bibfield
  {journal} {\bibinfo  {journal} {The Journal of Chemical Physics}\ }\textbf
  {\bibinfo {volume} {119}},\ \bibinfo {pages} {3840--3848} (\bibinfo {year}
  {2003}{\natexlab{c}})},\ \Eprint
  {https://arxiv.org/abs/https://pubs.aip.org/aip/jcp/article-pdf/119/7/3840/19010134/3840\_1\_online.pdf}
  {https://pubs.aip.org/aip/jcp/article-pdf/119/7/3840/19010134/3840\_1\_online.pdf}
  \BibitemShut {NoStop}%
\bibitem [{\citenamefont {Yagasaki}\ and\ \citenamefont
  {Saito}(2008)}]{YagasakiSaitoJCP20082DIR}%
  \BibitemOpen
  \bibfield  {author} {\bibinfo {author} {\bibfnamefont {T.}~\bibnamefont
  {Yagasaki}}\ and\ \bibinfo {author} {\bibfnamefont {S.}~\bibnamefont
  {Saito}},\ }\bibfield  {title} {\enquote {\bibinfo {title} {Ultrafast
  intermolecular dynamics of liquid water: A theoretical study on
  two-dimensional infrared spectroscopy},}\ }\href
  {https://doi.org/10.1063/1.2903470} {\bibfield  {journal} {\bibinfo
  {journal} {The Journal of Chemical Physics}\ }\textbf {\bibinfo {volume}
  {128}},\ \bibinfo {pages} {154521} (\bibinfo {year} {2008})},\ \Eprint
  {https://arxiv.org/abs/https://doi.org/10.1063/1.2903470}
  {https://doi.org/10.1063/1.2903470} \BibitemShut {NoStop}%
\bibitem [{\citenamefont {Yagasaki}\ and\ \citenamefont
  {Saito}(2011)}]{YagasakiSaitoJCP2011Relax}%
  \BibitemOpen
  \bibfield  {author} {\bibinfo {author} {\bibfnamefont {T.}~\bibnamefont
  {Yagasaki}}\ and\ \bibinfo {author} {\bibfnamefont {S.}~\bibnamefont
  {Saito}},\ }\bibfield  {title} {\enquote {\bibinfo {title} {A novel method
  for analyzing energy relaxation in condensed phases using nonequilibrium
  molecular dynamics simulations: Application to the energy relaxation of
  intermolecular motions in liquid water},}\ }\href
  {https://doi.org/10.1063/1.3587105} {\bibfield  {journal} {\bibinfo
  {journal} {The Journal of Chemical Physics}\ }\textbf {\bibinfo {volume}
  {134}},\ \bibinfo {pages} {184503} (\bibinfo {year} {2011})},\ \Eprint
  {https://arxiv.org/abs/https://doi.org/10.1063/1.3587105}
  {https://doi.org/10.1063/1.3587105} \BibitemShut {NoStop}%
\bibitem [{\citenamefont {Yagasaki}\ and\ \citenamefont
  {Saito}(2013)}]{Yagasaki_ARPC64}%
  \BibitemOpen
  \bibfield  {author} {\bibinfo {author} {\bibfnamefont {T.}~\bibnamefont
  {Yagasaki}}\ and\ \bibinfo {author} {\bibfnamefont {S.}~\bibnamefont
  {Saito}},\ }\bibfield  {title} {\enquote {\bibinfo {title} {Fluctuations and
  relaxation dynamics of liquid water revealed by linear and nonlinear
  spectroscopy},}\ }\href
  {https://doi.org/10.1146/annurev-physchem-040412-110150} {\bibfield
  {journal} {\bibinfo  {journal} {Annual Review of Physical Chemistry}\
  }\textbf {\bibinfo {volume} {64}},\ \bibinfo {pages} {55--75} (\bibinfo
  {year} {2013})},\ \Eprint
  {https://arxiv.org/abs/https://doi.org/10.1146/annurev-physchem-040412-110150}
  {https://doi.org/10.1146/annurev-physchem-040412-110150} \BibitemShut
  {NoStop}%
\bibitem [{\citenamefont {Imoto}, \citenamefont {Xantheas},\ and\ \citenamefont
  {Saito}(2013)}]{ImotXanteasSaitoJCP2013H2O}%
  \BibitemOpen
  \bibfield  {author} {\bibinfo {author} {\bibfnamefont {S.}~\bibnamefont
  {Imoto}}, \bibinfo {author} {\bibfnamefont {S.~S.}\ \bibnamefont
  {Xantheas}},\ and\ \bibinfo {author} {\bibfnamefont {S.}~\bibnamefont
  {Saito}},\ }\bibfield  {title} {\enquote {\bibinfo {title} {Ultrafast
  dynamics of liquid water: Frequency fluctuations of the \uppercase{OH}
  stretch and the \uppercase{HOH} bend},}\ }\href
  {https://doi.org/10.1063/1.4813071} {\bibfield  {journal} {\bibinfo
  {journal} {The Journal of Chemical Physics}\ }\textbf {\bibinfo {volume}
  {139}},\ \bibinfo {pages} {044503} (\bibinfo {year} {2013})},\ \Eprint
  {https://arxiv.org/abs/https://doi.org/10.1063/1.4813071}
  {https://doi.org/10.1063/1.4813071} \BibitemShut {NoStop}%
\bibitem [{\citenamefont {Imoto}, \citenamefont {Xantheas},\ and\ \citenamefont
  {Saito}(2015)}]{Imotobend-lib2015}%
  \BibitemOpen
  \bibfield  {author} {\bibinfo {author} {\bibfnamefont {S.}~\bibnamefont
  {Imoto}}, \bibinfo {author} {\bibfnamefont {S.~S.}\ \bibnamefont
  {Xantheas}},\ and\ \bibinfo {author} {\bibfnamefont {S.}~\bibnamefont
  {Saito}},\ }\bibfield  {title} {\enquote {\bibinfo {title} {Ultrafast
  dynamics of liquid water: Energy relaxation and transfer processes of the
  \uppercase{OH} stretch and the \uppercase{HOH} bend},}\ }\href
  {https://doi.org/10.1021/acs.jpcb.5b02589} {\bibfield  {journal} {\bibinfo
  {journal} {The Journal of Physical Chemistry B}\ }\textbf {\bibinfo {volume}
  {119}},\ \bibinfo {pages} {11068--11078} (\bibinfo {year} {2015})},\ \bibinfo
  {note} {pMID: 26042611},\ \Eprint
  {https://arxiv.org/abs/https://doi.org/10.1021/acs.jpcb.5b02589}
  {https://doi.org/10.1021/acs.jpcb.5b02589} \BibitemShut {NoStop}%
\bibitem [{\citenamefont {Tanimura}\ and\ \citenamefont
  {Kubo}(1989)}]{TK89JPSJ1}%
  \BibitemOpen
  \bibfield  {author} {\bibinfo {author} {\bibfnamefont {Y.}~\bibnamefont
  {Tanimura}}\ and\ \bibinfo {author} {\bibfnamefont {R.}~\bibnamefont
  {Kubo}},\ }\bibfield  {title} {\enquote {\bibinfo {title} {Time evolution of
  a quantum system in contact with a nearly
  \uppercase{G}aussian-\uppercase{M}arkoffian noise bath},}\ }\href
  {https://doi.org/10.1143/JPSJ.58.101} {\bibfield  {journal} {\bibinfo
  {journal} {Journal of the Physical Society of Japan}\ }\textbf {\bibinfo
  {volume} {58}},\ \bibinfo {pages} {101--114} (\bibinfo {year}
  {1989})}\BibitemShut {NoStop}%
\bibitem [{\citenamefont {Tanimura}(2012)}]{T12JCP}%
  \BibitemOpen
  \bibfield  {author} {\bibinfo {author} {\bibfnamefont {Y.}~\bibnamefont
  {Tanimura}},\ }\bibfield  {title} {\enquote {\bibinfo {title} {Reduced
  hierarchy equations of motion approach with \uppercase{D}rude plus
  \uppercase{B}rownian spectral distribution: Probing electron transfer
  processes by means of two-dimensional correlation spectroscopy},}\ }\href
  {https://doi.org/10.1063/1.4766931} {\bibfield  {journal} {\bibinfo
  {journal} {The Journal of Chemical Physics}\ }\textbf {\bibinfo {volume}
  {137}},\ \bibinfo {pages} {22A550} (\bibinfo {year} {2012})}\BibitemShut
  {NoStop}%
\bibitem [{\citenamefont {Tanaka}\ and\ \citenamefont
  {Tanimura}(2009)}]{TT09JPSJ}%
  \BibitemOpen
  \bibfield  {author} {\bibinfo {author} {\bibfnamefont {M.}~\bibnamefont
  {Tanaka}}\ and\ \bibinfo {author} {\bibfnamefont {Y.}~\bibnamefont
  {Tanimura}},\ }\bibfield  {title} {\enquote {\bibinfo {title} {Quantum
  dissipative dynamics of electron transfer reaction system: nonperturbative
  hierarchy equations approach},}\ }\href
  {https://doi.org/10.1143/JPSJ.78.073802} {\bibfield  {journal} {\bibinfo
  {journal} {Journal of the Physical Society of Japan}\ }\textbf {\bibinfo
  {volume} {78}},\ \bibinfo {pages} {073802} (\bibinfo {year}
  {2009})}\BibitemShut {NoStop}%
\bibitem [{\citenamefont {Tanaka}\ and\ \citenamefont
  {Tanimura}(2010)}]{TT10JCP}%
  \BibitemOpen
  \bibfield  {author} {\bibinfo {author} {\bibfnamefont {M.}~\bibnamefont
  {Tanaka}}\ and\ \bibinfo {author} {\bibfnamefont {Y.}~\bibnamefont
  {Tanimura}},\ }\bibfield  {title} {\enquote {\bibinfo {title} {Multistate
  electron transfer dynamics in the condensed phase: Exact calculations from
  the reduced hierarchy equations of motion approach},}\ }\href
  {https://doi.org/10.1063/1.3428674} {\bibfield  {journal} {\bibinfo
  {journal} {The Journal of Chemical Physics}\ }\textbf {\bibinfo {volume}
  {132}},\ \bibinfo {pages} {214502} (\bibinfo {year} {2010})}\BibitemShut
  {NoStop}%
\bibitem [{\citenamefont {Dijkstra}\ and\ \citenamefont
  {Tanimura}(2015)}]{DT15JCP}%
  \BibitemOpen
  \bibfield  {author} {\bibinfo {author} {\bibfnamefont {A.~G.}\ \bibnamefont
  {Dijkstra}}\ and\ \bibinfo {author} {\bibfnamefont {Y.}~\bibnamefont
  {Tanimura}},\ }\bibfield  {title} {\enquote {\bibinfo {title} {Linear and
  third- and fifth-order nonlinear spectroscopies of a charge transfer system
  coupled to an underdamped vibration},}\ }\href
  {https://doi.org/10.1063/1.4917025} {\bibfield  {journal} {\bibinfo
  {journal} {The Journal of Chemical Physics}\ }\textbf {\bibinfo {volume}
  {142}},\ \bibinfo {pages} {212423} (\bibinfo {year} {2015})}\BibitemShut
  {NoStop}%
\bibitem [{\citenamefont {Cainelli}, \citenamefont {Borrelli},\ and\
  \citenamefont {Tanimura}(2022)}]{CBT22JCP}%
  \BibitemOpen
  \bibfield  {author} {\bibinfo {author} {\bibfnamefont {M.}~\bibnamefont
  {Cainelli}}, \bibinfo {author} {\bibfnamefont {R.}~\bibnamefont {Borrelli}},\
  and\ \bibinfo {author} {\bibfnamefont {Y.}~\bibnamefont {Tanimura}},\
  }\bibfield  {title} {\enquote {\bibinfo {title} {Effect of mixed
  \uppercase{F}renkel and charge transfer states in time-gated fluorescence
  spectra of perylene bisimides \uppercase{H}-aggregates: Hierarchical
  equations of motion approach},}\ }\href {https://doi.org/10.1063/5.0102000}
  {\bibfield  {journal} {\bibinfo  {journal} {The Journal of Chemical Physics}\
  }\textbf {\bibinfo {volume} {157}},\ \bibinfo {pages} {084103} (\bibinfo
  {year} {2022})},\ \Eprint
  {https://arxiv.org/abs/https://doi.org/10.1063/5.0102000}
  {https://doi.org/10.1063/5.0102000} \BibitemShut {NoStop}%
\bibitem [{\citenamefont {Ohmine}\ and\ \citenamefont
  {Tanaka}(1993)}]{Ohmine_ChemRev93}%
  \BibitemOpen
  \bibfield  {author} {\bibinfo {author} {\bibfnamefont {I.}~\bibnamefont
  {Ohmine}}\ and\ \bibinfo {author} {\bibfnamefont {H.}~\bibnamefont
  {Tanaka}},\ }\bibfield  {title} {\enquote {\bibinfo {title} {Fluctuation,
  relaxations, and hydration in liquid water. hydrogen-bond rearrangement
  dynamics},}\ }\href {https://doi.org/10.1021/cr00023a011} {\bibfield
  {journal} {\bibinfo  {journal} {Chemical Reviews}\ }\textbf {\bibinfo
  {volume} {93}},\ \bibinfo {pages} {2545--2566} (\bibinfo {year} {1993})},\
  \Eprint {https://arxiv.org/abs/https://doi.org/10.1021/cr00023a011}
  {https://doi.org/10.1021/cr00023a011} \BibitemShut {NoStop}%
\bibitem [{\citenamefont {Ohmine}\ and\ \citenamefont
  {Saito}(1999)}]{OCSACR1999}%
  \BibitemOpen
  \bibfield  {author} {\bibinfo {author} {\bibfnamefont {I.}~\bibnamefont
  {Ohmine}}\ and\ \bibinfo {author} {\bibfnamefont {S.}~\bibnamefont {Saito}},\
  }\bibfield  {title} {\enquote {\bibinfo {title} {Water dynamics {{:}}
  fluctuation, relaxation, and chemical reactions in hydrogen bond network
  rearrangement},}\ }\href {https://doi.org/10.1021/ar970161g} {\bibfield
  {journal} {\bibinfo  {journal} {Accounts of Chemical Research}\ }\textbf
  {\bibinfo {volume} {32}},\ \bibinfo {pages} {741--749} (\bibinfo {year}
  {1999})}\BibitemShut {NoStop}%
\bibitem [{\citenamefont {Nibbering}\ and\ \citenamefont
  {Elsaesser}(2004)}]{Nibbering2004UltrafastVD}%
  \BibitemOpen
  \bibfield  {author} {\bibinfo {author} {\bibfnamefont {E.~T.~J.}\
  \bibnamefont {Nibbering}}\ and\ \bibinfo {author} {\bibfnamefont
  {T.}~\bibnamefont {Elsaesser}},\ }\bibfield  {title} {\enquote {\bibinfo
  {title} {Ultrafast vibrational dynamics of hydrogen bonds in the condensed
  phase.}}\ }\href {https://doi.org/10.1021/cr020694p} {\bibfield  {journal}
  {\bibinfo  {journal} {Chemical reviews}\ }\textbf {\bibinfo {volume} {104
  4}},\ \bibinfo {pages} {1887--1914} (\bibinfo {year} {2004})}\BibitemShut
  {NoStop}%
\bibitem [{\citenamefont {Bagchi}(2013)}]{bagchi_2013}%
  \BibitemOpen
  \bibfield  {author} {\bibinfo {author} {\bibfnamefont {B.}~\bibnamefont
  {Bagchi}},\ }\bibfield  {title} {\enquote {\bibinfo {title} {Water in
  biological and chemical processes: From structure and dynamics to
  function},}\ }\href {https://doi.org/10.1017/CBO9781139583947} {\ \bibinfo
  {series} {Cambridge Molecular Science} (\bibinfo {year} {2013}),\
  10.1017/CBO9781139583947}\BibitemShut {NoStop}%
\bibitem [{\citenamefont {Kraemer}\ \emph {et~al.}(2008)\citenamefont
  {Kraemer}, \citenamefont {Cowan}, \citenamefont {Paarmann}, \citenamefont
  {Huse}, \citenamefont {Nibbering}, \citenamefont {Elsaesser},\ and\
  \citenamefont {Miller}}]{ElsaesserDwaynePNAS2008}%
  \BibitemOpen
  \bibfield  {author} {\bibinfo {author} {\bibfnamefont {D.}~\bibnamefont
  {Kraemer}}, \bibinfo {author} {\bibfnamefont {M.~L.}\ \bibnamefont {Cowan}},
  \bibinfo {author} {\bibfnamefont {A.}~\bibnamefont {Paarmann}}, \bibinfo
  {author} {\bibfnamefont {N.}~\bibnamefont {Huse}}, \bibinfo {author}
  {\bibfnamefont {E.~T.~J.}\ \bibnamefont {Nibbering}}, \bibinfo {author}
  {\bibfnamefont {T.}~\bibnamefont {Elsaesser}},\ and\ \bibinfo {author}
  {\bibfnamefont {R.~J.~D.}\ \bibnamefont {Miller}},\ }\bibfield  {title}
  {\enquote {\bibinfo {title} {Temperature dependence of the two-dimensional
  infrared spectrum of liquid \uppercase{H}$_2$\uppercase{O}},}\ }\href
  {https://doi.org/10.1073/pnas.0705792105} {\bibfield  {journal} {\bibinfo
  {journal} {Proceedings of the National Academy of Sciences}\ }\textbf
  {\bibinfo {volume} {105}},\ \bibinfo {pages} {437--442} (\bibinfo {year}
  {2008})}\BibitemShut {NoStop}%
\bibitem [{\citenamefont {De~Marco}\ \emph {et~al.}(2016)\citenamefont
  {De~Marco}, \citenamefont {Fournier}, \citenamefont {Thämer}, \citenamefont
  {Carpenter},\ and\ \citenamefont {Tokmakoff}}]{Tokmakoff2016H2O}%
  \BibitemOpen
  \bibfield  {author} {\bibinfo {author} {\bibfnamefont {L.}~\bibnamefont
  {De~Marco}}, \bibinfo {author} {\bibfnamefont {J.~A.}\ \bibnamefont
  {Fournier}}, \bibinfo {author} {\bibfnamefont {M.}~\bibnamefont {Thämer}},
  \bibinfo {author} {\bibfnamefont {W.}~\bibnamefont {Carpenter}},\ and\
  \bibinfo {author} {\bibfnamefont {A.}~\bibnamefont {Tokmakoff}},\ }\bibfield
  {title} {\enquote {\bibinfo {title} {Anharmonic exciton dynamics and energy
  dissipation in liquid water from two-dimensional infrared spectroscopy},}\
  }\href {https://doi.org/10.1063/1.4961752} {\bibfield  {journal} {\bibinfo
  {journal} {The Journal of Chemical Physics}\ }\textbf {\bibinfo {volume}
  {145}},\ \bibinfo {pages} {094501} (\bibinfo {year} {2016})},\ \Eprint
  {https://arxiv.org/abs/https://aip.scitation.org/doi/pdf/10.1063/1.4961752}
  {https://aip.scitation.org/doi/pdf/10.1063/1.4961752} \BibitemShut {NoStop}%
\bibitem [{\citenamefont {Carpenter}\ \emph {et~al.}(2017)\citenamefont
  {Carpenter}, \citenamefont {Fournier}, \citenamefont {Biswas}, \citenamefont
  {Voth},\ and\ \citenamefont {Tokmakoff}}]{VothTokmakoff_St-BendJCP2017}%
  \BibitemOpen
  \bibfield  {author} {\bibinfo {author} {\bibfnamefont {W.~B.}\ \bibnamefont
  {Carpenter}}, \bibinfo {author} {\bibfnamefont {J.~A.}\ \bibnamefont
  {Fournier}}, \bibinfo {author} {\bibfnamefont {R.}~\bibnamefont {Biswas}},
  \bibinfo {author} {\bibfnamefont {G.~A.}\ \bibnamefont {Voth}},\ and\
  \bibinfo {author} {\bibfnamefont {A.}~\bibnamefont {Tokmakoff}},\ }\bibfield
  {title} {\enquote {\bibinfo {title} {Delocalization and stretch-bend mixing
  of the \uppercase{HOH} bend in liquid water},}\ }\href
  {https://doi.org/10.1063/1.4987153} {\bibfield  {journal} {\bibinfo
  {journal} {The Journal of Chemical Physics}\ }\textbf {\bibinfo {volume}
  {147}},\ \bibinfo {pages} {084503} (\bibinfo {year} {2017})},\ \Eprint
  {https://arxiv.org/abs/https://doi.org/10.1063/1.4987153}
  {https://doi.org/10.1063/1.4987153} \BibitemShut {NoStop}%
\bibitem [{\citenamefont {Lewis}\ \emph {et~al.}(2022)\citenamefont {Lewis},
  \citenamefont {Dereka}, \citenamefont {Zhang}, \citenamefont {Maginn},\ and\
  \citenamefont {Tokmakoff}}]{Tokmakoff2022}%
  \BibitemOpen
  \bibfield  {author} {\bibinfo {author} {\bibfnamefont {N.~H.~C.}\
  \bibnamefont {Lewis}}, \bibinfo {author} {\bibfnamefont {B.}~\bibnamefont
  {Dereka}}, \bibinfo {author} {\bibfnamefont {Y.}~\bibnamefont {Zhang}},
  \bibinfo {author} {\bibfnamefont {E.~J.}\ \bibnamefont {Maginn}},\ and\
  \bibinfo {author} {\bibfnamefont {A.}~\bibnamefont {Tokmakoff}},\ }\bibfield
  {title} {\enquote {\bibinfo {title} {From networked to isolated: Observing
  water hydrogen bonds in concentrated electrolytes with two-dimensional
  infrared spectroscopy},}\ }\href {https://doi.org/10.1021/acs.jpcb.2c03341}
  {\bibfield  {journal} {\bibinfo  {journal} {The Journal of Physical Chemistry
  B}\ }\textbf {\bibinfo {volume} {126}},\ \bibinfo {pages} {5305--5319}
  (\bibinfo {year} {2022})},\ \bibinfo {note} {pMID: 35829623},\ \Eprint
  {https://arxiv.org/abs/https://doi.org/10.1021/acs.jpcb.2c03341}
  {https://doi.org/10.1021/acs.jpcb.2c03341} \BibitemShut {NoStop}%
\bibitem [{\citenamefont {Chuntonov}, \citenamefont {Kumar},\ and\
  \citenamefont {Kuroda}(2014)}]{Kuroda_BendPCCP2014}%
  \BibitemOpen
  \bibfield  {author} {\bibinfo {author} {\bibfnamefont {L.}~\bibnamefont
  {Chuntonov}}, \bibinfo {author} {\bibfnamefont {R.}~\bibnamefont {Kumar}},\
  and\ \bibinfo {author} {\bibfnamefont {D.~G.}\ \bibnamefont {Kuroda}},\
  }\bibfield  {title} {\enquote {\bibinfo {title} {Non-linear infrared
  spectroscopy of the water bending mode: direct experimental evidence of
  hydration shell reorganization?}}\ }\href
  {https://doi.org/10.1039/C4CP00643G} {\bibfield  {journal} {\bibinfo
  {journal} {Phys. Chem. Chem. Phys.}\ }\textbf {\bibinfo {volume} {16}},\
  \bibinfo {pages} {13172--13181} (\bibinfo {year} {2014})}\BibitemShut
  {NoStop}%
\bibitem [{\citenamefont {Grechko}\ \emph {et~al.}(2018)\citenamefont
  {Grechko}, \citenamefont {Hasegawa}, \citenamefont {D'Angelo}, \citenamefont
  {Ito}, \citenamefont {Turchinovich}, \citenamefont {Nagata},\ and\
  \citenamefont {Bonn}}]{grechko2018}%
  \BibitemOpen
  \bibfield  {author} {\bibinfo {author} {\bibfnamefont {M.}~\bibnamefont
  {Grechko}}, \bibinfo {author} {\bibfnamefont {T.}~\bibnamefont {Hasegawa}},
  \bibinfo {author} {\bibfnamefont {F.}~\bibnamefont {D'Angelo}}, \bibinfo
  {author} {\bibfnamefont {H.}~\bibnamefont {Ito}}, \bibinfo {author}
  {\bibfnamefont {D.}~\bibnamefont {Turchinovich}}, \bibinfo {author}
  {\bibfnamefont {Y.}~\bibnamefont {Nagata}},\ and\ \bibinfo {author}
  {\bibfnamefont {M.}~\bibnamefont {Bonn}},\ }\bibfield  {title} {\enquote
  {\bibinfo {title} {Coupling between intra- and intermolecular motions in
  liquid water revealed by two-dimensional terahertz-infrared-visible
  spectroscopy},}\ }\href {https://doi.org/10.1038/s41467-018-03303-y}
  {\bibfield  {journal} {\bibinfo  {journal} {Nat Commun}\ }\textbf {\bibinfo
  {volume} {9}},\ \bibinfo {pages} {885} (\bibinfo {year} {2018})}\BibitemShut
  {NoStop}%
\bibitem [{\citenamefont {Vietze}\ \emph {et~al.}(2021)\citenamefont {Vietze},
  \citenamefont {Backus}, \citenamefont {Bonn},\ and\ \citenamefont
  {Grechko}}]{Bonn2DTZIFvis2021}%
  \BibitemOpen
  \bibfield  {author} {\bibinfo {author} {\bibfnamefont {L.}~\bibnamefont
  {Vietze}}, \bibinfo {author} {\bibfnamefont {E.~H.~G.}\ \bibnamefont
  {Backus}}, \bibinfo {author} {\bibfnamefont {M.}~\bibnamefont {Bonn}},\ and\
  \bibinfo {author} {\bibfnamefont {M.}~\bibnamefont {Grechko}},\ }\bibfield
  {title} {\enquote {\bibinfo {title} {Distinguishing different excitation
  pathways in two-dimensional \uppercase{T}erahertz-infrared-visible
  spectroscopy},}\ }\href {https://doi.org/10.1063/5.0047918} {\bibfield
  {journal} {\bibinfo  {journal} {The Journal of Chemical Physics}\ }\textbf
  {\bibinfo {volume} {154}},\ \bibinfo {pages} {174201} (\bibinfo {year}
  {2021})},\ \Eprint {https://arxiv.org/abs/https://doi.org/10.1063/5.0047918}
  {https://doi.org/10.1063/5.0047918} \BibitemShut {NoStop}%
\bibitem [{\citenamefont {Begušić}\ and\ \citenamefont
  {Blake}(2023)}]{Begusic2023}%
  \BibitemOpen
  \bibfield  {author} {\bibinfo {author} {\bibfnamefont {T.}~\bibnamefont
  {Begušić}}\ and\ \bibinfo {author} {\bibfnamefont {G.~A.}\ \bibnamefont
  {Blake}},\ }\bibfield  {title} {\enquote {\bibinfo {title} {Two-dimensional
  infrared-\uppercase{R}aman spectroscopy as a probe of water's
  tetrahedrality},}\ }\href {https://doi.org/10.1038/s41467-023-37667-7}
  {\bibfield  {journal} {\bibinfo  {journal} {Nature Communications}\ }\textbf
  {\bibinfo {volume} {14}},\ \bibinfo {pages} {1950} (\bibinfo {year}
  {2023})}\BibitemShut {NoStop}%
\bibitem [{\citenamefont {Hamm}\ and\ \citenamefont
  {Savolainen}(2012)}]{HammTHz2012}%
  \BibitemOpen
  \bibfield  {author} {\bibinfo {author} {\bibfnamefont {P.}~\bibnamefont
  {Hamm}}\ and\ \bibinfo {author} {\bibfnamefont {J.}~\bibnamefont
  {Savolainen}},\ }\bibfield  {title} {\enquote {\bibinfo {title}
  {Two-dimensional-\uppercase{R}aman-\uppercase{T}erahertz spectroscopy of
  water: Theory},}\ }\href {https://doi.org/10.1063/1.3691601} {\bibfield
  {journal} {\bibinfo  {journal} {The Journal of Chemical Physics}\ }\textbf
  {\bibinfo {volume} {136}},\ \bibinfo {pages} {094516} (\bibinfo {year}
  {2012})},\ \Eprint {https://arxiv.org/abs/https://doi.org/10.1063/1.3691601}
  {https://doi.org/10.1063/1.3691601} \BibitemShut {NoStop}%
\bibitem [{\citenamefont {Hamm}\ \emph {et~al.}(2012)\citenamefont {Hamm},
  \citenamefont {Savolainen}, \citenamefont {Ono},\ and\ \citenamefont
  {Tanimura}}]{HSOT12JCP}%
  \BibitemOpen
  \bibfield  {author} {\bibinfo {author} {\bibfnamefont {P.}~\bibnamefont
  {Hamm}}, \bibinfo {author} {\bibfnamefont {J.}~\bibnamefont {Savolainen}},
  \bibinfo {author} {\bibfnamefont {J.}~\bibnamefont {Ono}},\ and\ \bibinfo
  {author} {\bibfnamefont {Y.}~\bibnamefont {Tanimura}},\ }\bibfield  {title}
  {\enquote {\bibinfo {title} {Note: Inverted time-ordering in
  two-dimensional-\uppercase{R}aman-terahertz spectroscopy of water},}\ }\href
  {https://doi.org/10.1063/1.4729945} {\bibfield  {journal} {\bibinfo
  {journal} {The Journal of Chemical Physics}\ }\textbf {\bibinfo {volume}
  {136}},\ \bibinfo {pages} {236101} (\bibinfo {year} {2012})}\BibitemShut
  {NoStop}%
\bibitem [{\citenamefont {Savolainen}, \citenamefont {Ahmed},\ and\
  \citenamefont {Hamm}(2013)}]{Hamm2013PNAS}%
  \BibitemOpen
  \bibfield  {author} {\bibinfo {author} {\bibfnamefont {J.}~\bibnamefont
  {Savolainen}}, \bibinfo {author} {\bibfnamefont {S.}~\bibnamefont {Ahmed}},\
  and\ \bibinfo {author} {\bibfnamefont {P.}~\bibnamefont {Hamm}},\ }\bibfield
  {title} {\enquote {\bibinfo {title} {Two-dimensional
  \uppercase{R}aman--\uppercase{T}erahertz spectroscopy of water},}\ }\href
  {https://doi.org/10.1073/pnas.1317459110} {\bibfield  {journal} {\bibinfo
  {journal} {Proceedings of the National Academy of Sciences}\ }\textbf
  {\bibinfo {volume} {110}},\ \bibinfo {pages} {20402--20407} (\bibinfo {year}
  {2013})}\BibitemShut {NoStop}%
\bibitem [{\citenamefont {Hamm}(2014)}]{hamm2014}%
  \BibitemOpen
  \bibfield  {author} {\bibinfo {author} {\bibfnamefont {P.}~\bibnamefont
  {Hamm}},\ }\bibfield  {title} {\enquote {\bibinfo {title}
  {{{2D-\uppercase{R}aman-\uppercase{TH}z}} spectroscopy: {{A}} sensitive test
  of polarizable water models},}\ }\href {https://doi.org/10.1063/1.4901216}
  {\bibfield  {journal} {\bibinfo  {journal} {The Journal of Chemical Physics}\
  }\textbf {\bibinfo {volume} {141}},\ \bibinfo {pages} {184201} (\bibinfo
  {year} {2014})}\BibitemShut {NoStop}%
\bibitem [{\citenamefont {Abraham}\ \emph {et~al.}(2015)\citenamefont
  {Abraham}, \citenamefont {Murtola}, \citenamefont {Schulz}, \citenamefont
  {Páll}, \citenamefont {Smith}, \citenamefont {Hess},\ and\ \citenamefont
  {Lindahl}}]{ABRAHAM201519}%
  \BibitemOpen
  \bibfield  {author} {\bibinfo {author} {\bibfnamefont {M.~J.}\ \bibnamefont
  {Abraham}}, \bibinfo {author} {\bibfnamefont {T.}~\bibnamefont {Murtola}},
  \bibinfo {author} {\bibfnamefont {R.}~\bibnamefont {Schulz}}, \bibinfo
  {author} {\bibfnamefont {S.}~\bibnamefont {Páll}}, \bibinfo {author}
  {\bibfnamefont {J.~C.}\ \bibnamefont {Smith}}, \bibinfo {author}
  {\bibfnamefont {B.}~\bibnamefont {Hess}},\ and\ \bibinfo {author}
  {\bibfnamefont {E.}~\bibnamefont {Lindahl}},\ }\bibfield  {title} {\enquote
  {\bibinfo {title} {Gromacs: High performance molecular simulations through
  multi-level parallelism from laptops to supercomputers},}\ }\href
  {https://doi.org/https://doi.org/10.1016/j.softx.2015.06.001} {\bibfield
  {journal} {\bibinfo  {journal} {SoftwareX}\ }\textbf {\bibinfo {volume}
  {1-2}},\ \bibinfo {pages} {19--25} (\bibinfo {year} {2015})}\BibitemShut
  {NoStop}%
\bibitem [{\citenamefont {Berendsen}, \citenamefont {Grigera},\ and\
  \citenamefont {Straatsma}(1987)}]{doi:10.1021/j100308a038}%
  \BibitemOpen
  \bibfield  {author} {\bibinfo {author} {\bibfnamefont {H.~J.~C.}\
  \bibnamefont {Berendsen}}, \bibinfo {author} {\bibfnamefont {J.~R.}\
  \bibnamefont {Grigera}},\ and\ \bibinfo {author} {\bibfnamefont {T.~P.}\
  \bibnamefont {Straatsma}},\ }\bibfield  {title} {\enquote {\bibinfo {title}
  {The missing term in effective pair potentials},}\ }\href
  {https://doi.org/10.1021/j100308a038} {\bibfield  {journal} {\bibinfo
  {journal} {The Journal of Physical Chemistry}\ }\textbf {\bibinfo {volume}
  {91}},\ \bibinfo {pages} {6269--6271} (\bibinfo {year} {1987})},\ \Eprint
  {https://arxiv.org/abs/https://doi.org/10.1021/j100308a038}
  {https://doi.org/10.1021/j100308a038} \BibitemShut {NoStop}%
\bibitem [{GRO(2025)}]{GROMACS2025Manual}%
  \BibitemOpen
  \href {https://doi.org/10.5281/zenodo.16992569} {\enquote {\bibinfo {title}
  {Gromacs 2025.3 manual},}\ } (\bibinfo {year} {2025})\BibitemShut {NoStop}%
\bibitem [{\citenamefont {{Ferguson}}(1995)}]{1995JCoCh..16..501F}%
  \BibitemOpen
  \bibfield  {author} {\bibinfo {author} {\bibfnamefont {D.~M.}\ \bibnamefont
  {{Ferguson}}},\ }\bibfield  {title} {\enquote {\bibinfo {title}
  {{Parameterization and evaluation of a flexible water model}},}\ }\href
  {https://doi.org/10.1002/jcc.540160413} {\bibfield  {journal} {\bibinfo
  {journal} {Journal of Computational Chemistry}\ }\textbf {\bibinfo {volume}
  {16}},\ \bibinfo {pages} {501--511} (\bibinfo {year} {1995})}\BibitemShut
  {NoStop}%
\bibitem [{\citenamefont {Hasegawa}\ and\ \citenamefont
  {Tanimura}(2011)}]{HT11JPCB}%
  \BibitemOpen
  \bibfield  {author} {\bibinfo {author} {\bibfnamefont {T.}~\bibnamefont
  {Hasegawa}}\ and\ \bibinfo {author} {\bibfnamefont {Y.}~\bibnamefont
  {Tanimura}},\ }\bibfield  {title} {\enquote {\bibinfo {title} {A polarizable
  water model for intramolecular and intermolecular vibrational
  spectroscopies},}\ }\href {https://doi.org/10.1021/jp111308f} {\bibfield
  {journal} {\bibinfo  {journal} {The Journal of Physical Chemistry B}\
  }\textbf {\bibinfo {volume} {115}},\ \bibinfo {pages} {5545--5553} (\bibinfo
  {year} {2011})}\BibitemShut {NoStop}%
\bibitem [{\citenamefont {Babin}, \citenamefont {Leforestier},\ and\
  \citenamefont {Paesani}(2013)}]{Babin2013_MBpol_I}%
  \BibitemOpen
  \bibfield  {author} {\bibinfo {author} {\bibfnamefont {V.}~\bibnamefont
  {Babin}}, \bibinfo {author} {\bibfnamefont {C.}~\bibnamefont {Leforestier}},\
  and\ \bibinfo {author} {\bibfnamefont {F.}~\bibnamefont {Paesani}},\
  }\bibfield  {title} {\enquote {\bibinfo {title} {Development of a
  "first-principles" water potential with flexible monomers: Dimer potential
  energy surface, vrt spectrum, and second virial coefficient},}\ }\href
  {https://doi.org/10.1021/ct400863t} {\bibfield  {journal} {\bibinfo
  {journal} {Journal of Chemical Theory and Computation}\ }\textbf {\bibinfo
  {volume} {9}},\ \bibinfo {pages} {5395--5403} (\bibinfo {year}
  {2013})}\BibitemShut {NoStop}%
\bibitem [{\citenamefont {Babin}, \citenamefont {Medders},\ and\ \citenamefont
  {Paesani}(2014)}]{Babin2014_MBpol_II}%
  \BibitemOpen
  \bibfield  {author} {\bibinfo {author} {\bibfnamefont {V.}~\bibnamefont
  {Babin}}, \bibinfo {author} {\bibfnamefont {G.~R.}\ \bibnamefont {Medders}},\
  and\ \bibinfo {author} {\bibfnamefont {F.}~\bibnamefont {Paesani}},\
  }\bibfield  {title} {\enquote {\bibinfo {title} {Development of a
  "first-principles" water potential with flexible monomers. ii. trimer
  potential energy surface, third virial coefficient, and small clusters},}\
  }\href {https://doi.org/10.1021/ct500079y} {\bibfield  {journal} {\bibinfo
  {journal} {Journal of Chemical Theory and Computation}\ }\textbf {\bibinfo
  {volume} {10}},\ \bibinfo {pages} {1599--1607} (\bibinfo {year}
  {2014})}\BibitemShut {NoStop}%
\bibitem [{\citenamefont {Medders}, \citenamefont {Babin},\ and\ \citenamefont
  {Paesani}(2014)}]{Medders2014_MBpol_III}%
  \BibitemOpen
  \bibfield  {author} {\bibinfo {author} {\bibfnamefont {G.~R.}\ \bibnamefont
  {Medders}}, \bibinfo {author} {\bibfnamefont {V.}~\bibnamefont {Babin}},\
  and\ \bibinfo {author} {\bibfnamefont {F.}~\bibnamefont {Paesani}},\
  }\bibfield  {title} {\enquote {\bibinfo {title} {Development of a
  "first-principles" water potential with flexible monomers. iii. liquid phase
  properties},}\ }\href {https://doi.org/10.1021/ct5004115} {\bibfield
  {journal} {\bibinfo  {journal} {Journal of Chemical Theory and Computation}\
  }\textbf {\bibinfo {volume} {10}},\ \bibinfo {pages} {2906--2910} (\bibinfo
  {year} {2014})}\BibitemShut {NoStop}%
\bibitem [{\citenamefont {Maréchal}(2011)}]{IRexp2011}%
  \BibitemOpen
  \bibfield  {author} {\bibinfo {author} {\bibfnamefont {Y.}~\bibnamefont
  {Maréchal}},\ }\bibfield  {title} {\enquote {\bibinfo {title} {The molecular
  structure of liquid water delivered by absorption spectroscopy in the whole
  \uppercase{IR} region completed with thermodynamics data},}\ }\href
  {https://doi.org/https://doi.org/10.1016/j.molstruc.2011.07.054} {\bibfield
  {journal} {\bibinfo  {journal} {Journal of Molecular Structure}\ }\textbf
  {\bibinfo {volume} {1004}},\ \bibinfo {pages} {146--155} (\bibinfo {year}
  {2011})}\BibitemShut {NoStop}%
\bibitem [{\citenamefont {Wang}, \citenamefont {Venkatesh},\ and\ \citenamefont
  {Judd}(1993)}]{NIPS1993_43fa7f58}%
  \BibitemOpen
  \bibfield  {author} {\bibinfo {author} {\bibfnamefont {C.}~\bibnamefont
  {Wang}}, \bibinfo {author} {\bibfnamefont {S.}~\bibnamefont {Venkatesh}},\
  and\ \bibinfo {author} {\bibfnamefont {J.}~\bibnamefont {Judd}},\ }\bibfield
  {title} {\enquote {\bibinfo {title} {Optimal stopping and effective machine
  complexity in learning},}\ }in\ \href
  {https://proceedings.neurips.cc/paper_files/paper/1993/file/43fa7f58b7eac7ac872209342e62e8f1-Paper.pdf}
  {\emph {\bibinfo {booktitle} {Advances in Neural Information Processing
  Systems}}},\ Vol.~\bibinfo {volume} {6},\ \bibinfo {editor} {edited by\
  \bibinfo {editor} {\bibfnamefont {J.}~\bibnamefont {Cowan}}, \bibinfo
  {editor} {\bibfnamefont {G.}~\bibnamefont {Tesauro}},\ and\ \bibinfo {editor}
  {\bibfnamefont {J.}~\bibnamefont {Alspector}}}\ (\bibinfo  {publisher}
  {Morgan-Kaufmann},\ \bibinfo {year} {1993})\BibitemShut {NoStop}%
\end{thebibliography}%

\end{document}